\documentclass[prx,aps,twocolumn,notitlepage,superscriptaddress,showpacs,nofootinbib]{revtex4-2}

\usepackage{enumerate,appendix}
\usepackage{amsmath,amsthm, amssymb,commath}
\usepackage{color,calc,graphicx}
\usepackage[dvipsnames,svgnames,table,cmyk]{xcolor}
\usepackage[colorlinks]{hyperref}
\usepackage{optidef}
\hypersetup{
	colorlinks = true,
	urlcolor = {blue},
	citecolor = {blue},
	linkcolor= {blue}
}

\usepackage{graphicx}
\usepackage{latexsym}
\usepackage{bbm}

\usepackage[charter,cal=cmcal,sfscaled=false]{mathdesign}
\usepackage{booktabs}
\usepackage{multirow} 
\usepackage{dcolumn}
\usepackage{mathrsfs}

\def \be {\begin{equation}}
\def \ee {\end{equation}}

\newcommand{\tr}{\mathrm{Tr}}
\newcommand{\Tr}{\mathrm{Tr}}

\def \Re{\mathrm{Re}\,}

\def \del{\partial}

\def \cX{{\cal X}}

\def \sofc2{{\cal S}({\mathbb C}^2)}

\def\>{\rangle}
\def\<{\langle}

\newtheorem{theorem}{Theorem}
\newtheorem{lemma}[theorem]{Lemma}
\newtheorem{proposition}[theorem]{Proposition}
\newtheorem{corollary}[theorem]{Corollary}

\def\Label#1{\label{#1}\ [\ \text{#1}\ ]\ }
\def\Label{\label}

\begin{document}


\title{Multiparameter quantum metrology for general and adaptive strategies}

\author{Masahito Hayashi}
\email{hmasahito@cuhk.edu.cn}
\affiliation{School of Data Science, The Chinese University of Hong Kong,
Shenzhen, Longgang District, Shenzhen, 518172, China}
\affiliation{International Quantum Academy, Futian District, Shenzhen 518048, China}
\affiliation{Graduate School of Mathematics, Nagoya University, Nagoya, 464-8602, Japan}
\author{Yingkai Ouyang}
\email{y.ouyang@sheffield.ac.uk}
\affiliation{School of Mathematical and Physical Sciences, University of Sheffield, Sheffield, S3 7RH, United Kingdom}

\begin{abstract}  
We develop a conic-programming framework for multiparameter estimation of quantum processes under different classes of sensing strategies. Estimation precision depends on the choice of input probe state and measurement and, when a channel is used multiple times, on the arrangement of channel uses and intermediate controls. We specify different strategy classes through linear normalization constraints on generalized measurements, and formulate the local estimation problem using linear constraints and a linear objective function.
We show that the separable-cone program gives the exact precision limit under the local estimation conditions considered here, while cone relaxations yield Cramér–Rao-type lower bounds. For $N$ uses of an identical quantum channel, we prove an exact $1/N$ scaling law for an RLD-based positive-cone lower bound, even for general process-matrix testers that allow indefinite causal order. Under the stated conditions on the channel and its parameter dependence, the minimum weighted mean-square error also scales as $1/N$. This extends an RLD-based restriction on Heisenberg scaling from adaptive strategies to this broader class.
For fixed-order adaptive strategies, we show that each process-side conic bound equals the corresponding state-estimation bound optimized over sensor combs. We establish this equality by constructing a sensor comb from the process-side normalization operator. A noisy three-parameter field-sensing model illustrates the framework.
\end{abstract}

\maketitle

\section{Introduction}
Estimating physical parameters is a fundamental problem in science.
Parameters of interest include time, length, acceleration, and the
magnitudes and directions of electric, magnetic, and gravitational
fields.  Sensors, which are ubiquitous in modern life, estimate these
physical parameters, and one of the principal measures of their
performance is precision.  Quantum sensors exploit quantum systems and
quantum resources to improve estimation precision beyond what is
achievable with classical sensing strategies
\cite{toth2014quantum,sidhu2020geometric}.  Optimizing the precision of
quantum sensors is, however, highly nontrivial, particularly in noisy
and multiparameter settings \cite{escher2011general,D_and_D_2020}.
Quantum metrology provides the mathematical framework for studying
these problems.

A fundamental task in quantum metrology is channel estimation, where
one estimates physical parameters embedded within a quantum channel.
Optimal channel estimation involves finding an input probe state, an
arrangement of the channel uses and intermediate controls, and a final
measurement that jointly optimize the estimation precision
\cite{Imai,Hayashi11,zhou2021asymptotic}.  The optimization is especially
delicate in multiparameter problems because measurements that are
optimal for different parameters can be incompatible
\cite{PhysRevLett.126.120503,D_and_D_2020,HO}.

Different arrangements of the channel uses lead to different strategy
classes.  Parallel strategies use the channels without intermediate
adaptive controls, whereas adaptive strategies allow later operations
to depend coherently on earlier outputs.  Quantum networks and quantum
combs provide a systematic representation of these higher-order
strategies \cite{CDP08,Chiribella_2008,CDP09}.  The choice of strategy
class can change the attainable metrological precision, and strict
hierarchies among parallel, adaptive, and more general strategies have
been identified \cite{LHYY,KGAD,LHYY24}.

More general process strategies can also be considered, including
those that permit channel uses without a predetermined causal order.
The quantum switch is one example and has exhibited metrological
advantages in particular estimation tasks \cite{ZYC}.  Comparisons between strategy classes require their
admissibility conditions and resources to be specified precisely
\cite{Mothe2024}.

Several optimization frameworks have recently been developed for
quantum metrology with general channel-use strategies.  Semidefinite
and higher-order-process formulations have clarified the optimization
of parallel, adaptive, and more general strategies
\cite{LHYY,KGAD,LHYY24}.  Quantum-comb and tensor-network methods have
also enabled the analysis of adaptive protocols with multiple channel
uses \cite{Kurd_2025,Liu2024efficienttensor}.  Separately, conic
programming unifies several Cram\'er--Rao-type bounds for multiparameter
quantum-state and channel estimation and provides a systematic route
to optimizing the probe \cite{HO,Hayashi2024}.  These developments
motivate a formulation that specifies a process strategy by its linear
normalization conditions and relates the resulting process-side conic
program directly to a state-estimation problem.

Here, we present such a conic-programming approach to multiparameter
quantum metrology of quantum processes.  A strategy class is specified
by linear constraints on the sum of the outcome operators of a
generalized measurement, called its normalization operator.  This description accommodates adaptive fixed-order
strategies and, with the corresponding linear conditions, more general
process strategies, including strategies not restricted to a fixed
causal order.  Each conic program minimizes a linear objective function
of a block-operator variable.  The variable is subject to linear
normalization and local-unbiasedness constraints, together with a conic
constraint inherited from the conic formulation of quantum-state
estimation.

We write $N$ for the total number of channel uses. We establish three groups of results. First, for generalized measurements satisfying
prescribed linear normalization conditions, the optimal value of the separable-cone
program equals the minimum weighted mean-square error over locally unbiased estimators,
while larger cones give Cram\'er--Rao-type lower bounds. Second, for repeated uses of an identical quantum channel, we show that the positive-cone lower bound obeys an exact $1/N$ law even for general process-matrix testers, including those allowing indefinite causal order. These testers are normalized against quantum channels chosen independently in the different slots. Under the rank and support conditions of Corollary~\ref{cor:indef-operational-SQL}, the minimum weighted mean-square error over locally unbiased estimators is bounded above and below by positive multiples of $1/N$, thereby excluding Heisenberg scaling for these channel models. Third, for fixed-order adaptive strategies, the
optimal value of each process-side conic program ($h=1,\ldots,5$) equals the corresponding
state-estimation bound minimized over sensor combs. The proof constructs a sensor comb
from a process-side normalization operator.

Previous RLD-based analyses have established restrictions on the precision scaling of parallel channel-estimation strategies~\cite{Hayashi11} and sequential or adaptive strategies~\cite{KatariyaWilde2021}. Under suitable finiteness and regularity conditions on the channel RLD information, these restrictions rule out Heisenberg scaling. For a nontrivial noiseless unitary channel, the Choi operator has rank one and its parameter derivative extends outside its support, so the corresponding channel RLD Fisher information is not finite. Such channels can achieve Heisenberg scaling with entangled input probes, but they do not satisfy the full-rank or support conditions of Corollary~\ref{cor:indef-operational-SQL}. Here we extend this type of restriction to general process-matrix testers, including strategies with indefinite causal order. Our more precise multiparameter result identifies the full set of real RLD-type information matrices associated with the allowed $N$-use normalization operators: this set is exactly $N$ times the one-use set. The resulting positive-cone bound therefore satisfies an exact $1/N$ identity.

Recent studies have addressed related estimation problems using conic or semidefinite programming. Zhou and Zhang~\cite{ZhouZhang2026} formulated multiparameter estimation with general weight matrices using quantum testers and conic programming, including strategies with indefinite causal order. Tanggara \emph{et al.}~\cite{TanggaraEtAl2026} derived semidefinite-programming lower bounds for sequential sensing, including multiparameter problems and temporally correlated quantum processes. Wei, Li, and Pang~\cite{WeiLiPang2026} investigated asymptotic quantum Fisher information limits for single-parameter estimation with general indefinite causal order.

The present work emphasizes two exact relations for multiparameter estimation, formulated through conic optimization with a general weight matrix $G$. First, for general process-matrix testers, we determine the entire set of real RLD-type information matrices generated by $N$ identical channel uses: it is exactly $N$ times the one-use set. For $G>0$ and under the stated rank and support conditions, this yields an exact $1/N$ identity for the RLD-based positive-cone bound and a lower bound on the weighted mean-square error, even when indefinite causal order is allowed. Second, for fixed-order adaptive strategies, we prove that each of the five process-side conic bounds ($h=1,\ldots,5$) equals the corresponding state-estimation bound minimized over sensor combs. The equality holds for general positive-semidefinite weight matrices and follows from an explicit correspondence between process-side normalization operators and sensor combs. Thus the second result is an exact reduction of the optimization problem, beyond a conic formulation or a semidefinite lower bound alone.

We illustrate the framework numerically with a noisy
three-parameter field-sensing model and report the available
process-side conic bounds.  The numerical comparison is kept separate
from the general conic equivalences proved in the main theoretical
sections.

Section~\ref{sec:estimation} reviews conic bounds for multiparameter quantum-state
estimation. Section~\ref{sec:general-strategies} develops the generalized-measurement
conic programs and proves the scaling results for repeated channel uses.
Section~\ref{sec:adaptive-strategies} treats fixed-order adaptive strategies and
establishes the reduction to sensor-comb optimization. The remaining sections present the
field-sensing application, numerical results, and discussion.

\section{Quantum state estimation}
\label{sec:estimation}
\subsection{Review of existing bounds}
Here, we consider the conic programming framework for quantum state estimation \cite{HO}.
In quantum state estimation, we are given copies of an unknown state $\rho_{\bm{\theta}_0}$ from the {\em quantum model} $\{\rho_{\bm{\theta}} : \bm{\theta}=(\theta^1,\ldots,\theta^d)\in\Theta\subseteq\mathbb R^d\}$, and seek an estimator of the true parameter $\bm{\theta}_0$.
We assume that the family $\bm\theta\mapsto\rho_{\bm\theta}$
is differentiable on $\Theta$, and write
$\rho:=\rho_{\bm\theta_0}$ when evaluating conic programs
at the reference parameter $\bm\theta_0$.

We describe a measurement using a set of positive operators $\Pi = \{ \Pi_x : x\in \mathcal X\}$ labeled by a set $\mathcal X$, where the completeness condition $\sum_{x \in \mathcal X}\Pi_x = I $ holds.
Born's rule assigns the classical label $x$ the probability $p_\theta(x) = \tr(\Pi_x \rho_{\theta})$ when the measurement $\Pi$ is applied to the quantum state $\rho_\theta$.
The notation $\mathbb E_{\bm{\theta}}[f(x) |\Pi]$ denotes the expectation of a function $f$ of the classical label $x$ under the probability distribution determined by Born's rule.

Given a measurement $\Pi$ and an estimator $\hat{\bm{\theta}}$ that depends on the classical label $x$, we denote  
$\hat{\Pi}=(\Pi , \hat{{\bm{\theta}}})$ as an {\it estimator}. 
When the true parameter $\theta_0$ is equal to $\theta$, we define the mean-square error (MSE) matrix for the estimator $\hat{\Pi}$ as  
\begin{align}\nonumber
V_{{\bm{\theta}}}[\hat{\Pi}]
&=
\sum_{i,j=1}^d |i\>\<j|
 \mathbb E_{\bm{\theta}}\big[({\hat{\theta}^i}(x)-\theta^i)({\hat{\theta}^j}(x)-\theta^j)|\Pi\big] \notag.
\end{align}
In multiparameter quantum metrology, one aims to 
find an estimator $\hat{\Pi}=(\Pi , \hat{{\bm{\theta}}})$ that minimizes $\tr G V_{{\bm{\theta}}}[\hat{\Pi}]$, where $G$ is a weight matrix that quantifies the relative importance of the different parameters. Mathematically,
$G = \sum_{i,j=1}^d g_{i,j}|i\>\<j|$ is a size $d$ real positive semidefinite matrix. 


At a fixed parameter value $\bm{\theta}_0$, an estimator is locally unbiased if its mean equals $\bm{\theta}_0$ and the derivative of its mean with respect to the parameter is the identity matrix.  The first requirement is, for each $i$,
\begin{align}
\mathbb E_{\bm{\theta}_0}\big[{\hat{\theta}^i}(x)|\Pi\big]&
=\sum_{x\in\cX} {\hat{\theta}^i}(x) \tr{\big[\rho_{{\bm{\theta}_0}}\Pi_x\big]}
=\theta_0^i \label{MK}.
\end{align}
For the second, independent requirement, let $D_j:= \left.\frac{\partial}{\partial \theta^j}\rho_\theta\right|_{\bm{\theta}_0}$.  The derivative of the estimator mean at $\bm{\theta}_0$ must satisfy
\begin{align}
\left.\frac{\del}{\del\theta^j}\mathbb E_{\bm{\theta}}\big[{\hat{\theta}^i}(x)|\Pi\big]\right|_{\bm{\theta}=\bm{\theta}_0}&=
\sum_{x\in\cX} {\hat{\theta}^i}(x)\tr{   D_j \Pi_x}
=\delta_i^j \label{M1}.
\end{align}
Equations~\eqref{MK} and \eqref{M1} are the two separate requirements of local unbiasedness at $\bm{\theta}_0$; equality of the mean at one parameter value alone does not imply its derivative condition.

For the conic representation at $\bm\theta_0$, define the centered
estimator values $u^i(x):=\hat\theta^i(x)-\theta_0^i$.  For a
locally unbiased estimator, Eqs.~\eqref{MK} and \eqref{M1} give
\begin{align}
\mathbb E_{\bm\theta_0}[u^i(x)|\Pi]&=0,&
\sum_x u^i(x)\Tr(D_j\Pi_x)&=\delta_{ij}.
\label{eq:state-centered-constraints}
\end{align}
The derivative equality uses $\Tr D_j=0$; the MSE at
$\bm\theta_0$ is the second moment of $u$, not of the uncentered
estimate $\hat\theta$.

The fundamental precision limit \cite{HO} is
\be\label{qcrbound}
C_{\bm{\theta}}[G]:=
\min_{\hat{\Pi}\mathrm{\,:l.u.at\,}{\bm{\theta}}}
\Tr{ \big[G V_{\bm{\theta}}[\hat{\Pi}]\big]}, \notag
\ee
where `l.u.~at ${\bm{\theta}}$' indicates minimization over all locally unbiased estimators. 
An estimator satisfying only Eq.~\eqref{M1} can be shifted to satisfy Eq.~\eqref{MK} without changing the minimum.  The definition of $C_{\bm{\theta}}[G]$ may therefore impose only Eq.~\eqref{M1}.

Cram\'er--Rao (CR) type bounds~\cite{HO} give lower bounds on the weighted trace of the MSE matrix. The tight CR bound $C_{\bm{\theta}}[G]$ is the infimum of the weighted MSE over locally unbiased estimators based on a measurement of one copy \cite{HO}. A one-copy locally unbiased estimator may be applied independently to repeated copies and its outputs averaged, reducing its weighted MSE by a factor of the number of copies.
Collective measurements across many copies lead to a different asymptotic optimization,
characterized by the Holevo--Nagaoka (HN)
bound~\cite{holevo,nagaoka89,HM,Albarelli2019_PRL,sidhu2020tight}. In the many-copy limit, the weighted MSE scales inversely with the number of copies.
Collective measurements can achieve a smaller leading coefficient than independently repeated
one-copy measurements~\cite{HO}.

A semidefinite program (SDP) efficiently approximates the tight CR bound \cite{HO}, with the approximation's precision increasing with SDP's complexity. 
Other CR-type bounds are significantly simpler SDPs, such as the Nagaoka-Hayashi (NH) bound, the HN bound, and the symmetric logarithmic derivative (SLD) bound.
We clarified in Ref.~\cite{HO} that these disparate CR-type bounds are unified under a common conic programming framework. 

These conic programs use an operator on $\mathcal R \otimes \mathcal H$ as the optimization variable,
where $\mathcal R = \mathbb R^{d+1}$ is spanned by basis vectors $|0\>,|1\>,\dots, |d\>$
and $\cal H$ is the Hilbert space for our quantum state.
For an estimator $\hat\Pi=(\Pi,\hat\theta)$ at $\bm\theta_0$,
use the centered values $u^i(x)=\hat\theta^i(x)-\theta_0^i$ to define
an operator on $\mathcal R\otimes\mathcal H$:
\begin{align}
|w_x\rangle&:=|0\rangle+\sum_{i=1}^d u^i(x)|i\rangle,\notag\\
X(\hat\Pi)&:=\sum_{x\in\mathcal X}
|w_x\rangle\langle w_x|\otimes\Pi_x.
\Label{EQ3}
\end{align}

These conic programs share a common objective function, given by
\begin{align}
\Tr{G V_{\bm{\theta}_0}[\hat{\Pi}]} = \Tr ( G \otimes \rho_{\bm\theta_0}) X(\hat\Pi) \Label{o1}.
\end{align}
The completeness condition ${\sum_{x \in \mathcal X} \Pi_x = I_{\mathcal H}}$ implies that 
\begin{align}
\Tr_{{\cal R}} (|0\rangle \langle 0| \otimes I_{{\cal H}})  
X(\hat \Pi)&=
\sum_{x \in \mathcal X} \Pi_x =
I_{{\cal H}} \Label{c1},
\end{align}
where $\Tr_{{\cal R}}$ denotes the partial trace on system $\mathcal R$.
Hence, \eqref{c1} rewrites the completeness condition $\sum_{x \in \mathcal X} \Pi_x = I_{\mathcal H}$
as a linear constraint on $X$.
The derivative condition \eqref{M1} translates to the same linear constraint, since $\Tr D_j=0$:
\begin{align}
\Tr (
\frac{1}{2}(|0\rangle \langle i| +|i\rangle \langle 0 |)
\otimes D_j )  X(\hat \Pi) &=\delta_{i,j}  \Label{c2} , 
\end{align}
which is also a linear constraint on $X$.

The operator $X(\hat \Pi)$ has a tensor product structure; 
namely, $X(\hat \Pi)$ is an element of a cone generated by separable states on $\mathcal R \otimes \mathcal H$.
This cone $\mathcal S^1$ is a separable cone, which is
the convex hull of operators that are 
a tensor product of a real positive semidefinite matrix on $\mathcal R$ and 
a complex positive semidefinite operators on $\mathcal H$ with bounded norm.
Hence, we consider the minimization
\begin{align}
S_1:= \min_{X \in \mathcal S^1}
\{\Tr ( G \otimes \rho) X |
\eqref{c1},\eqref{c2} \hbox{ hold.}
\},
\notag
\end{align} 
which equals the tight CR type bound \cite{HO}.  Although only the
derivative condition is imposed on $X$, any recovered estimator can
be shifted to have mean $\bm\theta_0$; this shift leaves its derivative
unchanged and cannot increase the weighted MSE.
Minimization of 
$\Tr ( G \otimes \rho) X $ subject to \eqref{c1}, \eqref{c2} and over suitable cones that contain $\mathcal S^1$ allow us to recover the NH bound, the HN bound and the SLD bound. 
We motivate the constructions of these cones by taking note of some additional properties of the matrix $X(\hat \Pi)$.

First, for some real $A_{x,j,k}$ where $A_{x,j,k} = A_{x,k,j}$, we can write
\begin{align}
   &\sum_x \sum_{j,k} A_{x,j,k} |j\>\<k| \otimes \Pi_x\notag\\
=& \sum_{j,k} |j\>\<k| \otimes \left( \sum_{x} A_{x,j,k} \Pi_x \right) = \sum_{j,k} |j\>\<k| \otimes X_{j,k},
\end{align}
where we set $X_{j,k} = \sum_{x} A_{x,j,k} \Pi_x $. Note then that $X_{j,k}$ inherits its self-adjointness from the self-adjointness of $\Pi_x$.
Also $X_{j,k} = X_{k,j}$ because $A_{x,j,k} = A_{x,k,j}$.
Therefore, we introduce the vector space
$\mathcal B$ to have the following structure:
\begin{align}
{\cal B} &:=\Big\{\sum_{j=0}^d \sum_{k=0}^d
|k\rangle \langle j| \otimes X_{k,j} \Big|
X_{k,j} \in {\cal B}_{\rm sa}({\cal H}), X_{k,j}=X_{j,k} \Big\},\notag
\end{align}
where ${\cal B}_{\rm sa}({\cal H})$ denotes the set of self-adjoint (Hermitian) matrices on $\mathcal H$ with bounded norm.

We extend the space ${\cal B}$ to 
\begin{align}
{\cal B}''&:=\Big\{\sum_{j=0}^d \sum_{k=0}^d
|k\rangle \langle j| \otimes X_{k,j} \Big|
X_{k,0} \in {\cal B}_{\rm sa}({\cal H}), X_{k,j}=(X_{j,k})^\dagger \Big\} ,\notag
\end{align}
so as to allow blocks from the first to $d$th columns to be not necessarily Hermitian. To ensure that elements in $\cal B''$ remain Hermitian, we require that
the blocks $(j,k)$ and $(k,j)$ are related by an adjoint operator $\dagger$.

Based on the spaces $\cal B$ and $\cal B''$ we define cones over which we optimize $X$.
Denote ${\cal R}_C = \mathbb C^{d+1}$.
The subscript $C$ denotes that $\mathcal R_C$ is the complexification of the real space $\mathcal R = \mathbb R^{d+1}$.
Then we consider the cone 
\begin{align}
\mathcal S^2&:=\{X \in {\cal B}| \langle v| X|v\rangle \ge 0 \hbox{ for all } |v\> \in  \cal R_C \otimes {\cal H}\} .\notag
\end{align}

Let $\Gamma_{\cal R}$ denote partial transpose on the parameter
factor ${\cal R}_C$ in its standard basis.  We define the PPT cone by
\begin{align}
{\cal S}({\cal R}_C\otimes{\cal H})_{\rm PPT}
:=\{X=X^\dagger:\ X\geq0,\ X^{\Gamma_{\cal R}}\geq0\}.
\label{eq:state-PPT-cone}
\end{align}
The cone $\mathcal S^3$ is
${\cal S}({\cal R}_C\otimes{\cal H})_{\rm PPT}\cap{\cal B}''$.
Likewise, we define the set ${\cal S}({\cal R}_C\otimes {\cal H})_{\rm P}$ as the set of 
positive semi-definite self-adjoint operators on ${\cal R}_C \otimes {\cal H}$,
and define the cone
$\mathcal S^4$ as ${{\cal S}({\cal R}_C\otimes {\cal H})_{\rm P} \cap {\cal B}''}$.
Then, for $h=1,2,3,4$,
we define the conic programs
\begin{align}
S_h[\rho,(D_j) ]:=
 \min_{X \in {\cal S}^h}
\{\Tr ( G \otimes \rho) X |
\eqref{c1},\eqref{c2} \hbox{ hold.}
\}
\Label{o1-T}.
\end{align}
The relation
$\mathcal S^1\subseteq\mathcal S^2
\subseteq\mathcal S^3\subseteq\mathcal S^4$
follows because every operator in $\mathcal S^2$ has real symmetric
parameter blocks and is therefore invariant under
$\Gamma_{\cal R}$, while Eq.~\eqref{eq:state-PPT-cone}
requires positivity before and after partial transpose.  These
inclusions imply 
\begin{align}
S_1[\rho,(D_j) ] \ge S_2 [\rho,(D_j) ]\ge S_3 [\rho,(D_j) ]\ge S_4[\rho,(D_j) ].\label{IN1}
\end{align} 
 
In addition, given a
trace-class self-adjoint operator $T$,
we introduce a linear constraint on $X \in {\cal B}''$ as
\begin{align}
\Tr X( ( |j \rangle \langle i |-  |i \rangle \langle j |)\otimes T )=0\Label{c3}
\end{align}
for $i,j=1,2, \ldots, d$.
We correspondingly define 
${\cal B}_T'':=\{X \in {\cal B}''|
\eqref{c3} \hbox{ holds.}
\}.$
Next, given a density matrix $\rho$ on $\mathcal H$, 
we define ${\cal S}^{5}(\rho)$
as ${\cal S}(\mathcal R_C\otimes {\cal H})_{\rm P}\cap {\cal B}''_\rho$.
We consider the minimization:
\begin{align}
&S_5:= \min_{X \in {\cal S}^5(\rho)}
\{\Tr ( G \otimes \rho) X |
\eqref{c1},\eqref{c2} \hbox{ hold.}
\}.
\notag
\end{align}
The cones $\mathcal S^1,\mathcal S^2,\mathcal S^3,\mathcal S^4$ are independent of $\rho$; only the cone $\mathcal S^5(\rho)$ depends on $\rho$.
The set inclusions 
$ {\cal S}^2
\subset {\cal S}^5(\rho)
\subset {\cal S}^4,$
imply the inequalities 
\begin{align}
S_4 \le S_5 \le S_2. \label{IN2}
\end{align}
$S_h$ depends on the model $(\rho,(D_j)_j)$, namely, the probe state $\rho$ and its partial derivatives $D_j$.  For $h=1,2,3,4,5$, we therefore also write $S_h$ as
$S_h[\rho,(D_j)_j]$
 to emphasize the CR-type bounds' dependence on $\rho$ and $D_j$.
 Here $S_2$ equals the Nagaoka-Hayashi bound (NH bound) studied in Ref.~\cite{Nagaoka-generalization,Hayanoncomm,CSLA}, $S_4$ equals the SLD bound, and
$S_5$ equals the HN bound \cite{HO}.
The SLD bound is attainable in the single-parameter case $d=1$, and the hierarchy of bounds collapses: 
\begin{align}
S_1 = S_2 = S_3 =S_5= S_4.\label{IN3}
\end{align}
The multiparameter case with $d \ge 2$ on the other hand is not so trivial.

\subsection{A further relaxation and the RLD Fisher information}
We next introduce a further cone relaxation, separate from the preceding
family, to examine its connection with the RLD Fisher information.
This additional relaxation will also be used to analyze the scaling of
channel-estimation precision under repeated uses.  The cone
$\mathcal S^4$ requires the blocks $X_{k,0}$ to be Hermitian.
We remove this requirement while retaining positivity of the full block operator.  More precisely, define
\begin{align}
\mathcal S^6
:= {\cal S}({\cal R}_C\otimes{\cal H})_{\rm P}.
\label{eq:S6-cone}
\end{align}
Thus $\mathcal S^6$ allows non-Hermitian blocks $X_{k,0}$,
subject to $X_{0,k}=X_{k,0}^\dagger$, as required by the
self-adjointness of $X$.  With the same objective and the same
constraints \eqref{c1} and \eqref{c2}, set
\begin{align}
S_6[\rho,(D_j)_j]
:=\inf_{X\in\mathcal S^6}
\bigl\{\Tr (G\otimes\rho)X:
\eqref{c1},\eqref{c2}\ \hbox{hold}\bigr\}.
\label{eq:S6-program}
\end{align}
Since $\mathcal S^4\subseteq\mathcal S^6$ and
$\mathcal S^5(\rho)\subseteq\mathcal S^4$, the resulting bound
satisfies
\begin{align}
S_6\le S_4\le S_5\le S_2\le S_1.
\label{eq:S6-hierarchy}
\end{align}
In particular, the single-parameter equality in \eqref{IN3}
does not in general extend to $S_6$.

For clarity, consider a finite-dimensional model with $\rho>0$.
Write the block entries of $X$ as $X_{ij}$ and let
$A_i:=X_{i0}$.  Constraint \eqref{c1} gives $X_{00}=I$;
constraint \eqref{c2} becomes
\begin{align}
\operatorname{Re}\Tr(D_jA_i)=\delta_{ij}.
\label{eq:S6-unbiasedness}
\end{align}
Because $X\ge0$, its Schur complement with respect to $X_{00}$
implies
\begin{align}
[X_{ij}]_{i,j=1}^d\succeq[A_iA_j^\dagger]_{i,j=1}^d.
\label{eq:S6-Schur}
\end{align}
Conversely, any collection $(A_i)_i$ can be completed to a
positive block operator by taking equality in
\eqref{eq:S6-Schur}.  For $G\ge0$, the objective is therefore
minimized, at fixed $(A_i)_i$, by this completion.  It follows that
\begin{align}
S_6[\rho,(D_j)_j]
=\inf_{(A_i)_i}
\left\{\sum_{i,j=1}^d g_{ij}\Tr(\rho A_iA_j^\dagger):
\operatorname{Re}\Tr(D_jA_i)=\delta_{ij}\right\}.
\label{eq:S6-reduced}
\end{align}
Define the RLD logarithmic derivatives by
$L_j^{\rm R}:=\rho^{-1}D_j$ and their Fisher information matrix by
\begin{align}
(J_{\rm R})_{ij}
:=\Tr\bigl(\rho (L_i^{\rm R})^\dagger L_j^{\rm R}\bigr)
=\Tr(D_i\rho^{-1}D_j).
\label{eq:S6-RLD-information}
\end{align}
\begin{theorem}[Evaluation of the positive-cone bound using the RLD Fisher information]
\label{thm:S6-RLD-real-bound}
Let $\mathcal H$ be finite-dimensional, let $\rho>0$ be a
fixed density operator, and let $D_1,\ldots,D_d$ be fixed
Hermitian operators.  Suppose that $G$ is a real symmetric
positive semidefinite matrix and that
$F_{\rm R}:=\operatorname{Re}J_{\rm R}>0$, where $J_{\rm R}$
is defined by \eqref{eq:S6-RLD-information}.  Then
\begin{align}
S_6[\rho,(D_j)_j]
=\Tr\bigl[G(\operatorname{Re}J_{\rm R})^{-1}\bigr].
\label{eq:S6-RLD-real-bound}
\end{align}
The infimum in \eqref{eq:S6-reduced} is attained by
\begin{align}
A_i^*:=\sum_{k=1}^d(F_{\rm R}^{-1})_{ik}L_k^{\rm R},
\qquad i=1,\ldots,d.
\label{eq:S6-RLD-minimizers}
\end{align}
\end{theorem}

\begin{proof}
The minimization in \eqref{eq:S6-reduced} is over the
operators $A_1,\ldots,A_d$, with $\rho$ and $D_1,\ldots,D_d$
fixed.  On the real vector space of complex operators, define
\begin{align}
\langle A,B\rangle_\rho
:=\operatorname{Re}\Tr(A^\dagger\rho B).
\label{eq:S6-real-inner-product}
\end{align}
Since the $D_j$ are Hermitian and $\rho>0$, the constraints
\eqref{eq:S6-unbiasedness} become
\begin{align}
\langle L_j^{\rm R},A_i\rangle_\rho
=\operatorname{Re}\Tr(D_jA_i)=\delta_{ij}.
\label{eq:S6-RLD-inner-constraints}
\end{align}
The Gram matrix of $L_1^{\rm R},\ldots,L_d^{\rm R}$ is
\begin{align}
(F_{\rm R})_{ij}
&=\langle L_i^{\rm R},L_j^{\rm R}\rangle_\rho
=\operatorname{Re}\Tr(D_i\rho^{-1}D_j).
\label{eq:S6-real-Gram}
\end{align}
By $F_{\rm R}>0$, the operators $A_i^*$ in
\eqref{eq:S6-RLD-minimizers} are well defined.  They satisfy
all the constraints because
\begin{align}
\operatorname{Re}\Tr(D_jA_i^*)
&=\sum_{k=1}^d(F_{\rm R}^{-1})_{ik}(F_{\rm R})_{jk}
=\delta_{ij}.
\label{eq:S6-RLD-minimizer-feasible}
\end{align}
To prove optimality, write any feasible family as
$A_i=A_i^*+Z_i$.  Equation~\eqref{eq:S6-RLD-inner-constraints}
gives $\langle L_j^{\rm R},Z_i\rangle_\rho=0$ for all
$i,j$.  Since $A_i^*$ is a real linear combination of the
$L_k^{\rm R}$, all cross terms between $A_i^*$ and $Z_j$
vanish.  In addition,
\begin{align}
\langle A_i^*,A_j^*\rangle_\rho
=(F_{\rm R}^{-1})_{ij}.
\notag
\end{align}
As $G$ is real symmetric, the objective in
\eqref{eq:S6-reduced} is the corresponding weighted sum of
real inner products.  Therefore,
\begin{align}
\sum_{i,j=1}^d g_{ij}\Tr(\rho A_iA_j^\dagger)
&=\Tr(GF_{\rm R}^{-1})
  +\sum_{i,j=1}^d g_{ij}\langle Z_i,Z_j\rangle_\rho
\notag\\
&\ge\Tr(GF_{\rm R}^{-1}).
\label{eq:S6-RLD-optimality}
\end{align}
The inequality follows because $G\ge0$ and
$[\langle Z_i,Z_j\rangle_\rho]_{i,j}$ is a positive
semidefinite real Gram matrix.  Equality holds when
$Z_i=0$ for every $i$, proving
\eqref{eq:S6-RLD-real-bound} and the stated attainment.
\end{proof}

For a single parameter, the value in \eqref{eq:S6-RLD-real-bound}
coincides with the RLD bound.

\section{Channel estimation with general strategies}
\label{sec:general-strategies}

\subsection{Parametric process model}
\label{subsec:parametric-process-model}

We consider multiparameter estimation for a parameter-dependent quantum process with $N$ input-output slots.  For each $k=1,\ldots,N$, let ${\cal H}_{A,k}$ and ${\cal H}_{B,k}$ denote the input and output Hilbert spaces of the $k$-th slot, respectively.  We define the overall input and output spaces by
\begin{align}
{\cal K}_A
&:=
\bigotimes_{k=1}^{N}{\cal H}_{A,k},
&
{\cal K}_B
&:=
\bigotimes_{k=1}^{N}{\cal H}_{B,k}.
\end{align}
We fix the Choi convention here.  For a map
$\Lambda:\mathcal B(\mathcal H_A)\to\mathcal B(\mathcal H_B)$,
let $\mathcal H_{A'}$ be a copy of the input space with the chosen
orthonormal basis $\{|a\rangle\}_a$, and set
\begin{align}
|\Omega_A\rangle&:=\sum_a|a\rangle_A\otimes|a\rangle_{A'},\notag\\
T_\Lambda&:=(\Lambda\otimes\operatorname{id}_{A'})
(|\Omega_A\rangle\langle\Omega_A|).
\label{eq:Choi-convention}
\end{align}
We identify $A'$ with the input slot $A$ in Choi expressions and
order the resulting factors as output followed by input.  A quantum channel is a completely positive trace-preserving (CPTP) linear map. Its Choi
operator satisfies $T_\Lambda\ge0$ and $\Tr_B T_\Lambda=I_A$.  Transposes in link products are taken in fixed, mutually
consistent bases on the connected systems, including intermediate
output spaces.

Let $\bar T_\theta^{(N)}$ be the parameter-dependent Choi operator on ${\cal K}_B\otimes{\cal K}_A$.  The conic programs below depend locally on this family through its value and first derivatives at the true parameter $\theta_0$, which we denote by
\begin{align*}
\bar T
:=
\bar T_{\theta_0}^{(N)},\quad
\bar F_j^{(N)}
:=
\left.
\frac{\partial}{\partial\theta^j}
\bar T_\theta^{(N)}
\right|_{\theta=\theta_0},
\qquad
j=1,\ldots,d.
\end{align*}
The general formulation below depends only on the $N$ input-output slots and on the process family $\{\bar T_\theta^{(N)}\}_\theta$.   

\subsection{A repeated-channel process model}
\label{subsec:repeated-channel-model}

The application section uses the repeated-channel specialization recorded here before the strategy-dependent measurement constraints are introduced.  This model consists of $m$ types of parameter-dependent channels
$\Lambda_\theta^1,\ldots,\Lambda_\theta^m$,
where each channel type is used $n$ times.  The total number of channel slots is therefore
$N=mn$.
The Choi operator $T_\theta^a$ represents $\Lambda_\theta^a$ for each $a=1,\ldots,m$.  We choose the slot labeling so that the $n$ uses of channel type $a$ occupy slots
$(a-1)n+1,\ldots,an$.
Equivalently, if $a(k)$ denotes the channel type assigned to the $k$-th slot, then
$a(k)=a$
for
$k=(a-1)n+1,\ldots,an$.
With this convention, the overall Choi operator is
\begin{align}
\bar T_\theta^{(N)}
=
\bigotimes_{a=1}^{m}
\left(T_\theta^a\right)^{\otimes n}.
\label{eq:repeated-channel-Choi}
\end{align}
Equivalently, writing $\bar T_{k,\theta}:=T_\theta^{a(k)}$ gives
$\bar T_\theta^{(N)}=\bigotimes_{k=1}^{N}\bar T_{k,\theta}$.
The derivative of the overall Choi operator is then
\begin{align}
\bar F_j^{(N)}
=
\sum_{k=1}^{N}\bar F_{j,(k)},
\label{eq:overall-Choi-derivative}
\end{align}
where
\begin{align}
\bar F_{j,(k)}
:=
\left.
\left(
\bigotimes_{u=1}^{k-1}\bar T_{u,\theta}
\right)
\otimes
\frac{\partial\bar T_{k,\theta}}{\partial\theta^j}
\otimes
\left(
\bigotimes_{v=k+1}^{N}\bar T_{v,\theta}
\right)
\right|_{\theta=\theta_0}.
\end{align}
This repeated-channel family is used later in the application, while the general formulation also covers other slot assignments and process families without the tensor-product form in Eq.~\eqref{eq:repeated-channel-Choi}.  We now return to the general $N$-slot process.

\subsection{Generalized measurements and strategy classes}
\label{subsec:generalized-measurements}

We describe a generalized measurement on the $N$-slot process by a family of positive semidefinite operators
\begin{align}
M=\{M_x\}_{x\in{\cal X}}
\end{align}
on ${\cal K}_B\otimes{\cal K}_A$.  Its normalization operator is defined by
\begin{align}
R^{(N)}
:=
\sum_{x\in{\cal X}}M_x.
\label{eq:GPOVM-normalization}
\end{align}
$\bar T_\theta^{(N)}$ assigns outcome $x$ the probability
\begin{align}
p_\theta(x)
:=
\operatorname{Tr}
\left[
\bar T_\theta^{(N)}M_x
\right].
\label{eq:GPOVM-probability}
\end{align}

Let $\mathfrak P$ denote the class of normalized process Choi operators on ${\cal K}_B\otimes{\cal K}_A$ associated with the $N$ input-output slots under consideration.  A family $M=\{M_x\}_{x\in{\cal X}}$ defines a normalized generalized measurement on $\mathfrak P$ when its normalization operator satisfies
\begin{align}
\operatorname{Tr}
\left[
R^{(N)}\Lambda
\right]
=
1
\qquad
\text{for every }\Lambda\in\mathfrak P.
\label{eq:process-normalization}
\end{align}
Indeed, defining $p_\Lambda(x):=\operatorname{Tr}[\Lambda M_x]$, Eq.~\eqref{eq:process-normalization} implies
\begin{align}
\sum_{x\in{\cal X}}p_\Lambda(x)
=
\sum_{x\in{\cal X}}
\operatorname{Tr}[\Lambda M_x]
=
\operatorname{Tr}
\left[
\Lambda R^{(N)}
\right]
=1.
\end{align}

The process and strategy classes considered in this paper impose the overall trace normalization
\begin{align}
\operatorname{Tr}R^{(N)}
=
\dim{\cal K}_B.
\label{AD}
\end{align}
We then introduce a linear map $F$ from the Hermitian operators on ${\cal K}_B\otimes{\cal K}_A$ to a linear space $V$, together with a fixed element $C\in V$, such that
\begin{align}
F\bigl(R^{(N)}\bigr)
=
C
\label{NM}
\end{align}
is sufficient, together with Eq.~\eqref{AD}, to guarantee the normalization condition \eqref{eq:process-normalization}.  In the remainder of this section, positivity of the outcome operators together with Eqs.~\eqref{NM} and \eqref{AD} defines the prescribed strategy class.  The map $F$ and the element $C$ are fixed by the input-output structure of the $N$ slots and by the allowed class of generalized measurements, and they do not depend on the unknown value of $\theta$.

The Choi family $\{\bar T_\theta^{(N)}\}_\theta$ specifies what is estimated,
whereas $F$, $C$, and Eq.~\eqref{AD} specify which generalized measurements
are allowed. Different strategy classes lead to different choices of
$F$ and $C$.

For two channel-use slots $A_1\to B_1$ and $A_2\to B_2$,
we first distinguish the causal-order restrictions on the admissible
testers.  A fixed-order adaptive tester specifies either
$1\prec 2$ or $2\prec 1$ and satisfies the corresponding
recursive comb normalization.  A causally separable tester is a
convex mixture of testers with these definite orders.  A general
two-slot process-matrix tester is not required to admit either a
single definite order or such a convex decomposition.  Instead, its
positive normalization operator $R^{(2)}$ must satisfy
\begin{align}
\Tr[R^{(2)}(T_1\otimes T_2)]=1
\label{eq:two-slot-all-CPTP}
\end{align}
for every pair of CPTP Choi operators $T_1,T_2$.
The two maps are allowed to be chosen independently.  Every definite-order comb and every convex mixture of
such combs satisfies this requirement.  Unlike the comb recursion,
Eq.~\eqref{eq:two-slot-all-CPTP} imposes no fixed signalling direction
between the slots and also permits causally nonseparable processes
when the slot spaces admit them.

For any subsystem $Y$, let ${}_Y Z$ denote the map that traces out
$Y$ and replaces it by $I_Y/d_Y$, with tensor factors restored to
their original order:
\begin{align}
 {}_Y Z:=\frac{I_Y}{d_Y}\otimes\Tr_Y Z.
\label{eq:two-slot-replacement-map}
\end{align}
The general two-slot normalization conditions are
\begin{align}
 R^{(2)}&\geq0,\qquad \Tr R^{(2)}=d_{B_1}d_{B_2},\nonumber\\
 {}_{A_2B_2}R^{(2)}&={}_{B_1A_2B_2}R^{(2)},\nonumber\\
 {}_{A_1B_1}R^{(2)}&={}_{A_1B_1B_2}R^{(2)},\nonumber\\
 R^{(2)}&={}_{B_1}R^{(2)}+{}_{B_2}R^{(2)}-{}_{B_1B_2}R^{(2)}.
 \label{eq:indef-two-slot-conditions}
\end{align}
These are the two-slot general-process conditions of
Ref.~\cite{BMQ}.  They are necessary and sufficient for
$\Tr[R^{(2)}(T_1\otimes T_2)]=1$ for every pair of independent
CPTP Choi operators $T_1,T_2$, using the Choi normalization
$\Tr_{B_k}T_k=I_{A_k}$.  The first two replacement identities
eliminate components incompatible with normalization when one slot
is varied and the other is fixed at the completely depolarizing
channel; the final identity eliminates the component where both
slots vary in the trace-preserving directions.  Unlike either
fixed-order comb recursion, these conditions do not designate a
first or second slot.  The corresponding affine part of
$F(R^{(2)})=C$ can thus be taken to consist of these three
replacement identities, with the trace condition imposed separately
by Eq.~\eqref{AD}.  The adaptive fixed-order conditions are treated
explicitly in the next section.

\subsection{Conic-programming bounds for generalized measurements}
\label{subsec:general-conic-programs}

We now extend the conic-programming formulation of
Ref.~\cite{Hayashi2024} to generalized measurements on the
$N$-slot process.  The principal modification is that the
completeness condition for an ordinary POVM is replaced by the general
normalization conditions \eqref{NM} and \eqref{AD}.

Given a generalized measurement
$M=\{M_x\}_{x\in{\cal X}}$ and an estimating function
$\hat\theta(x)$, we denote the resulting estimator by
$\hat M
:=
(M,\hat\theta)$.
At the reference point $\bm\theta_0$, put
$u^i(x)=\hat\theta^i(x)-\theta_0^i$ as in the state-estimation
construction.  With $|w_x\rangle$ defined as in
Eq.~\eqref{EQ3}, the corresponding block operator is
\begin{align}
X(\hat M):=\sum_{x\in{\cal X}}
|w_x\rangle\langle w_x|\otimes M_x.
\label{eq:GPOVM-X}
\end{align} 
The $(0,0)$ block of $X(\hat M)$ is the normalization operator of the
generalized measurement:
\begin{align}
R_X
:=
\operatorname{Tr}_{\cal R}
\left[
\left(
|0\rangle\langle0|
\otimes
I_{{\cal K}_B\otimes{\cal K}_A}
\right)
X(\hat M)
\right]
=
\sum_{x\in{\cal X}}M_x
=
R^{(N)}.
\label{eq:GPOVM-zero-block}
\end{align}
Consequently, Eqs.~\eqref{NM} and \eqref{AD} are equivalent to the
following linear conditions on $X(\hat M)$:
\begin{align}
F(R_X)
=
C,\quad 
\operatorname{Tr}R_X
=
\dim{\cal K}_B.
\label{c1-G} 
\end{align}
We refer to Eq.~\eqref{c1-G} as the generalized-measurement
normalization constraints.

At the true parameter value $\theta_0$, the derivative part of local unbiasedness gives
\begin{align}
\operatorname{Tr}
\left[
W_{i,j}X(\hat M)
\right]
=
\delta_{i,j},
\qquad
i,j=1,\ldots,d,
\label{c2-G}
\end{align}
where
\begin{align}
W_{i,j}
:=
\frac{1}{2}
\left(
|0\rangle\langle i|
+
|i\rangle\langle0|
\right)
\otimes
\bar F_j^{(N)}.
\label{eq:Wij-general}
\end{align}
Here, replacing $\hat\theta^i$ by $u^i$ leaves the derivative
constraint unchanged: normalization of $M$ for every process in the
parametric family gives
$\sum_x\Tr[\bar F_j^{(N)}M_x]=0$.
For a locally unbiased estimator the weighted mean-square error is
\begin{align}
\operatorname{Tr}
\left[
G V_{\bm\theta_0}[\hat M]
\right]
=
\operatorname{Tr}
\left[
(G\otimes\bar T)X(\hat M)
\right].
\label{o1-G}
\end{align}
Thus, the normalization and derivative constraints in
Eqs.~\eqref{c1-G} and \eqref{c2-G}, and the objective function,
are linear in $X(\hat M)$.  The remaining mean condition of local
unbiasedness is enforced when recovering the estimator, by shifting
its output values.  The estimation problem minimizes over generalized-measurement estimators. In the conic formulation, the estimator is represented by a block operator; for $h\geq2$ the feasible cone is enlarged to provide lower bounds.

For the normalization conditions specified by $F$ and $C$, we define
the minimum weighted MSE over locally unbiased estimators by
\begin{align}
C_{\bm\theta}^{F,C}[G]
:=
\inf_{\hat M\,:\,{\rm l.u.\ at}\ \bm\theta}
\operatorname{Tr}
\left[
G V_{\bm\theta}[\hat M]
\right],
\label{eq:general-operational-bound}
\end{align}
where the infimum is taken over all locally unbiased estimators
$\hat M=(M,\hat\theta)$ whose normalization operator satisfies
Eqs.~\eqref{NM} and \eqref{AD}.

The quantity $C_{\bm\theta}^{F,C}[G]$ is the infimum of the weighted MSE over the locally
unbiased estimators in the specified generalized-measurement class. To express this
estimation problem as an optimization over block operators, we use the separable cone of
Section~\ref{sec:estimation}; its larger cone relaxations provide Cram\'er--Rao-type lower
bounds. We denote the optimal values of these conic programs by $J_h^{F,C}[G]$. Thus
$C_{\bm\theta}^{F,C}[G]$ is defined by an optimization over estimators, whereas
$J_h^{F,C}[G]$ is defined by an optimization over block operators. The superscript $F,C$
specifies the normalization conditions, and $G$ is the weight matrix. Shorter notation for
particular $N$-use strategy classes is introduced at first use. Using the cones of
Section~\ref{sec:estimation}, we define
\begin{align}
J_h^{F,C}[G]:= 
\min_{X \in {\cal S}_{BA}^h}
\{\Tr [X (G \otimes \bar T)]
|X \hbox{ satisfies }
\hbox{\eqref{c1-G}, \eqref{c2-G}.} \}
\label{eq:Jh-general}
\end{align}
for $h=1,2,3,4$.
Here, ${\cal S}_{BA}^h$ denotes the cone obtained from
${\cal S}^h$ in Section~\ref{sec:estimation} by replacing the state
space ${\cal H}$ with ${\cal K}_B\otimes{\cal K}_A$.
For $h=5$, 
we define ${\cal S}^5_{BA}(\bar T):=
{\cal S} \left( {\cal R}_C \otimes {\cal K}_B \otimes {\cal K}_A \right)_{\rm P} \cap {\cal B}_{\bar T}''$,
and define the conic program
\begin{align}
J_5^{F,C}[G]:= \min_{X \in {\cal S}_{BA}^5(\bar T)}
\{\Tr [X (G \otimes \bar T)]|X \hbox{ satisfies }
\hbox{\eqref{c1-G}, \eqref{c2-G}.} \}
\end{align}

The following result extends the exact conic representation of the
tight precision bound in Ref.~\cite{Hayashi2024} from the
single-channel completeness condition to the general normalization
conditions \eqref{NM} and \eqref{AD}.

\begin{theorem}
\label{thm:general-GPOVM-conic}
The optimal value of the separable-cone program equals the minimum weighted MSE over locally unbiased generalized-measurement estimators:
\begin{align}
J_1^{F,C}[G]
=
C_{\bm\theta}^{F,C}[G].
\label{eq:J1-operational-equivalence}
\end{align}
Replacing the separable cone by the larger cones gives the associated CR-type lower bounds, which
satisfy
\begin{align}
J_1^{F,C}[G]
\geq
J_2^{F,C}[G]
\geq
J_3^{F,C}[G]
\geq
J_4^{F,C}[G],
\label{eq:J-hierarchy-1234}
\end{align}
and
\begin{align}
J_4^{F,C}[G]
\leq
J_5^{F,C}[G]
\leq
J_2^{F,C}[G].
\label{eq:J-hierarchy-5}
\end{align}
\end{theorem}

Here $C$ in $F(R)=C$ is fixed normalization data, whereas $C_{\bm\theta}^{F,C}[G]$ denotes a precision limit; the different uses are distinguished by their arguments.

Theorem~\ref{thm:general-GPOVM-conic} separates two choices that play different roles.  Equations~\eqref{NM} and \eqref{AD} specify the prescribed strategy class, whereas the cone ${\cal S}_{BA}^h$ specifies the level of conic relaxation.  The separable-cone program has optimal value $C_{\bm\theta}^{F,C}[G]$, while the larger cones give the stated Cram\'er--Rao-type lower bounds.  The proof establishes the equality in Eq.~\eqref{eq:J1-operational-equivalence} by constructing the correspondence in both directions and then derives the remaining inequalities from the cone inclusions.

\begin{proof}
We first map every locally unbiased generalized-measurement estimator to a feasible separable-cone variable with the same objective value.  For every locally unbiased generalized-measurement estimator
$\hat M=(M,\hat\theta)$, the operator $X(\hat M)$ in
Eq.~\eqref{eq:GPOVM-X} belongs to the separable cone
${\cal S}_{BA}^1$.  Equations~\eqref{eq:GPOVM-zero-block},
\eqref{c1-G}, and \eqref{c2-G} show that $X(\hat M)$ satisfies the
normalization and derivative constraints in the definition of
$J_1^{F,C}[G]$.  Since $X(\hat M)\in{\cal S}_{BA}^1$, these properties place
$X(\hat M)$ in the optimization domain of $J_1^{F,C}[G]$.  Equation~\eqref{o1-G} shows
that its objective value equals the weighted mean-square error of $\hat M$.  Hence,
\begin{align}
J_1^{F,C}[G]
\leq
C_{\bm\theta}^{F,C}[G].
\end{align}

We next recover a locally unbiased generalized-measurement estimator from an arbitrary feasible separable-cone variable without increasing its objective value.  Let $X\in{\cal S}_{BA}^1$ satisfy the normalization and
derivative constraints \eqref{c1-G} and \eqref{c2-G}.
The definition of the separable cone gives the decomposition
\begin{align}
X
=
\sum_{\alpha\in{\cal A}}
A_\alpha\otimes Q_\alpha,
\qquad
A_\alpha\geq0,
\quad
Q_\alpha\geq0,
\label{eq:separable-decomposition-X}
\end{align}
where each $A_\alpha$ is a real positive semidefinite operator on
${\cal R}$ and each $Q_\alpha$ is a positive semidefinite operator on
${\cal K}_B\otimes{\cal K}_A$.
By spectrally decomposing the operators $A_\alpha$ and refining the
index set, we may write
$X
=
\sum_{\alpha\in{\cal A}}
|v_\alpha\rangle\langle v_\alpha|
\otimes Q_\alpha$,
where $|v_\alpha\rangle\in{\cal R}$ is a real vector.

We expand each vector as
$|v_\alpha\rangle
=
a_\alpha|0\rangle
+
\sum_{i=1}^{d}
b_{\alpha,i}|i\rangle$ for 
$a_\alpha,b_{\alpha,i}\in{\mathbb R}$.  To recover an estimator, we
separate these terms according to whether their coefficient along
$|0\rangle$ is nonzero.  Let
\begin{align}
{\cal A}_+
&:=
\left\{
\alpha\in{\cal A}:a_\alpha\neq0
\right\},
&
{\cal A}_0
&:=
\left\{
\alpha\in{\cal A}:a_\alpha=0
\right\}.
\end{align}
Each $\alpha\in{\cal A}_+$ determines
\begin{align}
M_\alpha
:=
a_\alpha^2Q_\alpha,\quad
u^i(\alpha)
:=
\frac{b_{\alpha,i}}{a_\alpha},
\qquad
i=1,\ldots,d.
\label{eq:estimator-from-separable}
\end{align}
Then,
\begin{align}
&|v_\alpha\rangle\langle v_\alpha|
\otimes Q_\alpha
\notag\\
&=
\left(
|0\rangle
+
\sum_{i=1}^{d}
u^i(\alpha)|i\rangle
\right)
\left(
\langle0|
+
\sum_{i=1}^{d}
u^i(\alpha)\langle i|
\right)
\otimes M_\alpha
\label{eq:separable-term-estimator-form}
\end{align}
for every $\alpha\in{\cal A}_+$.

We define
\begin{align}
X'
:=
\sum_{\alpha\in{\cal A}_+}
|v_\alpha\rangle\langle v_\alpha|
\otimes Q_\alpha.
\label{eq:X-prime-estimator-part}
\end{align}
The terms indexed by ${\cal A}_0$ have no $(0,0)$ or $(0,i)$
blocks because $a_\alpha=0$. Therefore,
\begin{align}
\operatorname{Tr}_{\cal R}
\left[
\left(
|0\rangle\langle0|
\otimes
I_{{\cal K}_B\otimes{\cal K}_A}
\right)
X'
\right]
&=
R_X,
\label{eq:X-prime-zero-block}
\\
\operatorname{Tr}
\left[
W_{i,j}X'
\right]
&=
\operatorname{Tr}
\left[
W_{i,j}X
\right]
=
\delta_{i,j}.
\label{eq:X-prime-local-unbiasedness}
\end{align}
Hence, $X'$ satisfies the same normalization and derivative
constraints as $X$; the estimator's mean is adjusted below.

Both $G\otimes\bar T$ and
$|v_\alpha\rangle\langle v_\alpha|\otimes Q_\alpha$
are positive semidefinite.  Consequently, every discarded term has a nonnegative
contribution to the objective:
\begin{align}
\operatorname{Tr}
\left[
(G\otimes\bar T)
\left(
|v_\alpha\rangle\langle v_\alpha|
\otimes Q_\alpha
\right)
\right]
\geq0
\label{eq:discarded-term-nonnegative}
\end{align}
for every $\alpha\in{\cal A}_0$. It follows that
\begin{align}
\operatorname{Tr}
\left[
(G\otimes\bar T)X'
\right]
\leq
\operatorname{Tr}
\left[
(G\otimes\bar T)X
\right].
\label{eq:X-prime-objective}
\end{align}

Equations~\eqref{eq:estimator-from-separable} and
\eqref{eq:separable-term-estimator-form} identify $X'$ as the block
operator built from the offsets $u^i(\alpha)$ and the outcome
operators $M_\alpha$.  In particular,
\begin{align}
\sum_{\alpha\in{\cal A}_+}M_\alpha=R_X.
\label{eq:constructed-GPOVM-normalization}
\end{align}
Thus Eq.~\eqref{c1-G} makes $\{M_\alpha\}_{\alpha\in{\cal A}_+}$
an admissible generalized measurement.  Equation~
\eqref{eq:X-prime-local-unbiasedness} proves the derivative
condition, but does not by itself impose the required mean at
$\bm\theta_0$.  To impose it, let
\begin{align}
m^i&:=\sum_{\alpha\in{\cal A}_+}
\Tr(\bar T M_\alpha)u^i(\alpha),\notag\\
\hat\theta^i(\alpha)&:=\theta_0^i+u^i(\alpha)-m^i.
\label{eq:recovered-estimator-centered}
\end{align}
Because $\sum_\alpha\Tr(\bar T M_\alpha)=1$,
this estimator has mean $\theta_0^i$.  The constant shift does
not alter its derivative, so it is locally unbiased.  Its
weighted MSE satisfies
\begin{align}
&\Tr[G V_{\bm\theta_0}[\hat M]]\notag\\
&\quad\stackrel{(a)}{=}
\Tr[(G\otimes\bar T)X']-m^{\mathsf T}Gm\notag\\
&\quad\stackrel{(b)}{\leq}
\Tr[(G\otimes\bar T)X']
\stackrel{(c)}{\leq}\Tr[(G\otimes\bar T)X].
\label{eq:recovered-estimator-MSE}
\end{align}
Here (a) expands the centered second moment, (b) uses $G\geq0$,
and (c) is Eq.~\eqref{eq:X-prime-objective}.

Therefore, every $X\in{\cal S}_{BA}^1$ satisfying
Eqs.~\eqref{c1-G} and \eqref{c2-G} determines a locally unbiased
generalized-measurement estimator whose weighted mean-square error is
no larger than $\operatorname{Tr}[(G\otimes\bar T)X]$.  Taking the
infimum over all such operators $X$ gives
$C_{\bm\theta}^{F,C}[G]
\leq
J_1^{F,C}[G]$.
Together with the opposite inequality established above, this proves
Eq.~\eqref{eq:J1-operational-equivalence}.

It remains to prove the hierarchy of relaxed bounds.  Eqs.~\eqref{eq:J-hierarchy-1234} and
\eqref{eq:J-hierarchy-5} follow from the cone inclusions
${\cal S}_{BA}^1
\subset
{\cal S}_{BA}^2
\subset
{\cal S}_{BA}^3
\subset
{\cal S}_{BA}^4
$ and
${\cal S}_{BA}^2
\subset
{\cal S}_{BA}^5(\bar T)
\subset
{\cal S}_{BA}^4$.
\end{proof}

\subsection{Normalization of process-matrix strategies without a fixed causal order}
\label{subsec:indefinite-order-normalization}

Equation~\eqref{eq:indef-two-slot-conditions} characterized the
normalization operators of the general two-slot process-matrix class.
We extend this characterization to $N$ slots by requiring
normalization for arbitrary quantum channels inserted independently in
each slot. A fixed-order comb obeys recursive partial-trace
conditions, and causally separable testers are convex mixtures of
fixed-order testers. The general class also permits causally
nonseparable processes when the slot spaces admit them.

More precisely, in this subsection the class $\mathfrak P$ in
Eq.~\eqref{eq:process-normalization} consists of products
$\bigotimes_{k=1}^N T_k$ of arbitrary independently chosen CPTP
Choi operators, $\Tr_{B_k}T_k=I_{A_k}$.  Tensor factors are
identified with the ordering on ${\cal K}_B\otimes{\cal K}_A$
by the canonical permutation.  For every such independent choice of CPTP Choi operators
$T_1,\ldots,T_N$, the normalization requirement is
\begin{align}
\Tr\left[R^{(N)}\bigotimes_{k=1}^N T_k\right]=1.
\label{eq:indef-universal-condition}
\end{align}
The next theorem replaces the requirement~\eqref{eq:indef-universal-condition}, imposed
for every independent choice of quantum channels, by finitely many explicit linear equations on
$R^{(N)}$. We will then use these equations to specify the normalization map $F$ and
right-hand side $C$ in Eq.~\eqref{NM}. The independent-slot requirement also enters the
scaling proof: under the rank or support hypotheses of Theorem~\ref{thm:indef-S6-scaling}
and Proposition~\ref{prop:indef-S6-singular-extension}, perturbing the quantum channels in two
distinct slots separately forces the cross-slot contributions to the real RLD-type
information matrix to vanish.

For a subsystem $Y$ of the total Hilbert space, introduce the
replacement map
\begin{align}
{}_{Y}Z
:=
\frac{I_Y}{\dim{\cal H}_Y}\otimes\Tr_Y Z,
\label{eq:indef-replacement-map}
\end{align}
where the tensor factors on the right are returned to their
original order.  For each slot $k$ define the commuting orthogonal
projections, with respect to the Hilbert--Schmidt inner product,
\begin{align}
P_k Z:={}_{A_kB_k}Z,
\quad
Q_k Z:=Z-{}_{B_k}Z.
\label{eq:indef-PQ}
\end{align}
The map $P_k$ retains the component proportional to the identity
on $A_kB_k$, while $Q_k$ retains the component with zero partial
trace over $B_k$. They satisfy $P_kQ_k=Q_kP_k=0$, and maps
belonging to different slots commute. We use them to select the
sectors in which at least one slot has a $Q_k$ component and every
other slot has either a $P_\ell$ or a $Q_\ell$ component. The
corresponding projection is
\begin{align}
\Pi_N
&:=\prod_{k=1}^N(P_k+Q_k)-\prod_{k=1}^N P_k\nonumber\\
&=\sum_{\emptyset\ne S\subseteq\{1,\ldots,N\}}
\left(\prod_{k\in S}Q_k\right)
\left(\prod_{\ell\notin S}P_\ell\right).
\label{eq:indef-PiN}
\end{align}
All products here denote composition of superoperators; in the
sum, the slots in $S$ carry $Q_k$ and the others carry $P_\ell$.

\begin{theorem}[Normalization of general process-matrix testers with $N$ slots]
\label{thm:indef-N-normalization}
Let $R^{(N)}$ be a positive semidefinite operator on
${\cal K}_B\otimes{\cal K}_A$.  The following are equivalent:
\begin{enumerate}
\item For all CPTP Choi operators $T_k$ on
${\cal H}_{B,k}\otimes{\cal H}_{A,k}$,
\begin{align}
\Tr\left[R^{(N)}\bigotimes_{k=1}^N T_k\right]=1.
\label{eq:indef-all-CPTP}
\end{align}
\item The following linear normalization conditions hold:
\begin{align}
\Tr R^{(N)}=\dim{\cal K}_B,
\quad
\Pi_N R^{(N)}=0.
\label{eq:indef-normalization}
\end{align}
\end{enumerate}
Equivalently, the second equation in
\eqref{eq:indef-normalization} can be replaced by the family
\begin{align}
\left(\prod_{k\in S}Q_k\right)
\left(\prod_{\ell\notin S}P_\ell\right)R^{(N)}=0
\quad
(\emptyset\ne S\subseteq\{1,\ldots,N\}).
\label{eq:indef-subset-conditions}
\end{align}
\end{theorem}

\begin{proof}
Write $d_{B_k}:=\dim{\cal H}_{B,k}$, and let
$T_k^{(0)}:=I_{A_kB_k}/d_{B_k}$ be the strictly positive Choi
operator of the completely depolarizing channel.  Every CPTP
Choi operator has the form $T_k=T_k^{(0)}+V_k$, where $V_k$ is
Hermitian and $\Tr_{B_k}V_k=0$.  Conversely, for every Hermitian
$V_k$ satisfying this trace condition,
$T_k^{(0)}+t_kV_k$ is CPTP for all sufficiently small real $t_k$.

Suppose first that Eq.~\eqref{eq:indef-all-CPTP} holds.  Substitution
of $T_k=T_k^{(0)}$ for every $k$ gives
$\Tr R^{(N)}=\prod_k d_{B_k}=\dim{\cal K}_B$.
For any nonempty subset $S$ choose
$T_k=T_k^{(0)}+t_kV_k$ for $k\in S$, leaving the remaining
slots at $T_\ell^{(0)}$.  The left-hand side of
Eq.~\eqref{eq:indef-all-CPTP} is a multiaffine polynomial in
the real variables $(t_k)_{k\in S}$ that is identically one in
a neighborhood of the origin.  Its coefficient of
$\prod_{k\in S}t_k$ must therefore vanish.  Thus
\begin{align}
\Tr\left[
R^{(N)}
\left(\bigotimes_{k\in S}V_k\otimes
\bigotimes_{\ell\notin S}T_\ell^{(0)}\right)
\right]=0.
\label{eq:indef-coefficient}
\end{align}
Here and below the product is reordered into the total
input-output convention.  The Hermitian $V_k$ with
$\Tr_{B_k}V_k=0$ span the real range of $Q_k$,
while $T_\ell^{(0)}$ spans the real range of $P_\ell$
on each remaining slot.  Since the projections are
Hilbert--Schmidt self-adjoint, Eq.~\eqref{eq:indef-coefficient}
for all such $V_k$ is equivalent to
Eq.~\eqref{eq:indef-subset-conditions}.

Conversely, suppose the trace equality and
Eq.~\eqref{eq:indef-subset-conditions} hold.
For arbitrary CPTP Choi operators, expand
$\bigotimes_k(T_k^{(0)}+V_k)$ as a sum indexed by subsets
$S\subseteq\{1,\ldots,N\}$.  Every term indexed by a nonempty
$S$ has zero Hilbert--Schmidt inner product with $R^{(N)}$
by Eq.~\eqref{eq:indef-subset-conditions}.  The term for
$S=\emptyset$ is
$\Tr R^{(N)}/\prod_kd_{B_k}=1$.
This proves Eq.~\eqref{eq:indef-all-CPTP}.

Finally, the summands in Eq.~\eqref{eq:indef-PiN} are
mutually orthogonal projections.  Consequently, their sum
annihilates $R^{(N)}$ if and only if every summand does.
\end{proof}

Theorem~\ref{thm:indef-N-normalization} shows that the normalization
requirement~\eqref{eq:indef-universal-condition} is equivalent to the linear
conditions~\eqref{eq:indef-normalization}. In the generalized-measurement conic programs
of Subsection~\ref{subsec:general-conic-programs}, this means that the abstract constraint
$F(R^{(N)})=C$ of Eq.~\eqref{NM} is specified by $F=\Pi_N$ and $C=0$. Positivity of
$R^{(N)}$ and the separate trace condition~\eqref{AD} are retained.

For $N=2$, Eq.~\eqref{eq:indef-subset-conditions} reduces to
$Q_1P_2R^{(2)}=P_1Q_2R^{(2)}=Q_1Q_2R^{(2)}=0$.
Using the definitions of $P_k,Q_k$, these are exactly the three
replacement-map identities in
Eq.~\eqref{eq:indef-two-slot-conditions} stated earlier.
Thus the arbitrary-$N$ characterization genuinely extends the
known two-slot general-process conditions.

The theorem concerns normalization on independently chosen
quantum channels in the individual slots, even if the model being
estimated uses identical channel instances.  Requiring
normalization only on tensor powers of one channel would give
a different condition.  Moreover, the class described here
is the positive process-matrix tester class; the normalization
equations alone do not assert implementation by any specified
family of quantum circuits.

\subsection{Exact scaling of the positive-cone bound for multiparameter channel estimation}
\label{subsec:indef-S6-scaling}

Consider a differentiable family of quantum channels with $d$ real
parameters.  At the parameter value under consideration, write
$T=T_{\bm\theta}$ and $D_i=\partial_iT_{\bm\theta}$ for
$i=1,\ldots,d$.  All $N$ slots contain the same channel, but
normalization is required for arbitrary \emph{independent} CPTP
maps, as in Theorem~\ref{thm:indef-N-normalization}.  Set
\begin{align}
T_N=T^{\otimes N},\quad
D_{i,N}=\sum_{k=1}^N
T^{\otimes(k-1)}\otimes D_i\otimes T^{\otimes(N-k)}.
\label{eq:indef-S6-TN-DN}
\end{align}

For the repeated-channel model, we write
\begin{align*}
J_{h,\mathrm{gen}}^{(N)}(G):=J_h^{\Pi_N,0}[G],\qquad h=1,\ldots,5.
\end{align*}
These are the values of the programs in Eq.~\eqref{eq:Jh-general}
and its $h=5$ counterpart with $\bar T=T^{\otimes N}$,
the derivatives of $T^{\otimes N}$, and the general tester
normalization characterized above. Each one-use conic feasible
point can be repeated independently, giving
\begin{align}
J_{h,\mathrm{gen}}^{(N)}(G)\leq\frac{J_{h,\mathrm{gen}}^{(1)}(G)}{N},
\qquad h=1,\ldots,5.
\label{eq:indef-general-repetition-bound}
\end{align}
To see this for the relaxed cones as well, first center the
one-use parameter blocks so that $\Tr(TX_{i0})=0$.
Let $A_N$ map $|0\rangle^{\otimes N}$ to $|0\rangle$,
each single-slot vector $|0,\ldots,i,\ldots,0\rangle$ to
$|i\rangle/N$, and all other parameter-basis vectors to zero.
Then $(A_N\otimes I)\widetilde X^{\otimes N}(A_N^\dagger\otimes I)$,
where $\widetilde X$ is centered, is feasible for the $N$-slot
program: real congruence preserves positivity and PPT, and the
$h=5$ trace-symmetry is preserved since the cross-slot means vanish.
The same centering removes cross-slot terms from the objective,
leaving $1/N$ of the one-use value.
Taking infima proves Eq.~\eqref{eq:indef-general-repetition-bound}.

Let ${\cal S}_{BA}^6:={\cal S}({\cal R}_C\otimes{\cal K}_B
\otimes{\cal K}_A)_{\rm P}$ be the channel counterpart of
Eq.~\eqref{eq:S6-cone}. Denote by $J_{6,\mathrm{gen}}^{(N)}(G):=J_6^{\Pi_N,0}[G]$ the
infimum of the generalized-measurement conic program with this
cone, constraints \eqref{c1-G} and \eqref{c2-G}, and the
indefinite-order normalization $F=\Pi_N$, $C=0$ from
Eq.~\eqref{eq:indef-normalization}, for a real symmetric
weight matrix $G>0$.  This is a positive-cone relaxation,
not an assertion of operational attainability.

Write
\begin{align}
\mathfrak W_N:=\{R\geq0:\ \Tr R=(\dim{\cal H}_B)^N,
\ \Pi_NR=0\}.
\label{eq:indef-S6-WN}
\end{align}
For a density operator $\sigma$ on the input Hilbert space,
define the real symmetric matrix
\begin{align}
F(\sigma)_{ij}
:=\Re\Tr\!\left[(I_B\otimes\sigma)D_iT^{-1}D_j\right].
\label{eq:indef-S6-information}
\end{align}
The real symmetric matrix $F(\sigma)$ is the real RLD-type
information matrix associated with the input state $\sigma$.
In Eq.~\eqref{NM}, $F$ denotes the separately defined linear
normalization map (here $F=\Pi_N$).
Equation~\eqref{eq:S6-RLD-real-bound} evaluates the
positive-cone bound for a fixed state model, after minimizing
over its parameter blocks $A_i$.  In channel estimation, the
normalization operator is also optimized.  For one slot it
has the form $I_B\otimes\sigma$, so the channel bound requires
the further minimization of $\Tr[G F(\sigma)^{-1}]$ over
input density operators $\sigma$ for which $F(\sigma)>0$.
This second minimization is distinct from the one in
\eqref{eq:S6-reduced} and appears explicitly in
\eqref{eq:indef-S6-scaling} below.

\begin{theorem}[Exact $1/N$ scaling for multiple parameters]
\label{thm:indef-S6-scaling}
Assume $T>0$ and that $D_1,\ldots,D_d$ are linearly independent
over $\mathbb R$.  For each $R\in\mathfrak W_N$ set
\begin{align}
F^{(N)}(R)_{ij}
:=\Re\Tr\!\left[R D_{i,N}T_N^{-1}D_{j,N}\right].
\label{eq:indef-S6-N-information}
\end{align}
Then the information matrices generated by normalization
operators in $\mathfrak W_N$ satisfy the set equality
\begin{align}
\{F^{(N)}(R):R\in\mathfrak W_N\}
=\{NF(\sigma):\sigma\geq0,\ \Tr\sigma=1\}.
\label{eq:indef-S6-matrix-set}
\end{align}
Moreover, for every positive integer $N$,
\begin{align}
J_{6,\mathrm{gen}}^{(N)}(G)
&=\inf_{\sigma\geq0,\ \Tr\sigma=1\atop F(\sigma)>0}
\Tr\!\left[G\bigl(NF(\sigma)\bigr)^{-1}\right]
=\frac{J_{6,\mathrm{gen}}^{(1)}(G)}{N}.
\label{eq:indef-S6-scaling}
\end{align}
\end{theorem}

For any admissible normalization operator $R$, contributions from distinct channel-use slots vanish, and the remaining terms give \(F^{(N)}(R)=NF(\bar\sigma)\) for some input density operator \(\bar\sigma\). Conversely, every \(NF(\sigma)\) is obtained from the product normalization operator \(R=(I_B\otimes\sigma)^{\otimes N}\). This proves the set equality in Eq.~\eqref{eq:indef-S6-matrix-set}.
For distinct slots $k\ne\ell$, perturb
their CPTP Choi operators independently as $T+sD_i$ and $T+tD_j$, leaving all other slots
at $T$. For sufficiently small $s,t$ these are valid CPTP Choi operators under the
full-rank or supported-singular hypotheses. Since the normalization trace in
Eq.~\eqref{eq:indef-universal-condition} remains equal to one, its mixed derivative with
respect to $s$ and $t$ vanishes. This removes the cross-slot terms in $F^{(N)}(R)$; the
surviving terms come from one slot at a time. The detailed calculation is in
Lemma~\ref{lem:indef-S6-slot-additivity}, especially
Eq.~\eqref{eq:indef-S6-cross-derivative}.
For every $R\in\mathfrak W_N$, some input density operators
$\sigma_1,\ldots,\sigma_N$ therefore satisfy
\begin{align}
F^{(N)}(R)&=\sum_{k=1}^N F(\sigma_k)
=NF(\bar\sigma),\quad
\bar\sigma:=\frac1N\sum_{k=1}^N\sigma_k.
\label{eq:indef-S6-main-text-additivity}
\end{align}
Conversely, for every density operator $\sigma$, a product
normalization operator realizes $NF(\sigma)$.  These two directions
give Eq.~\eqref{eq:indef-S6-matrix-set}.  Combining this equality
with fixed-$R$ quadratic minimization gives the scaling formula:
\begin{align}
J_{6,\mathrm{gen}}^{(N)}(G)
&\stackrel{(a)}{=}
\inf_{\substack{R\in\mathfrak W_N\\F^{(N)}(R)>0}}
\Tr[G F^{(N)}(R)^{-1}]\notag\\
&\stackrel{(b)}{=}
\inf_{\substack{\sigma\geq0,\ \Tr\sigma=1\\F(\sigma)>0}}
\Tr[G(NF(\sigma))^{-1}]\notag\\
&\stackrel{(c)}{=}\frac{J_{6,\mathrm{gen}}^{(1)}(G)}{N}.
\label{eq:indef-S6-main-text-scaling-chain}
\end{align}
Here (a) is the fixed-normalization minimization established in
Lemma~\ref{lem:indef-S6-information-reduction}; (b) uses the
matrix-set identity; and (c) uses $(NF)^{-1}=N^{-1}F^{-1}$
and the one-slot variational formula.  The same mechanism applies
under the support assumptions of
Proposition~\ref{prop:indef-S6-singular-extension}, with $T^+$
in place of $T^{-1}$. Detailed proofs of the scaling theorem and its singular extension are in Appendix~\ref{subsec:proof-indef-S6-scaling}; the operational corollary is proved immediately after its statement below.

Theorem~\ref{thm:indef-S6-scaling} assumes $T>0$.
The next proposition gives an extension to singular CPTP Choi
operators when a support condition and information-matrix
nondegeneracy hold. Let $P$ be the support projection of
$T\geq0$. The support condition is
\begin{align}
D_i=PD_iP\qquad (i=1,\ldots,d).
\label{eq:indef-S6-support}
\end{align}

\begin{proposition}[Extension to singular Choi operators]
\label{prop:indef-S6-singular-extension}
Let $T\geq0$ be a CPTP Choi operator, and let $P$ be its
support projection. Assume $D_i=PD_iP$ for each $i$, as in
Eq.~\eqref{eq:indef-S6-support}, that $D_1,\ldots,D_d$ are
linearly independent over $\mathbb R$, and that $F(\sigma)>0$
for some input density operator $\sigma$, where $F(\sigma)$
is defined by
Eq.~\eqref{eq:indef-S6-information} with $T^{-1}$ replaced
by the Moore--Penrose inverse $T^+$.
Define $F^{(N)}(R)$ likewise by
Eq.~\eqref{eq:indef-S6-N-information} with $T_N^{-1}$
replaced by $T_N^+$.
Then the matrix-set identity~\eqref{eq:indef-S6-matrix-set}
and the exact scaling~\eqref{eq:indef-S6-scaling} hold with
these definitions, for every $N\geq1$ and $G>0$.
\end{proposition}

For $d=1$, let $G=(g)$ with $g>0$ and set
$K:=\Tr_B(D_1T^{-1}D_1)\ge0$ (using $T^+$ instead of
$T^{-1}$ in the supported singular case).  Then
$F(\sigma)=\Tr(\sigma K)$ and hence
\begin{align}
\sup_{\sigma\ge0,\ \Tr\sigma=1}F(\sigma)
&=\lambda_{\max}(K)=\|K\|_\infty,
\notag\\
J_{6,\mathrm{gen}}^{(1)}(g)&=\frac{g}{\|K\|_\infty},
\quad
J_{6,\mathrm{gen}}^{(N)}(g)=\frac{g}{N\|K\|_\infty}.
\label{eq:indef-S6-single-parameter-norm}
\end{align}
The one-parameter channel RLD Fisher information has this
operator-norm expression in Hayashi~\cite{Hayashi11}, which
also proves its additivity for parallel channel uses under the
corresponding support condition.  The present theorem establishes the equality of the sets of real
RLD-type information matrices
\eqref{eq:indef-S6-matrix-set} for arbitrary $d$ and the
independently normalized general process-matrix tester class.

The positive-cone identity, together with the general
repetition bound~\eqref{eq:indef-general-repetition-bound},
now gives the precision scaling of the operational optimum and the
other conic bounds.

\begin{corollary}[Operational standard quantum limit for general process testers]
\label{cor:indef-operational-SQL}
Consider generalized measurements normalized on independent CPTP
Choi operators in $N$ slots.  Suppose $G>0$, the derivatives
$D_1,\ldots,D_d$ are linearly independent over $\mathbb R$, and
either $T>0$, or the support condition~\eqref{eq:indef-S6-support}
holds with $F(\sigma)>0$ for some input density operator $\sigma$.
Then the optimal locally unbiased weighted MSE satisfies
\begin{align}
\frac{J_{6,\mathrm{gen}}^{(1)}(G)}{N}\le J_{1,\mathrm{gen}}^{(N)}(G)
\le\frac{J_{1,\mathrm{gen}}^{(1)}(G)}{N}, \qquad N\ge1.
\label{eq:indef-operational-SQL}
\end{align}
More generally, each conic bound satisfies
\begin{align}
\frac{J_{6,\mathrm{gen}}^{(1)}(G)}{N}\le J_{h,\mathrm{gen}}^{(N)}(G)
\le\frac{J_{h,\mathrm{gen}}^{(1)}(G)}{N},
\qquad h=1,\ldots,5.
\label{eq:indef-operational-SQL-relaxations}
\end{align}
In particular, $J_{h,\mathrm{gen}}^{(N)}(G)=\Theta(N^{-1})$ for
$h=1,\ldots,5$, including the operational value $J_{1,\mathrm{gen}}^{(N)}(G)$.
Thus the $1/N^2$ Heisenberg MSE scaling is excluded under
these hypotheses, including for general process testers.
\end{corollary}

\begin{proof}
The inequalities follow from the cone inclusions, the exact
positive-cone scaling theorem (or its supported-singular extension),
and Eq.~\eqref{eq:indef-general-repetition-bound}. The one-use constants
are positive and finite under the stated hypotheses: $F(\sigma)$ is
uniformly bounded, and a full-Schmidt-rank probe with an
informationally complete POVM yields a finite locally unbiased
one-use estimator when the derivatives are real-linearly independent.
\end{proof}

The exclusion of Heisenberg scaling comes from the assumptions on the channel, not from excluding entangled input states or imposing a fixed causal order. In particular, parallel strategies with entangled probes belong to the general process-tester class. Nontrivial unitary channel families have rank-one Choi operators and do not satisfy the full-rank hypothesis; their parameter derivatives also fail the fixed-support condition in the nontrivial directions. Such channels may exhibit Heisenberg scaling and lie outside the regime of Corollary~\ref{cor:indef-operational-SQL}.

The scalar consequence also constrains joint multiparameter scaling.
For a nonzero $c\in\mathbb R^d$, set
$D_c:=\sum_i c_iD_i$ and consider the one-parameter submodel
$\bm\theta(t)=\bm\theta_0+tc$.  Assume $D_c\ne0$ and
$D_c=PD_cP$, where $P$ projects onto the support of $T$.
Any locally unbiased $d$-parameter estimator $\hat{\bm\theta}$
induces a locally unbiased estimator
$\hat t:=c^{\mathsf T}(\hat{\bm\theta}-\bm\theta_0)/(c^{\mathsf T}c)$
for this submodel: its mean at $t=0$ is zero and the derivative
of its mean with respect to $t$ equals one.  Write
$V_{\bm\theta_0}^{(N)}$ for the MSE matrix of
$\hat{\bm\theta}$ when $N$ channel uses are available.
Applying the one-parameter case of
Corollary~\ref{cor:indef-operational-SQL} gives a constant
$\kappa_c>0$, independent of $N$, such that
\begin{align}
\frac{c^{\mathsf T}V_{\bm\theta_0}^{(N)}c}
{(c^{\mathsf T}c)^2}
=\operatorname{Var}_{\bm\theta_0}(\hat t)
\geq\frac{\kappa_c}{N}.
\label{eq:supported-direction-scalar-SQL}
\end{align}
Here the scalar corollary applies because $D_c$ is nonzero and
supported on $P$; a full-rank input state makes its scalar
information strictly positive.  Thus $O(N^{-2})$ MSE in every
parameter direction is possible only if
$D_c\ne PD_cP$ for every nonzero $c$.  This is a necessary,
not a sufficient, condition for simultaneous Heisenberg scaling.

\section{Channel estimation with adaptive strategies}
\label{sec:adaptive-strategies}

We now specialize the general framework of Section~\ref{sec:general-strategies} to adaptive strategies with the fixed causal order 
$1\prec2\prec\cdots\prec N$.
The $N$ input-output slots and the parameter-dependent process family $\{\bar T_\theta^{(N)}\}_\theta$ are those introduced in the preceding section. We restrict the admissible generalized measurements by requiring their normalization operators to satisfy the recursive causal conditions associated with this fixed order.
This specialization applies both to the general $N$-slot process and to the repeated-channel model with $N=mn$ introduced in Subsection~\ref{subsec:repeated-channel-model}.

\subsection{Adaptive normalization conditions}
\label{subsec:adaptive-normalization}

To give the abstract linear condition \eqref{NM} explicitly for
adaptive strategies with the fixed causal order, for
$k=N,\ldots,1$, let
\begin{align}
R^{(k)}
:=
\operatorname{Tr}_{(A,k+1),(B,k+1),\ldots,(A,N),(B,N)}
R^{(N)},
\label{eq:reduced-adaptive-normalization}
\end{align}
where $R^{(N)}$ itself is used when $k=N$.  Here, $R^{(k)}$ acts on
\begin{align}
{\cal H}_{B,k}
\otimes
{\cal H}_{A,k}
\otimes
{\cal H}_{B,k-1}
\otimes
{\cal H}_{A,k-1}
\otimes\cdots\otimes
{\cal H}_{B,1}
\otimes
{\cal H}_{A,1}.
\end{align}
The normalization conditions for an adaptive strategy with the
prescribed causal order are
\begin{align}
R^{(k)}
=
\frac{I_{(B,k)}}{d_{B,k}}
\otimes
\operatorname{Tr}_{(B,k)}R^{(k)},
\qquad
k=1,\ldots,N,
\label{eq:adaptive-causal-normalization}
\end{align}
together with the total-trace normalization \eqref{AD}
\cite{Chiribella2008QuantumCircuit,Chiribella2009QuantumNetworks}.
Equation~\eqref{eq:adaptive-causal-normalization} imposes the causal
normalization at each stage, whereas Eq.~\eqref{AD} fixes the overall
normalization.  Thus, Eqs.~\eqref{eq:adaptive-causal-normalization}
and \eqref{AD} give the concrete specialization of the general linear
normalization condition \eqref{NM} for the adaptive fixed-order class
considered in this section.

For later use, define the linear map $F_N^{\mathrm{ad}}$ by
\begin{align}
F_N^{\mathrm{ad}}(R)
:=\left(
R^{(k)}-\frac{I_{(B,k)}}{d_{B,k}}
\otimes\operatorname{Tr}_{(B,k)}R^{(k)}
\right)_{k=1}^{N},
\qquad C_N^{\mathrm{ad}}:=0.
\label{eq:adaptive-FC-notation}
\end{align}
The equation $F_N^{\mathrm{ad}}(R)=C_N^{\mathrm{ad}}$, together with
the trace condition~\eqref{AD}, specifies the adaptive normalization.
For this strategy class, we abbreviate the corresponding conic values as
\begin{align*}
J_{h,\mathrm{ad}}^{(N)}(G):=J_h^{F_N^{\mathrm{ad}},C_N^{\mathrm{ad}}}[G]
\end{align*}
for each cone index $h$ considered below.

These conditions reduce for $N=1$ to
\begin{align}
R^{(1)}
=
\frac{I_{(B,1)}}{d_{B,1}}
\otimes
\operatorname{Tr}_{(B,1)}R^{(1)},
\quad
\operatorname{Tr}R^{(1)}
=
d_{B,1}.
\label{AD-simple}
\end{align}
Thus, in the single-use case, the normalization operator is
proportional to the identity operator on the output system, while its
overall scale is determined by the trace condition.

The recursive conditions above specialize the general linear
normalization framework to the fixed causal order
$1\prec2\prec\cdots\prec N$.  They should be distinguished from
the general process-matrix normalization developed in
Subsection~\ref{subsec:indefinite-order-normalization}, which
does not prescribe a definite order.  In particular, the explicit
two-slot equations are given in
Eq.~\eqref{eq:indef-two-slot-conditions}, and their extension
to arbitrary $N$ is Theorem~\ref{thm:indef-N-normalization}.
The present section concerns only the fixed-order adaptive class. 

\subsection{Sensor-comb representation}
\label{subsec:sensor-comb}

The adaptive generalized measurement introduced above describes the complete estimation procedure directly through its outcome operators $\{M_x\}_{x\in{\cal X}}$. We now separate this procedure into an adaptive sensing operation and a final measurement. 
In this paper, a \emph{sensor comb} is a deterministic quantum comb that connects the $N$ process slots in the fixed causal order $1\prec2\prec\cdots\prec N$ and produces a final quantum system ${\cal H}_{B,N}\otimes{\cal H}_{\rm aux}$. It represents the input preparation and all intermediate adaptive operations, but does not include the final POVM used to produce the classical estimate.

We use the comb \cite{CDP08,Chiribella_2008,CDP09,AY}
\begin{align}
L\in
Comb\bigl[
(\emptyset,{\cal H}_{1,A}),
({\cal H}_{1,B},{\cal H}_{2,A}),
\ldots,
({\cal H}_{N-1,B},
{\cal H}_{N,A}\otimes{\cal H}_{\rm aux})
\bigr]
\label{eq:sensor-comb-class}
\end{align}
to model our adaptive quantum sensor.
The comb $L$ is a positive semidefinite operator on
\begin{align}
{\cal H}_{1,A}
\otimes
{\cal H}_{1,B}
\otimes
{\cal H}_{2,A}
\otimes\cdots\otimes
{\cal H}_{N-1,B}
\otimes
{\cal H}_{N,A}
\otimes
{\cal H}_{\rm aux}
\end{align}
that satisfies the additional linear constraints
\begin{align}
\operatorname{Tr}L
&=
\prod_{k=1}^{N-1}d_{B,k}
=
\frac{\dim{\cal K}_B}{d_{B,N}},
\label{AD0T}
\end{align}
and, for $k=1,\ldots,N-1$,
\begin{align}
&
\operatorname{Tr}_{
{\rm aux},
(A,k+1),\ldots,(A,N),
(B,k+1),\ldots,(B,N-1)
}
L
\notag\\
&=
\frac{I_{(B,k)}}{d_{B,k}}
\otimes
\operatorname{Tr}_{
{\rm aux},
(A,k+1),\ldots,(A,N),
(B,k),\ldots,(B,N-1)
}
L.
\label{ADT}
\end{align}
Equation~\eqref{ADT} imposes the causal normalization condition at
each intermediate stage of the adaptive strategy, while
Eq.~\eqref{AD0T} fixes the overall normalization.

For $N=1$, Eq.~\eqref{ADT} is absent and Eq.~\eqref{AD0T} reduces to
\begin{align}
\operatorname{Tr}L=1.
\end{align}
Thus, in this case, $L$ is a density operator on
${\cal H}_{1,A}\otimes{\cal H}_{\rm aux}$.

In the following discussion, we abbreviate the comb class in
Eq.~\eqref{eq:sensor-comb-class} as $Comb$.
We use the link-product convention associated with the unnormalized
Choi definition in Eq.~\eqref{eq:Choi-convention}.  The link product
includes the transpose on all Hilbert spaces that are connected and
subsequently traced out \cite{Chiribella2009QuantumNetworks}.  Linking the process Choi
operator with the sensor comb contracts all internal slots and leaves
the final output system.

Each $L\in Comb$ induces the parameter-dependent output state
\begin{align}
\rho_\theta^L
:=&
\bar T_\theta^{(N)}*L
\notag\\
:=&
\operatorname{Tr}_{
(A,1),\ldots,(A,N),
(B,1),\ldots,(B,N-1)
}
\Bigg[
\notag\\
&
\left(
\left(\bar T_\theta^{(N)}\right)^{
T_{(A,1),\ldots,(A,N),(B,1),\ldots,(B,N-1)}
}
\otimes I_{\rm aux}
\right)
\left(
I_{(B,N)}\otimes L
\right)
\Bigg].
\label{eq:adaptive-output-state}
\end{align}
Here, the transpose is taken in the bases used to define the Choi
operators.  The resulting operator $\rho_\theta^L$ is a state on
${\cal H}_{B,N}\otimes{\cal H}_{\rm aux}$.

At the true parameter value $\theta_0$, we write
\begin{align}
\rho^L
&:=
\rho_{\theta_0}^L
=
\bar T*L,
\label{eq:adaptive-output-state-true}
\\
D_j^L
&:=
\left.
\frac{\partial\rho_\theta^L}{\partial\theta^j}
\right|_{\theta=\theta_0}
=
\bar F_j^{(N)}*L,
\qquad
j=1,\ldots,d.
\label{eq:adaptive-output-derivatives}
\end{align}
Thus, every sensor comb $L\in Comb$ induces the quantum-state model
\begin{align}
\left(
\bar T*L,
\left(
\bar F_j^{(N)}*L
\right)_{j=1}^{d}
\right)
\label{eq:adaptive-induced-state-model}
\end{align}
on the final output system
${\cal H}_{B,N}\otimes{\cal H}_{\rm aux}$.

\subsection{Conic-programming bounds for adaptive strategies}
\label{subsec:adaptive-conic-bounds}

For identical repeated uses of a channel, the positive-cone
relaxation $h=6$ requires no new adaptive optimization.  Let
$J_{6,\mathrm{ad}}^{(N)}(G)$ denote the program with cone
${\cal S}_{BA}^6$ from Eq.~\eqref{eq:S6-cone} and the adaptive
normalization \eqref{eq:adaptive-X-normalization}.  Under the
assumptions of Theorem~\ref{thm:indef-S6-scaling}, or of its supported
singular extension in Proposition~\ref{prop:indef-S6-singular-extension},
the adaptive normalization operators form a subset of
$\mathfrak W_N$.  Conversely, the product operator
$R_\sigma^{(N)}=(I_B\otimes\sigma)^{\otimes N}$ from
Eq.~\eqref{eq:indef-S6-product} satisfies the adaptive recursive
normalization \eqref{eq:adaptive-causal-normalization} and the
trace condition \eqref{AD}.  Indeed, its reductions are
\begin{align}
(R_\sigma^{(N)})^{(k)}
=d_B^{N-k}(I_B\otimes\sigma)^{\otimes k},
\label{eq:adaptive-S6-product-reductions}
\end{align}
where $d_B=\dim{\cal H}_B$ and the factors follow the fixed slot
order.  For $\sigma>0$ with $F(\sigma)>0$, the fixed-normalization
quadratic minimization in
Lemma~\ref{lem:indef-S6-quadratic-minimization}, together with
Eq.~\eqref{eq:indef-S6-product-realization}, constructs a feasible
positive-cone variable with objective
$\operatorname{Tr}[G(NF(\sigma))^{-1}]$.  Singular input states are
included in the infimum by taking
$\sigma_\varepsilon=(1-\varepsilon)\sigma+
\varepsilon I_A/d_A$ and using continuity wherever $F(\sigma)>0$.
Consequently,
\begin{align}
J_{6,\mathrm{gen}}^{(N)}(G)
&\stackrel{(a)}{\leq}J_{6,\mathrm{ad}}^{(N)}(G)
\stackrel{(b)}{\leq}
\inf_{\substack{\sigma\geq0,\ \Tr\sigma=1\\F(\sigma)>0}}
\Tr[G(NF(\sigma))^{-1}]\notag\\
&\stackrel{(c)}{=}J_{6,\mathrm{gen}}^{(N)}(G)
=\frac{J_{6,\mathrm{gen}}^{(1)}(G)}{N}.
\label{eq:adaptive-S6-exact}
\end{align}
Here (a) uses the inclusion of adaptive testers in the general
process-matrix tester class; (b) uses the adaptive product
realization and fixed-normalization minimization; and (c) is
Theorem~\ref{thm:indef-S6-scaling}, or
Proposition~\ref{prop:indef-S6-singular-extension} in the supported
singular case.  Thus the $h=6$ value is already known exactly for
identical repeated channels, so the reduction below addresses
$h=1,\ldots,5$.  This conclusion is not asserted for an arbitrary
nonidentical $N$-slot process family.

For a fixed sensor comb $L\in Comb$, the parameter-dependent output
state is $\rho_\theta^L=\bar T_\theta^{(N)}*L$.  Hence, at the true
parameter value $\theta_0$, the induced state-estimation model is
\begin{align}
\left(\rho^L,(D_j^L)_j\right)
=
\left(
\bar T*L,
\left(\bar F_j^{(N)}*L\right)_j
\right).
\label{eq:adaptive-induced-model}
\end{align}
The fixed comb $L$ yields the corresponding state-estimation bounds
$S_h[
\bar T*L,
(\bar F_j^{(N)}*L)_j]$ for $h=1,\ldots,5$.
The remaining optimization is over the sensor comb.  To compare this
two-stage description with the direct generalized-measurement
formulation, we now write the adaptive process-side programs
explicitly.

We first state the process-side cone and equality constraints for the
adaptive fixed-order strategy.  For $h=1,\ldots,4$, let
${\mathfrak F}^{\rm ad}_h$ be the set of all operators
$X\in{\cal S}_{BA}^h$ whose normalization block
$R_X:=X_{0,0}$ satisfies
\begin{align}
\begin{aligned}
R_X^{(k)}
&=
\frac{I_{(B,k)}}{d_{B,k}}
\otimes
\operatorname{Tr}_{(B,k)}R_X^{(k)},
\qquad k=1,\ldots,N,
\\
\operatorname{Tr}R_X
&=
\dim{\cal K}_B,
\end{aligned}
\label{eq:adaptive-X-normalization}
\end{align}
and whose parameter blocks satisfy the local-unbiasedness conditions
\eqref{c2-G}.
Here, $R_X^{(k)}$ denotes the reduction of $R_X$ defined in
Eq.~\eqref{eq:reduced-adaptive-normalization}.

For the adaptive fixed-order strategy, the process-side conic programs
are therefore
\begin{align}
J_{h,\mathrm{ad}}^{(N)}(G)=
\min_{X \in {\cal S}_{BA}^h}
\Bigl\{
\operatorname{Tr}[(G\otimes\bar T)X]
\ \Big|\
X \hbox{ satisfies }
\hbox{\eqref{eq:adaptive-X-normalization}, \eqref{c2-G}.} 
\Bigr\}
\label{eq:adaptive-Jh-explicit}
\end{align}
for $h=1,\ldots,4$, and 
the same relation holds by replacing 
${\cal S}_{BA}^h$ by 
${\cal S}_{BA}^5(\bar T)$ for $h=5$.

The following theorem reduces each adaptive process-side conic program in Eq.~\eqref{eq:adaptive-Jh-explicit} to the optimization of the corresponding state-estimation bound over adaptive sensor combs.

\begin{theorem}
\label{thm:adaptive-comb-reduction}
For the adaptive fixed-order conic programs defined in
Eq.~\eqref{eq:adaptive-Jh-explicit}, we have, for
$h=1,\ldots,5$,
\begin{align}
J_{h,\mathrm{ad}}^{(N)}(G)
=
\inf_{L\in Comb}
S_h
\left[
\bar T*L,
\left(\bar F_j^{(N)}*L\right)_j
\right].
\label{thm-rhs}
\end{align}
When the infimum on the right-hand side is attained, it may be
replaced by a minimum.
\end{theorem}

Theorem~\ref{thm:adaptive-comb-reduction} identifies the process-side conic programs with the optimization of the corresponding state-estimation bounds over all adaptive sensor combs. Thus, the optimization over a generalized measurement on the entire parameter-dependent process can be separated conceptually into two parts: the choice of an adaptive sensor comb and the evaluation of a state-estimation bound for the output model induced by that comb.

The identity for $h=1$ has a direct operational meaning. The process-side quantity $J_{1,\mathrm{ad}}^{(N)}(G)$ is the minimum weighted mean-square error over locally unbiased adaptive estimators, while the right-hand side first fixes an adaptive sensing operation and then optimizes the final measurement on its output state. The theorem therefore shows that the generalized-measurement description and the sensor-comb description give the same operational precision limit.

The same reduction applies for $h=2,3,4,5$ to the corresponding Cram\'er--Rao-type bounds. In particular, the hierarchy among the process-side conic programs is inherited from the hierarchy of the state-estimation bounds for each induced output model.

The proof of Theorem~\ref{thm:adaptive-comb-reduction}
uses feasible-point constructions in both directions.
For a fixed sensor comb $L$, the adjoint of its link map
sends each feasible state-side block operator to a feasible
process-side operator with the same objective.  Conversely,
a process-side operator with positive-definite normalization
block $R=X_{00}$ admits a sensor comb $L_R$ and an invertible
operator $V_R$ satisfying
\begin{align}
V_RV_R^\dagger&=R,\qquad
V_R^\dagger\Lambda V_R=\Lambda*L_R.
\label{eq:adaptive-main-text-factorization}
\end{align}
The inverse congruence
\begin{align}
Y=(I_{\cal R}\otimes V_R^{-1})X
(I_{\cal R}\otimes V_R^{-\dagger})
\label{eq:adaptive-main-text-inverse-congruence}
\end{align}
then gives a feasible state-side operator with the same
objective.  Full-rank approximation extends the comparison
to the original feasible domain, yielding the equality
\eqref{thm-rhs}.  Appendix~\ref{subsec:proof-adaptive-comb-reduction}
proves the factorization and verifies normalization, cone
membership, local unbiasedness, and objective preservation
in both directions.

\section{Application to noisy three-parameter field sensing}
\label{sec:application-field-sensing}
We apply the preceding framework to a noisy three-parameter field-sensing
model.  The numerical illustration compares the process-side bounds
$J_{2,\mathrm{ad}}^{(N)}(G)$ and $J_{5,\mathrm{ad}}^{(N)}(G)$ for the same adaptive fixed-order estimation problem.
Thus, the two quantities use the same physical channel family and the
same adaptive normalization conditions; they differ only in the cone
imposed on the process-side optimization variable.

\subsection{Collective noisy-field channel}
\label{subsec:noisy-field-sensing-channels}
Let
\begin{align}
\theta=(\theta_1,\theta_2,\theta_3)\in\mathbb R^3
\end{align}
be the unknown field parameter.  A single channel use acts on the
permutation-symmetric subspace of $q$ qubits, which has dimension
$q+1$.  On this subspace, define the collective spin operators by
\begin{align}
J_a^{(q)}:=\frac{1}{2}\sum_{\ell=1}^{q}\tau_a^{(\ell)},
\qquad a=x,y,z,
\label{eq:collective-spin-generators}
\end{align}
where $\tau_x,\tau_y,\tau_z$ are the Pauli operators and
$\tau_a^{(\ell)}$ acts on the $\ell$-th qubit.  The field Hamiltonian
is
\begin{align}
H_\theta^{(q)}
:=\theta_1J_x^{(q)}+\theta_2J_y^{(q)}+\theta_3J_z^{(q)}.
\label{eq:field-sensing-Hamiltonian}
\end{align}

To model collective energy loss, we introduce the numerical lowering
matrix used in the implementation,
\begin{align}
A^{(q)}
:=\frac{1}{2}\sum_{\ell=1}^{q}\tau_-^{(\ell)},
\qquad
\tau_-:=|0\rangle\langle1|.
\label{eq:collective-lowering-numerical}
\end{align}
The symbol $A^{(q)}$ is used here to distinguish this implementation-specific
normalization from other conventions for collective ladder operators.
We take the jump operator to be
\begin{align}
L_\gamma^{(q)}:=\gamma A^{(q)}.
\label{eq:jump-operator-gamma}
\end{align}
For an operator $A$, write
\begin{align}
{\cal D}_{A}(\rho)
:=A\rho A^\dagger
-\frac{1}{2}\left(A^\dagger A\rho+\rho A^\dagger A\right).
\label{eq:collective-damping-dissipator}
\end{align}
The Lindblad generator used in the calculation is then
\begin{align}
{\cal L}_{\theta,\gamma}^{(q)}(\rho)
&:=-i[H_\theta^{(q)},\rho]
+{\cal D}_{L_\gamma^{(q)}}(\rho)
\notag\\
&=-i[H_\theta^{(q)},\rho]
+\gamma^2{\cal D}_{A^{(q)}}(\rho).
\label{eq:field-sensing-Liouvillian}
\end{align}
Thus, $\gamma$ is the amplitude multiplying the jump operator, rather
than the coefficient multiplying the dissipator.  The corresponding
single-use channel at interaction time $t=1$ is
\begin{align}
\Lambda_{\theta,\gamma}^{(q)}
:=\exp\!\left({\cal L}_{\theta,\gamma}^{(q)}\right).
\label{eq:field-sensing-channel}
\end{align}

\subsection{Local model with identical channel uses}
\label{subsec:application-repeated-identical-channel}
The numerical model consists of $N$ identical uses of
$\Lambda_{\theta,\gamma}^{(q)}$.  If $T_{\theta,\gamma}^{(q)}$ denotes
the single-use Choi operator, the complete process Choi operator is
\begin{align}
\bar T_{\theta,\gamma}^{(N,q)}
:=\left(T_{\theta,\gamma}^{(q)}\right)^{\otimes N}.
\label{eq:application-N-use-Choi}
\end{align}
The estimation point and weight matrix are
\begin{align}
\theta_0=(0,0,0),
\qquad
G=I_3.
\label{eq:application-local-data}
\end{align}
The process operator and its three local derivatives are
\begin{align}
\bar T
&:=\bar T_{\theta_0,\gamma}^{(N,q)},
\\
\bar F_{j,\gamma}^{(N,q)}
&:=\left.
\frac{\partial}{\partial\theta_j}
\bar T_{\theta,\gamma}^{(N,q)}
\right|_{\theta=\theta_0},
\qquad j=1,2,3.
\label{eq:application-N-use-derivatives}
\end{align}
Here, $q$ denotes the number of qubits in one collective channel block,
whereas $N$ denotes the number of identical channel uses and hence the
number of adaptive slots.  We use $q$ rather than the theoretical
repetition index $n$ of Subsection~\ref{subsec:repeated-channel-model}.
The archived numerical filenames use the label $n=q$.

The input and output dimensions of one symmetric-subspace channel are
both $q+1$.  Therefore,
\begin{align}
\dim{\cal K}_A=\dim{\cal K}_B=(q+1)^N.
\end{align}
Since $d=3$ and $\dim{\cal R}_C=4$, the process-side variable $X$ is a
matrix of dimension
\begin{align}
\dim X
=4(q+1)^{2N}.
\label{eq:application-X-dimension}
\end{align}
For example, $X$ is $64\times64$ only when $q=1$ and $N=2$.

\subsection{Process-side bounds under adaptive normalization}
\label{subsec:application-process-bounds}
Both $J_{2,\mathrm{ad}}^{(N)}(G)$ and $J_{5,\mathrm{ad}}^{(N)}(G)$ minimize, for every displayed pair $(q,N)$,
\begin{align}
\operatorname{Tr}[(G\otimes\bar T)X]
\end{align}
subject to the same total-trace condition, the same local-unbiasedness
conditions, and the same recursive normalization conditions for an
$N$-slot adaptive fixed-order strategy.  The quantity $J_{2,\mathrm{ad}}^{(N)}(G)$ uses the
cone ${\cal S}_{BA}^2$, whereas $J_{5,\mathrm{ad}}^{(N)}(G)$ uses
${\cal S}_{BA}^5(\bar T)$, including its additional antisymmetric trace
conditions.  Consequently, the two curves compare distinct conic
relaxations of one adaptive process-estimation problem.

Theorem~\ref{thm:adaptive-comb-reduction} expresses each process-side bound as the infimum of the corresponding state-side
bound over adaptive sensor combs.  The optimizing sensor comb may
depend on the chosen cone, and the numerical study evaluates the two
bounds separately under the same adaptive normalization.

\subsection{Numerical implementation}
\label{subsec:numerical-implementation}
The channel action in Eq.~\eqref{eq:field-sensing-channel} was evaluated
by an iteratively truncated Taylor expansion of the Liouvillian
evolution.  At most $100$ terms were retained, and the iteration was
stopped when the norm of the newest term fell below $10^{-10}$.  The
output was rescaled to preserve the input trace, and the resulting Choi
operator was Hermitian-symmetrized.

Analytically, the derivative of the tensor-power process satisfies
\begin{align}
\left.\frac{\partial}{\partial\theta_j}
\left(T_{\theta,\gamma}^{(q)}\right)^{\otimes N}
\right|_{\theta=\theta_0}
=
\sum_{k=1}^{N}
T^{\otimes(k-1)}\otimes
\left.\frac{\partial T_{\theta,\gamma}^{(q)}}{\partial\theta_j}
\right|_{\theta=\theta_0}
\otimes T^{\otimes(N-k)},
\label{eq:application-exact-tensor-derivative}
\end{align}
where $T:=T_{\theta_0,\gamma}^{(q)}$.
The process derivatives used in the conic programs were first-order
forward finite differences of the complete $N$-use Choi operator.  With
$\delta=10^{-10}$ and with $e_j$ denoting the $j$-th coordinate vector,
we used
\begin{align}
\bar F_{j,\gamma}^{(N,q)}
\simeq
\frac{
\bar T_{\theta_0+\delta e_j,\gamma}^{(N,q)}
-
\bar T_{\theta_0,\gamma}^{(N,q)}
}{\delta}.
\label{eq:application-finite-difference}
\end{align}
Thus, the perturbed single-use Choi operator was tensor-powered before
the finite difference was formed, and every one of the $N$ identical
channel factors contributes to the derivative.

The semidefinite programs were formulated using the \texttt{Julia} language in \texttt{JuMP} and solved with \texttt{SCS} using its indirect linear-system solver.  The maximum iteration count
was $20000$, and both the absolute and relative solver tolerances were
$10^{-6}$.  These solver tolerances are distinct from the
$10^{-10}$ finite-difference step and Taylor-series stopping threshold.
There are two distinct outstanding checks. First, the surviving calculation does not include a finite-difference step-size
convergence study for derivative accuracy. Second, trace rescaling and Hermitian
symmetrization do not by themselves certify complete positivity and
trace preservation of the numerically constructed channel.  A step-size study of the derivatives and a direct audit of Choi positivity and partial-trace residuals are therefore separate numerical-validation tasks; neither follows from the SDP tolerance.

\section{Numerical results}
\label{sec:numerical-results}
\begin{figure*}[htbp]
\centering
\includegraphics[width=0.9\textwidth]{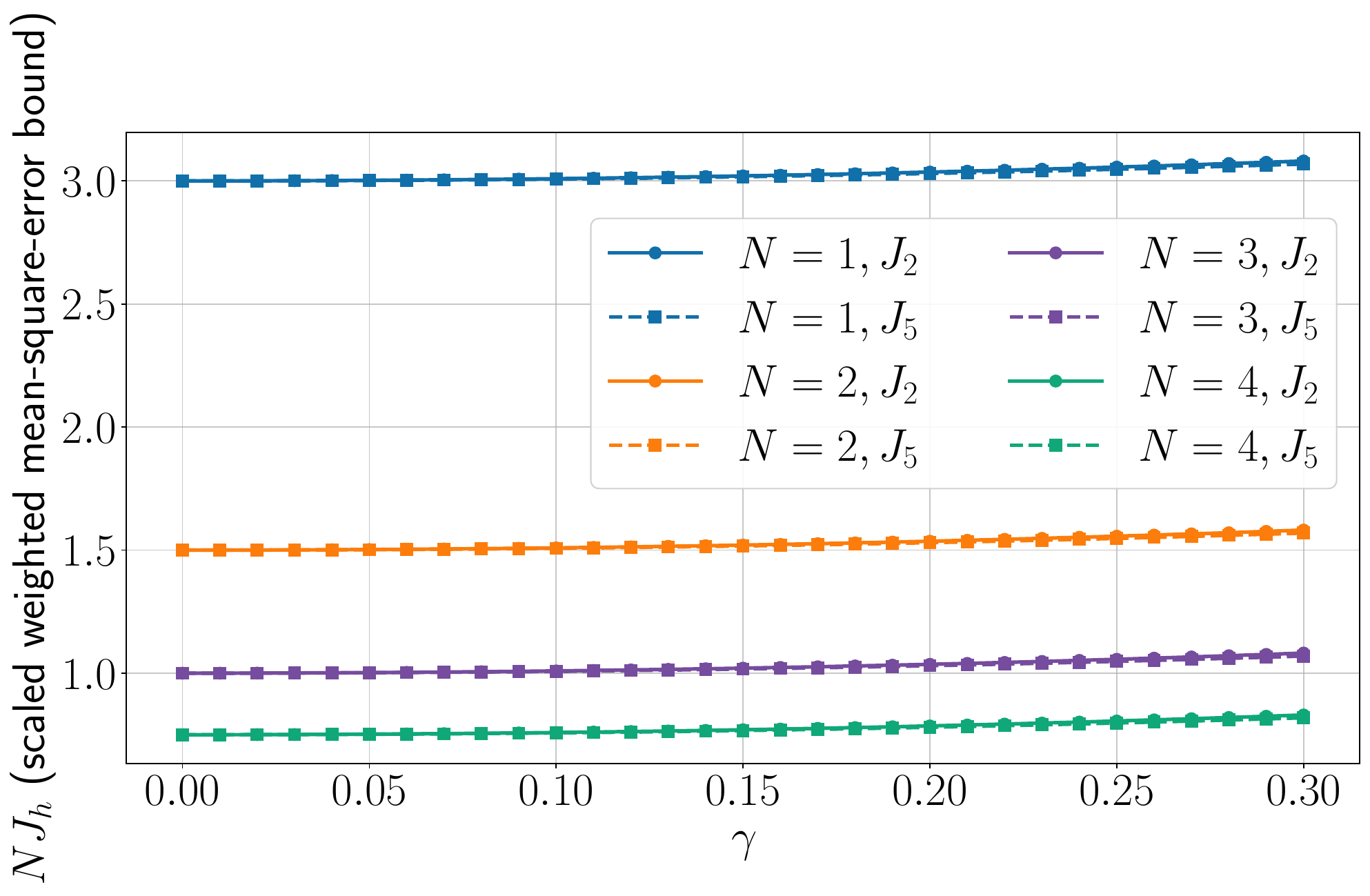}
\caption{Scaled process-side bounds $N J_{2,\mathrm{ad}}^{(N)}(G)$ and $N J_{5,\mathrm{ad}}^{(N)}(G)$ as functions of
the jump-operator amplitude $\gamma$ for $q=1$ and
$N=1,2,3,4$.  Solid lines show $N J_{2,\mathrm{ad}}^{(N)}(G)$ and dashed lines show $N J_{5,\mathrm{ad}}^{(N)}(G)$.}
\label{fig:n1}
\end{figure*}

\begin{figure*}[htbp]
\centering
\includegraphics[width=0.9\textwidth]{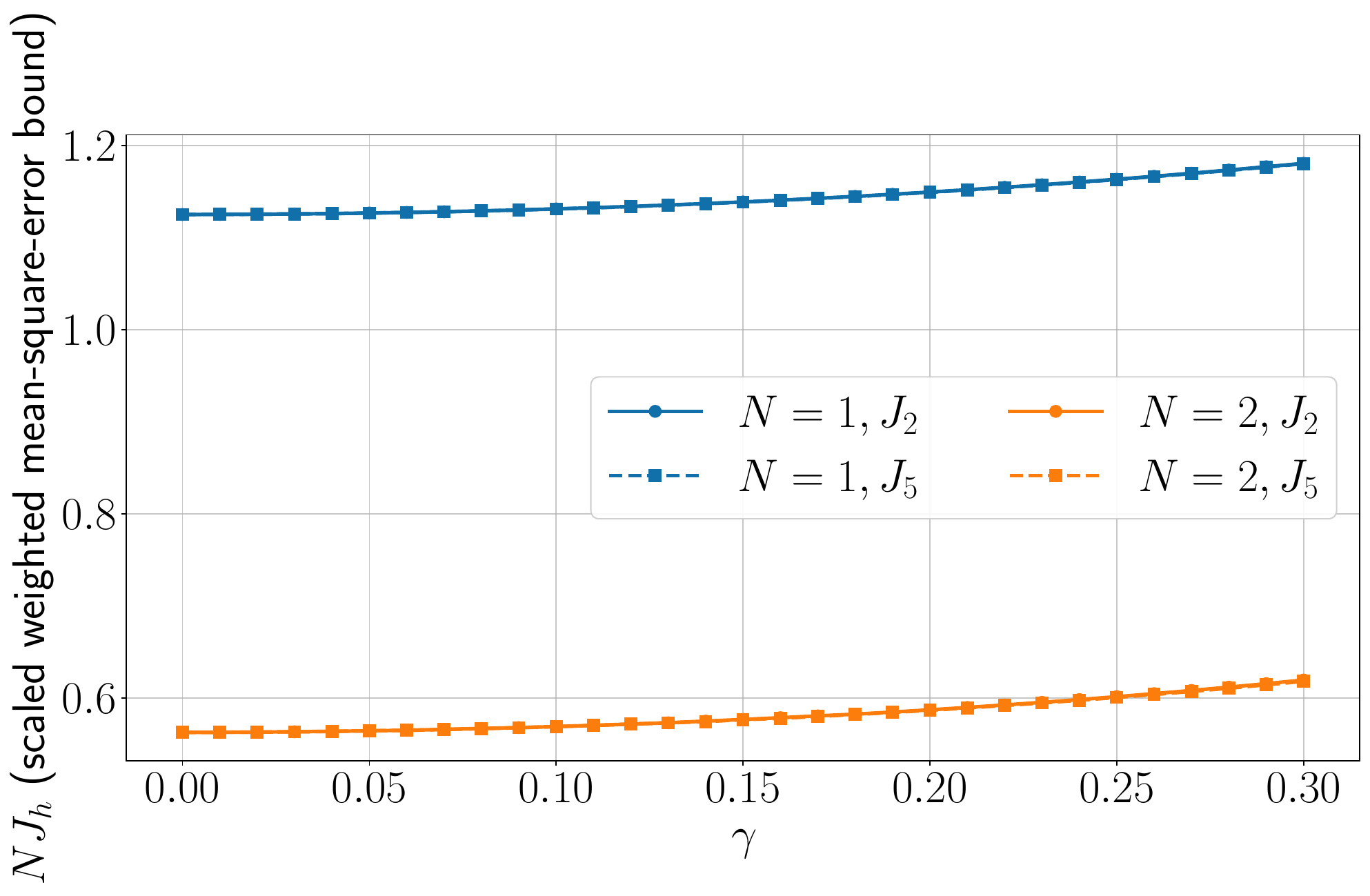}
\caption{Scaled process-side bounds $N J_{2,\mathrm{ad}}^{(N)}(G)$ and $N J_{5,\mathrm{ad}}^{(N)}(G)$ as functions of
the jump-operator amplitude $\gamma$ for $q=2$ and $N=1,2$.  Solid
lines show $N J_{2,\mathrm{ad}}^{(N)}(G)$ and dashed lines show $N J_{5,\mathrm{ad}}^{(N)}(G)$.}
\label{fig:n2}
\end{figure*}

\begin{figure*}[htbp]
\centering
\includegraphics[width=0.9\textwidth]{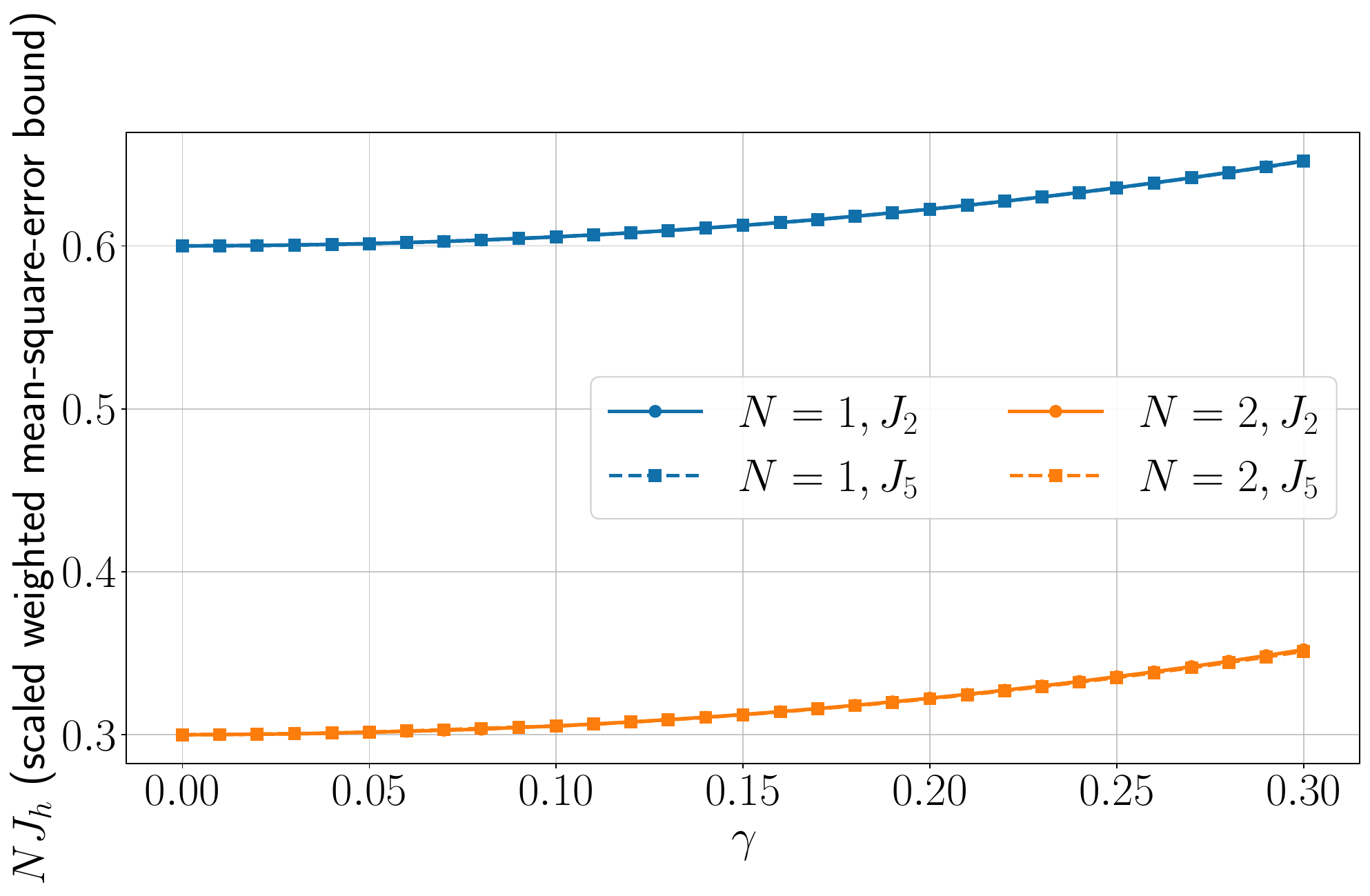}
\caption{Scaled process-side bounds $N J_{2,\mathrm{ad}}^{(N)}(G)$ and $N J_{5,\mathrm{ad}}^{(N)}(G)$ as functions of
the jump-operator amplitude $\gamma$ for $q=3$ and $N=1,2$.  Solid
lines show $N J_{2,\mathrm{ad}}^{(N)}(G)$ and dashed lines show $N J_{5,\mathrm{ad}}^{(N)}(G)$.}
\label{fig:n3}
\end{figure*}

Figures~\ref{fig:n1}, \ref{fig:n2}, and~\ref{fig:n3} show the scaled
quantities $N J_{2,\mathrm{ad}}^{(N)}(G)$ and $N J_{5,\mathrm{ad}}^{(N)}(G)$ as functions of $\gamma$.  The factor
$N$ is applied after optimization for presentation and is not part of
the conic objective.  The data use $q=1,2,3$ and, respectively,
$N=1,2,3,4$ for $q=1$ and $N=1,2$ for $q=2,3$.

Across the recorded grid $0\leq\gamma\leq0.3$, both bounds generally
increase with the jump-operator amplitude.  The two values remain
close for every displayed $(q,N)$.  The largest recorded absolute
difference between the unscaled bounds is approximately
$1.25\times10^{-2}$, and the largest relative difference is
approximately $1.41\%$.  Small violations of the exact ordering
$J_{5,\mathrm{ad}}^{(N)}(G)\leq J_{2,\mathrm{ad}}^{(N)}(G)$ occur in the recorded outputs, with the largest reversal
being approximately $2.87\times10^{-4}$.  Because the programs were
solved approximately and the collated tables do not retain run-level
primal residuals, dual residuals, or duality gaps, these violations are
regarded as unresolved numerical inaccuracies.  The close agreement shows only that the two process-side
relaxations yield numerically similar values for this example.  

\section{Discussion}
\label{sec:discussions}
The conic formulation separates the operational optimization over
generalized measurements from tractable relaxations of its block
operator cone. For an arbitrary number of independently normalized
channel-use slots, explicit linear conditions characterize the
general process-matrix tester class. This class contains fixed-order
testers and, when the slot spaces admit them, causally nonseparable
strategies with indefinite causal order. Independent variation of quantum channels
also eliminates cross-slot contributions to the real RLD-type
information matrix used in the positive-cone analysis.

Our principal scaling result extends an RLD-based restriction previously available for sequential or adaptive channel estimation to the general process-matrix tester class, including indefinite causal order. The set of real RLD-type information matrices is exactly $N$ times the one-use set, giving $J_{6,\mathrm{gen}}^{(N)}(G)=J_{6,\mathrm{gen}}^{(1)}(G)/N$. Under the rank and support hypotheses of Corollary~\ref{cor:indef-operational-SQL}, the optimal locally unbiased weighted MSE $J_{1,\mathrm{gen}}^{(N)}(G)$ and the intermediate conic bounds $J_{h,\mathrm{gen}}^{(N)}(G)$, $h=2,\ldots,5$, have order $1/N$. Thus, indefinite causal order does not restore Heisenberg scaling for the channel families covered by these assumptions. This result does not rule out Heisenberg scaling for unitary channels or other models outside the stated regime. The recent single-parameter asymptotic analysis of general indefinite causal order~\cite{WeiLiPang2026} addresses a complementary question; the present contribution is an exact multiparameter information-matrix-set characterization under the conditions stated above.

For fixed-order adaptive strategies, our second principal result is an exact reduction: for each $h=1,\ldots,5$ and a general positive-semidefinite weight matrix $G$, the process-side conic bound equals the corresponding state-estimation bound minimized over sensor combs. The proof constructs a sensor comb from a positive-definite process-side normalization operator and uses full-rank approximation to cover the general feasible domain.

The field-sensing calculation compares $J_{2,\mathrm{ad}}^{(N)}(G)$ and $J_{5,\mathrm{ad}}^{(N)}(G)$ under
one adaptive fixed-order normalization and finds similar values on
the recorded grid. A comparison with the operational value $J_{1,\mathrm{ad}}^{(N)}(G)$ or
between parallel and adaptive strategy classes would require
separate calculations.

The recorded values also contain small violations of the exact
hierarchy $J_{5,\mathrm{ad}}^{(N)}(G)\leq J_{2,\mathrm{ad}}^{(N)}(G)$.  These violations are treated as unresolved
numerical inaccuracies because the collated data do not contain the
run-level residuals and duality gaps needed for a sharper diagnosis.
Further numerical validation should separately examine finite-difference
step-size stability and the complete positivity and trace-preservation
residuals of the constructed Choi operators.
\section*{Acknowledgement}
M.H. was supported in part by the General R\&D Projects of the 1+1+1
CUHK--CUHK(SZ)--GDST Joint Collaboration Fund (Grant No.~GRDP2025-022),
the Guangdong Provincial Quantum Science Strategic Initiative
(Grant No.~GDZX2505003), and
the National Natural Science Foundation of China (Grant No. 62171212).
YO acknowledges support from EPSRC (Grant No. EP/W028115/1).
Microsoft Copilot GPT 5.6 model was used during the final preparation of the manuscript, after the authors had established the main results and the overall framework of the paper, to assist in organizing the presentation of the arguments and improving clarity. The authors reviewed and verified all suggested revisions and take full responsibility for the content of the manuscript.

\appendix

\section{Proofs of the exact scaling results}
\label{subsec:proof-indef-S6-scaling}

The scaling proofs have two main targets: the reduction of
$J_{6,\mathrm{gen}}^{(N)}(G)$ to the information matrices $F^{(N)}(R)$ in
Eq.~\eqref{eq:indef-S6-inf-reduction}, and the matrix-set identity
\eqref{eq:indef-S6-matrix-set}. The full-rank approximation and
fixed-$R$ quadratic minimization establish the first target.
The one-slot decomposition and product construction establish
the second. Combining them proves the exact scaling formula
\eqref{eq:indef-S6-scaling}.

\subsection{Full-rank approximation and constrained quadratic minimization}

\begin{lemma}[Full-rank approximation of normalization blocks]
\label{lem:indef-S6-full-rank}
The infimum of the positive-cone conic program defining
$J_{6,\mathrm{gen}}^{(N)}(G)$ is equal to the infimum over its feasible
operators $X$ satisfying $X_{00}>0$.
\end{lemma}

\begin{proof}
Let $R_0=I_{{\cal K}_B\otimes{\cal K}_A}/\dim{\cal K}_A$.
For any independent CPTP Choi operators $Z_1,\ldots,Z_N$,
\begin{align}
\Tr\!\left[R_0\bigotimes_{k=1}^NZ_k\right]
&\stackrel{(a)}{=}\frac{\prod_k\Tr Z_k}{\dim{\cal K}_A}
\stackrel{(b)}{=}1.
\label{eq:indef-S6-full-rank-R0}
\end{align}
Here (a) uses the definition of $R_0$ and the tensor-product
trace formula; (b) follows from $\Tr Z_k=\dim{\cal H}_{A,k}$.
Thus $R_0>0$ and $R_0\in\mathfrak W_N$ by
Theorem~\ref{thm:indef-N-normalization}.

Fix a feasible $X\geq0$, write $R=X_{00}$ and $A_i=X_{i0}$,
and choose $0<\epsilon<1$. Set
\begin{align}
H_\epsilon&:=\operatorname{diag}\bigl(
\sqrt{1-\epsilon},(1-\epsilon)^{-1/2}I_d\bigr),\nonumber\\
X_\epsilon&:=(H_\epsilon\otimes I)X(H_\epsilon\otimes I)
+\epsilon |0\rangle\langle0|\otimes R_0.
\label{eq:indef-S6-block-regularization}
\end{align}
Both summands are positive semidefinite, so $X_\epsilon\geq0$.
Its relevant blocks are
\begin{align}
(X_\epsilon)_{00}&=(1-\epsilon)R+\epsilon R_0>0,\notag\\
(X_\epsilon)_{i0}&=A_i,\qquad
(X_\epsilon)_{ij}=\frac{X_{ij}}{1-\epsilon}\quad(i,j\geq1).
\label{eq:indef-S6-regularized-blocks}
\end{align}
The first identity preserves normalization because
$R,R_0\in\mathfrak W_N$ and this set is convex. The second
preserves every local-unbiasedness constraint. The third gives
\begin{align}
\sum_{i,j}g_{ij}\Tr\bigl(T_N(X_\epsilon)_{ij}\bigr)
&\stackrel{(a)}{=}\frac{1}{1-\epsilon}
\sum_{i,j}g_{ij}\Tr(T_NX_{ij})\notag\\
&\xrightarrow[\epsilon\downarrow0]{}
\sum_{i,j}g_{ij}\Tr(T_NX_{ij}).
\label{eq:indef-S6-full-rank-objective-limit}
\end{align}
Here (a) uses Eq.~\eqref{eq:indef-S6-regularized-blocks}.
The restricted feasible set is contained in the original one.
Conversely, every original feasible objective value is a limit
of restricted feasible values by the displayed construction.
Therefore the two infima agree.
\end{proof}

\begin{lemma}[Constrained quadratic minimization on a support space]
\label{lem:indef-S6-quadratic-minimization}
Let $T\geq0$ act on a finite-dimensional Hilbert space, with
support projection $P$, and let $R>0$ act on the same space.
Let $D_1,\ldots,D_d$ be linearly independent Hermitian operators
satisfying $D_i=PD_iP$.  For a real symmetric matrix $G>0$, set
\begin{align}
H_{ij}:=\Re\Tr(RD_iT^+D_j).
\label{eq:indef-S6-support-Gram}
\end{align}
Then $H>0$, and the positive block-matrix problem
\begin{align}
\inf_{X\geq0}\ &\sum_{i,j=1}^d g_{ij}\Tr(TX_{ij}),\notag\\
\text{subject to}\quad &X_{00}=R,\qquad
\Re\Tr(D_jX_{i0})=\delta_{ij}
\label{eq:indef-S6-abstract-minimization}
\end{align}
has value $\Tr(GH^{-1})$.  Its infimum is attained.
\end{lemma}

\begin{proof}
Each Step states the formula to be proved, which is then used in subsequent Steps.

\medskip
\noindent\textbf{Step 1 (Block reduction).}
Writing $A_i=X_{i0}$, we prove
\begin{align}
&\inf_{\substack{X\geq0,\ X_{00}=R\\
       \Re\Tr(D_jX_{i0})=\delta_{ij}}}
  \sum_{i,j}g_{ij}\Tr(TX_{ij})\notag\\
&\quad =\inf_{\substack{(A_i)_i\\
       \Re\Tr(D_jA_i)=\delta_{ij}}}
  \sum_{i,j}g_{ij}\Tr(TA_iR^{-1}A_j^\dagger).
\label{eq:indef-S6-step1-goal}
\end{align}
Put $A=(A_1,\ldots,A_d)^{\mathsf T}$ and
$Y=[X_{ij}]_{i,j=1}^d$. Since $X_{00}=R>0$,
we can write
\begin{align}
S&:=Y-AR^{-1}A^\dagger,\notag\\
S_{ij}&=X_{ij}-A_iR^{-1}A_j^\dagger.
\label{eq:indef-S6-Schur-blocks}
\end{align}
The block factorization
\begin{align}
\begin{pmatrix}R&A^\dagger\\A&Y\end{pmatrix}
&=\begin{pmatrix}I&0\\AR^{-1}&I\end{pmatrix}
\begin{pmatrix}R&0\\0&S\end{pmatrix}
\begin{pmatrix}I&R^{-1}A^\dagger\\0&I\end{pmatrix}
\label{eq:indef-S6-supported-Schur}
\end{align}
shows that $X\geq0$ is equivalent to $S\geq0$.
For fixed $A_i$, the objective separates as follows:
\begin{align}
&\sum_{i,j}g_{ij}\Tr(TX_{ij})\notag\\
&\ \stackrel{(a)}{=}
\sum_{i,j}g_{ij}\Tr\bigl[T(A_iR^{-1}A_j^\dagger+S_{ij})\bigr]
\notag\\
&\ =\sum_{i,j}g_{ij}\Tr(TA_iR^{-1}A_j^\dagger)
 +\sum_{i,j}g_{ij}\Tr(TS_{ij})\notag\\
&\ \stackrel{(b)}{=}
\sum_{i,j}g_{ij}\Tr(TA_iR^{-1}A_j^\dagger)
 +\sum_{i,j}g_{ij}\Tr(TS_{ji})\notag\\
&\ \stackrel{(c)}{=}
\sum_{i,j}g_{ij}\Tr(TA_iR^{-1}A_j^\dagger)
 +\Tr[(G\otimes T)S].
\label{eq:indef-S6-Schur-objective-decomposition}
\end{align}
Here (a) substitutes the definition of $S_{ij}$ in
Eq.~\eqref{eq:indef-S6-Schur-blocks}; (b) interchanges
$i$ and $j$ in the second sum and uses $g_{ij}=g_{ji}$;
and (c) uses the block-trace identity
$\Tr[(G\otimes T)S]=\sum_{i,j}g_{ij}\Tr(TS_{ji})$.
The second term is nonnegative, as follows:
\begin{align}
&\Tr[(G\otimes T)S]\notag\\
&\ \stackrel{(a)}{=}
\Tr[(G\otimes T)^{1/2}S(G\otimes T)^{1/2}]\notag\\
&\ \stackrel{(b)}{\geq}0.
\label{eq:indef-S6-Schur-nonnegative}
\end{align}
Here (a) uses cyclicity of the trace; (b) follows
because $S\geq0$ and $G\otimes T\geq0$.
Moreover, $S=0$ gives a positive $X$ for every $A$ by
Eq.~\eqref{eq:indef-S6-supported-Schur}, so the infimum
over $S\geq0$ of the second term is zero.
By Eq.~\eqref{eq:indef-S6-Schur-objective-decomposition},
the reduced objective is
\begin{align}
\sum_{i,j}g_{ij}\Tr(TA_iR^{-1}A_j^\dagger),
\label{eq:indef-S6-reduced-quadratic}
\end{align}
with constraints $\Re\Tr(D_jA_i)=\delta_{ij}$. This proves
Eq.~\eqref{eq:indef-S6-step1-goal}.

\medskip
\noindent\textbf{Step 2 (Hilbert--Schmidt formulation).}
For supported operators $A_i=PA_i$, define
\begin{align}
Q_i&:=T^{1/2}A_iR^{-1/2},& E_j&:=(T^+)^{1/2}D_jR^{1/2}.
\label{eq:indef-S6-gram-representative}
\end{align}
Let $\langle U,V\rangle_{\mathbb R}:=\Re\Tr(U^\dagger V)$
on the real space of operators with $PU=U$. We prove the following two identities:
\begin{align}
\Re\Tr(D_jA_i)&=\langle E_j,Q_i\rangle_{\mathbb R},
\notag\\
&\sum_{i,j}g_{ij}\Tr(TA_iR^{-1}A_j^\dagger)
\notag\\
&\quad=\sum_{i,j}g_{ij}\langle Q_i,Q_j\rangle_{\mathbb R}.
\label{eq:indef-S6-step2-goal}
\end{align}
To apply these identities to the right-hand side of
Eq.~\eqref{eq:indef-S6-step1-goal}, we first show that
replacing each $A_i$ by
$PA_i$ preserves both the constraints and the objective:
\begin{align}
\Re\Tr(D_jPA_i)
&\stackrel{(a)}{=}\Re\Tr(D_jA_i),\notag\\
\Tr[T(PA_i)R^{-1}(PA_j)^\dagger]
&\stackrel{(b)}{=}\Tr(TA_iR^{-1}A_j^\dagger).
\label{eq:indef-S6-left-support-reduction}
\end{align}
Here (a) uses $D_jP=D_j$, and (b) uses
$TP=PT=T$. We may thus impose $A_i=PA_i$.
The definitions in Eq.~\eqref{eq:indef-S6-gram-representative}
now give the first identity in
Eq.~\eqref{eq:indef-S6-step2-goal}:
\begin{align}
\langle E_j,Q_i\rangle_{\mathbb R}
&\stackrel{(a)}{=}
\Re\Tr\bigl(R^{1/2}D_jPA_iR^{-1/2}\bigr)\notag\\
&\stackrel{(b)}{=}\Re\Tr(D_jA_i)
=\delta_{ij}.
\label{eq:indef-S6-HS-constraints}
\end{align}
Here (a) expands $E_j^\dagger Q_i$ and uses
$(T^+)^{1/2}T^{1/2}=P$; (b) uses cyclicity of the trace,
$D_jP=D_j$, and $R^{-1/2}R^{1/2}=I$.
The second identity in
Eq.~\eqref{eq:indef-S6-step2-goal} follows from
\begin{align}
&\sum_{i,j}g_{ij}\Tr(TA_iR^{-1}A_j^\dagger)\notag\\
&\ \stackrel{(a)}{=}
\sum_{i,j}g_{ij}\Tr(Q_j^\dagger Q_i)\notag\\
&\ \stackrel{(b)}{=}
\sum_{i,j}g_{ij}\Re\Tr(Q_i^\dagger Q_j)\notag\\
&\ =\sum_{i,j}g_{ij}\langle Q_i,Q_j\rangle_{\mathbb R}.
\label{eq:indef-S6-HS-objective}
\end{align}
Here (a) expands $Q_j^\dagger Q_i$ and uses cyclicity
of the trace, while (b) uses $g_{ij}=g_{ji}\in\mathbb R$:
interchanging $i,j$ and taking the real part leaves the weighted sum unchanged. The change of
variables is reversible on this support space, as the identities
\begin{align}
(T^+)^{1/2}Q_iR^{1/2}
&=(T^+)^{1/2}T^{1/2}A_i=A_i,\notag\\
T^{1/2}(T^+)^{1/2}Q_i&=PQ_i=Q_i
\label{eq:indef-S6-HS-inverse}
\end{align}
show. Thus Eq.~\eqref{eq:indef-S6-step2-goal} gives an equivalent
Hilbert--Schmidt optimization, with constraints
\eqref{eq:indef-S6-HS-constraints}.

\medskip
\noindent\textbf{Step 3 (Gram matrix).}
We prove
\begin{align}
H_{ij}&=\langle E_i,E_j\rangle_{\mathbb R},
\qquad H>0.
\label{eq:indef-S6-step3-goal}
\end{align}
The Gram matrix of a family $(E_i)_{i=1}^d$ in this real Hilbert--Schmidt
space is, by definition, the matrix whose $(i,j)$ entry is
$\langle E_i,E_j\rangle_{\mathbb R}$. We now identify this matrix with $H$.
Expanding the inner product gives
\begin{align}
\langle E_i,E_j\rangle_{\mathbb R}
&=\Re\Tr\bigl(R^{1/2}D_iT^+D_jR^{1/2}\bigr)
\notag\\
&\stackrel{(a)}{=}\Re\Tr(RD_iT^+D_j)
=H_{ij}.
\label{eq:indef-S6-gram-identity}
\end{align}
Here (a) is cyclicity of the trace.
For $c\in\mathbb R^d$, put $D_c=\sum_i c_iD_i$.
Then
\begin{align}
c^{\mathsf T}Hc
&\stackrel{(a)}{=}
\left\langle\sum_i c_iE_i,\sum_j c_jE_j\right\rangle_{\mathbb R}
\notag\\
&=\left\|\sum_i c_iE_i\right\|_2^2\notag\\
&\stackrel{(b)}{=}
\|(T^+)^{1/2}D_cR^{1/2}\|_2^2.
\label{eq:indef-S6-nondegeneracy}
\end{align}
Here (a) uses Eq.~\eqref{eq:indef-S6-gram-identity},
and (b) substitutes the definition of $E_i$.
If $c\ne0$, linear independence of the Hermitian
operators $D_i$ implies $D_c\ne0$. Since $D_c=PD_cP$,
$(T^+)^{1/2}$ is invertible on $\operatorname{Ran}P$,
and $R^{1/2}$ is invertible, the last norm is positive.
This completes the proof of Eq.~\eqref{eq:indef-S6-step3-goal}.

\medskip
\noindent\textbf{Step 4 (Feasible candidate).}
Since Eq.~\eqref{eq:indef-S6-step3-goal} gives $H>0$,
we can define
\begin{align}
Q_i^*:=\sum_{k=1}^d(H^{-1})_{ik}E_k.
\label{eq:indef-S6-HS-minimizer}
\end{align}
Our goal is the pair of identities
\begin{align}
\langle E_j,Q_i^*\rangle_{\mathbb R}&=\delta_{ij},\notag\\
\langle Q_i^*,Q_j^*\rangle_{\mathbb R}&=(H^{-1})_{ij}.
\label{eq:indef-S6-step4-goal}
\end{align}
The first identity follows from
\begin{align}
\langle E_j,Q_i^*\rangle_{\mathbb R}
&\stackrel{(a)}{=}\sum_k(H^{-1})_{ik}H_{jk}
\notag\\
&\stackrel{(b)}{=}\delta_{ij}.
\label{eq:indef-S6-HS-feasibility}
\end{align}
Here (a) follows from the definition of $Q_i^*$ and
Eq.~\eqref{eq:indef-S6-gram-identity}; (b) follows from
$H^{\mathsf T}=H$ and $H^{-1}H=I$.
To prove the second identity in
Eq.~\eqref{eq:indef-S6-step4-goal}, calculate
\begin{align}
\langle Q_i^*,Q_j^*\rangle_{\mathbb R}
&\stackrel{(a)}{=}
\sum_{k,l}(H^{-1})_{ik}H_{kl}(H^{-1})_{jl}
\notag\\
&\stackrel{(b)}{=}(H^{-1})_{ij}.
\label{eq:indef-S6-HS-minimizer-Gram}
\end{align}
Here (a) again uses Eq.~\eqref{eq:indef-S6-gram-identity},
and (b) uses symmetry of $H^{-1}$ and $H^{-1}H=I$.
This establishes Eq.~\eqref{eq:indef-S6-step4-goal}.

\medskip
\noindent\textbf{Step 5 (Optimality).}
For any family $(Q_i)_i$ with $PQ_i=Q_i$ satisfying the
constraints in Eq.~\eqref{eq:indef-S6-HS-constraints}, put
$Z_i:=Q_i-Q_i^*$.
We prove the lower bound
\begin{align}
\sum_{i,j}g_{ij}\langle Q_i,Q_j\rangle_{\mathbb R}
&\geq\Tr(GH^{-1}).
\label{eq:indef-S6-step5-goal}
\end{align}
The constraints for $(Q_i)_i$ are
Eq.~\eqref{eq:indef-S6-HS-constraints}; those for
$(Q_i^*)_i$ follow from Eq.~\eqref{eq:indef-S6-HS-feasibility}.
Thus $\langle E_k,Z_j\rangle_{\mathbb R}=0$
for all $j,k$. Consequently,
\begin{align}
\langle Q_i^*,Z_j\rangle_{\mathbb R}
&\stackrel{(a)}{=}
\sum_k(H^{-1})_{ik}\langle E_k,Z_j\rangle_{\mathbb R}
\notag\\
&\stackrel{(b)}{=}0.
\label{eq:indef-S6-HS-orthogonality}
\end{align}
Here (a) expands $Q_i^*$ and (b) uses the homogeneous constraints on $Z_j$. The same identity holds
with the arguments exchanged because the real inner product is symmetric. The objective therefore has
the decomposition
\begin{align}
&\sum_{i,j}g_{ij}\langle Q_i,Q_j\rangle_{\mathbb R}\notag\\
&\ \stackrel{(a)}{=}
\sum_{i,j}g_{ij}\bigl(\langle Q_i^*,Q_j^*\rangle_{\mathbb R}
 +\langle Z_i,Z_j\rangle_{\mathbb R}\bigr)\notag\\
&\ \stackrel{(b)}{=}
\Tr(GH^{-1})
 +\sum_{i,j}g_{ij}\langle Z_i,Z_j\rangle_{\mathbb R}
\notag\\
&\ \stackrel{(c)}{=}
\Tr(GH^{-1})
 +\sum_a\left\|\sum_i(G^{1/2})_{ai}Z_i\right\|_2^2
\notag\\
&\ \stackrel{(d)}{\geq}\Tr(GH^{-1}).
\label{eq:indef-S6-fixed-R-optimality}
\end{align}
Here (a) substitutes $Q_i=Q_i^*+Z_i$ and removes the
cross terms by Eq.~\eqref{eq:indef-S6-HS-orthogonality};
(b) uses Eq.~\eqref{eq:indef-S6-HS-minimizer-Gram}
and the symmetry of $G$ and $H^{-1}$;
(c) uses $G=(G^{1/2})^{\mathsf T}G^{1/2}$ with
$G^{1/2}$ real symmetric; and (d) uses nonnegativity
of the squared norms. Equality holds at $Q_i=Q_i^*$, establishing
Eq.~\eqref{eq:indef-S6-step5-goal}.

\medskip
\noindent\textbf{Step 6 (Attainment).}
Let $A_i^*:=(T^+)^{1/2}Q_i^*R^{1/2}$, and define
$X^*$ by $X^*_{00}=R$, $X^*_{i0}=A_i^*$,
$X^*_{0i}=(A_i^*)^\dagger$, and
$X^*_{ij}=A_i^*R^{-1}(A_j^*)^\dagger$ for $i,j\geq1$.
We prove
\begin{align}
X^*&\geq0,\quad
\Re\Tr(D_jX^*_{i0})=\delta_{ij},\notag\\
\sum_{i,j}g_{ij}\Tr(TX^*_{ij})&=\Tr(GH^{-1}).
\label{eq:indef-S6-step6-goal}
\end{align}
The inverse transformation in
Eq.~\eqref{eq:indef-S6-HS-inverse} gives
\begin{align}
A_i^*&=(T^+)^{1/2}Q_i^*R^{1/2}\notag\\
&\stackrel{(a)}{=}
\sum_k(H^{-1})_{ik}(T^+)^{1/2}E_kR^{1/2}
\notag\\
&\stackrel{(b)}{=}
\sum_k(H^{-1})_{ik}T^+D_kR.
\label{eq:indef-S6-fixed-R-minimizer}
\end{align}
Here (a) substitutes the definition of $Q_i^*$ and (b) substitutes the definition of $E_k$. The matrix
$X^*$ has $S=0$, so it is positive by
Eq.~\eqref{eq:indef-S6-supported-Schur}.
By Eqs.~\eqref{eq:indef-S6-step2-goal}
and~\eqref{eq:indef-S6-step4-goal}, it satisfies
$\Re\Tr(D_jX^*_{i0})=\delta_{ij}$.
For its objective, Eqs.~\eqref{eq:indef-S6-step2-goal}
and~\eqref{eq:indef-S6-step4-goal} give
\begin{align}
&\sum_{i,j}g_{ij}\Tr(TX^*_{ij})\notag\\
&\ \stackrel{(a)}{=}
\sum_{i,j}g_{ij}\Tr(TA_i^*R^{-1}(A_j^*)^\dagger)
\notag\\
&\ \stackrel{(b)}{=}
\sum_{i,j}g_{ij}\langle Q_i^*,Q_j^*\rangle_{\mathbb R}
\notag\\
&\ \stackrel{(c)}{=}\Tr(GH^{-1}).
\label{eq:indef-S6-step6-attained-value}
\end{align}
Here (a) uses the definition of $X^*_{ij}$;
(b) uses Eq.~\eqref{eq:indef-S6-step2-goal}; and
(c) uses Eq.~\eqref{eq:indef-S6-step4-goal}.
This proves Eq.~\eqref{eq:indef-S6-step6-goal}.
Together with the lower bound in
Eq.~\eqref{eq:indef-S6-step5-goal}, it shows that the
infimum is attained and equals $\Tr(GH^{-1})$.
\end{proof}

\subsection{Reduction to the real information matrix}

\begin{lemma}[Reduction to the real information matrix]
\label{lem:indef-S6-information-reduction}
Suppose $T\geq0$ is a CPTP Choi operator, with support
projection $P$, and $D_1,\ldots,D_d$ are linearly independent
Hermitian derivatives satisfying $D_i=PD_iP$.
Define $F^{(N)}(R)$ by
Eq.~\eqref{eq:indef-S6-N-information} with $T_N^+$ in place
of $T_N^{-1}$ when $T$ is singular.  Then
\begin{align}
J_{6,\mathrm{gen}}^{(N)}(G)
=\inf_{\substack{R\in\mathfrak W_N\\F^{(N)}(R)>0}}
\Tr\!\left[G F^{(N)}(R)^{-1}\right].
\label{eq:indef-S6-inf-reduction}
\end{align}
\end{lemma}

\begin{proof}
We first fix a positive normalization operator $R$, then extend the infimum to singular $R$.

\medskip
\noindent\textbf{Step 1 (Fixed positive normalization).}
For $R\in\mathfrak W_N$ with $R>0$, we prove
\begin{align}
&\inf_{\substack{X\geq0,\ X_{00}=R\\
\Re\Tr(D_{j,N}X_{i0})=\delta_{ij}}}
\sum_{i,j}g_{ij}\Tr(T_NX_{ij})\notag\\
&\quad=\Tr\!\left[G F^{(N)}(R)^{-1}\right].
\label{eq:indef-S6-information-fixed-R}
\end{align}
We verify the assumptions of Lemma~\ref{lem:indef-S6-quadratic-minimization}.
The support projection of $T_N=T^{\otimes N}$ is $P_N=P^{\otimes N}$.
By Eq.~\eqref{eq:indef-S6-TN-DN} and $D_i=PD_iP$, we have
$T_N\geq0$ and $D_{i,N}=P_ND_{i,N}P_N$.
Moreover, tracing out all slots other than any slot $k$ gives
\begin{align}
\Tr_{\mathrm{rest}(k)}D_{i,N}
&=(\Tr T)^{N-1}D_i.
\label{eq:indef-S6-derivative-independent}
\end{align}
Indeed, the term containing $D_i$ in slot $k$ gives the displayed right-hand side,
whereas every other term vanishes because $\Tr D_i=0$ for derivatives of CPTP Choi operators.
Since $\Tr T>0$, this identity and the real linear independence of $D_1,\ldots,D_d$
imply that $D_{1,N},\ldots,D_{d,N}$ are real linearly independent.

We now identify the matrix $H$ in Eq.~\eqref{eq:indef-S6-support-Gram}
under the substitutions $T\mapsto T_N$ and $D_i\mapsto D_{i,N}$,
keeping $R$ and $G$ unchanged. To specify the operators whose Gram matrix is involved, set
\begin{align}
E_{i,N}(R)&:=(T_N^+)^{1/2}D_{i,N}R^{1/2}.
\label{eq:indef-S6-information-gram-vectors}
\end{align}
The entries of the matrix $H$ in that lemma become
\begin{align}
H_{ij}
&\stackrel{(a)}{=}
\Re\Tr(RD_{i,N}T_N^+D_{j,N})\notag\\
&\stackrel{(b)}{=}F^{(N)}(R)_{ij}\notag\\
&\stackrel{(c)}{=}
\langle E_{i,N}(R),E_{j,N}(R)\rangle_{\mathbb R}.
\label{eq:indef-S6-information-gram-identification}
\end{align}
Here (a) substitutes $T_N,D_{i,N}$ into Eq.~\eqref{eq:indef-S6-support-Gram};
(b) uses the definition~\eqref{eq:indef-S6-N-information} with $T_N^+$;
and (c) follows by expanding Eq.~\eqref{eq:indef-S6-information-gram-vectors}
and using cyclicity of the trace. The inner product $\langle U,V\rangle_{\mathbb R}:=\Re\Tr(U^\dagger V)$ is the one
used in Lemma~\ref{lem:indef-S6-quadratic-minimization}.
Thus $F^{(N)}(R)$ is the Gram matrix of the explicitly specified family
$E_{1,N}(R),\ldots,E_{d,N}(R)$, and it is positive definite by
Lemma~\ref{lem:indef-S6-quadratic-minimization}.
Applying that lemma with this identification proves Eq.~\eqref{eq:indef-S6-information-fixed-R}.
Together with Lemma~\ref{lem:indef-S6-full-rank}, this yields
\begin{align}
J_{6,\mathrm{gen}}^{(N)}(G)
&=\inf_{\substack{R\in\mathfrak W_N\\R>0}}
\Tr\!\left[G F^{(N)}(R)^{-1}\right].
\label{eq:indef-S6-information-positive-R}
\end{align}

\medskip
\noindent\textbf{Step 2 (Singular normalization operators).}
We prove that the infimum in Eq.~\eqref{eq:indef-S6-information-positive-R}
remains unchanged when its domain is enlarged to $R\in\mathfrak W_N$ with $F^{(N)}(R)>0$:
\begin{align}
&\inf_{\substack{R\in\mathfrak W_N\\R>0}}
\Tr\!\left[G F^{(N)}(R)^{-1}\right]\notag\\
&\quad=\inf_{\substack{R\in\mathfrak W_N\\F^{(N)}(R)>0}}
\Tr\!\left[G F^{(N)}(R)^{-1}\right].
\label{eq:indef-S6-information-all-R}
\end{align}
Fix $R\in\mathfrak W_N$ with $F^{(N)}(R)>0$ and set
\begin{align}
R_0&=I_{{\cal K}_B\otimes{\cal K}_A}/\dim{\cal K}_A,\notag\\
R_\epsilon&=(1-\epsilon)R+\epsilon R_0\quad(0<\epsilon<1).
\end{align}
As in Lemma~\ref{lem:indef-S6-full-rank}, $R_\epsilon\in\mathfrak W_N$ and $R_\epsilon>0$.
By the linearity of Eq.~\eqref{eq:indef-S6-N-information} in $R$,
\begin{align}
F^{(N)}(R_\epsilon)
&=(1-\epsilon)F^{(N)}(R)+\epsilon F^{(N)}(R_0)\notag\\
&\longrightarrow F^{(N)}(R)\quad(\epsilon\downarrow0).
\label{eq:indef-S6-information-regularization}
\end{align}
Since the limit is positive definite, continuity of matrix inversion gives
\begin{align}
\lim_{\epsilon\downarrow0}
\Tr\!\left[G F^{(N)}(R_\epsilon)^{-1}\right]
&=\Tr\!\left[G F^{(N)}(R)^{-1}\right].
\label{eq:indef-S6-information-continuity}
\end{align}
The right-hand infimum in Eq.~\eqref{eq:indef-S6-information-all-R}
is no larger than the left-hand one because its domain is larger.
Conversely, Eq.~\eqref{eq:indef-S6-information-continuity} shows that every admissible
value on the right can be approximated by values admitted on the left.
This establishes Eq.~\eqref{eq:indef-S6-information-all-R}; combining it with
Eq.~\eqref{eq:indef-S6-information-positive-R} proves
Eq.~\eqref{eq:indef-S6-inf-reduction}.
\end{proof}

\subsection{One-slot decomposition and realization}

\begin{lemma}[Supported one-slot decomposition and realization]
\label{lem:indef-S6-slot-additivity}
Let $T\geq0$ be a CPTP Choi operator with support projection
$P$, and let $D_i=PD_iP$ be Hermitian derivatives of a CPTP
model.  In the information matrices $F(\sigma)$ and
$F^{(N)}(R)$ use $T^+$ and $T_N^+$, respectively; for
$T>0$ these are the inverse matrices already defined.
Then the following two statements hold.
\begin{enumerate}
\item Every $R\in\mathfrak W_N$ determines input density
operators $\sigma_1,\ldots,\sigma_N$ satisfying
\begin{align}
F^{(N)}(R)=\sum_{k=1}^NF(\sigma_k).
\label{eq:indef-S6-additivity}
\end{align}
\item For every input density operator $\sigma$, the product
operator $R_\sigma^{(N)}=(I_B\otimes\sigma)^{\otimes N}$
belongs to $\mathfrak W_N$ and satisfies
\begin{align}
F^{(N)}(R_\sigma^{(N)})=NF(\sigma).
\label{eq:indef-S6-product-realization}
\end{align}
\end{enumerate}
\end{lemma}

\begin{proof}
We prove part (1) by separating the terms in the same slot
from those in distinct slots. Part (2) then follows by using
a product normalization operator.

\medskip
\noindent\textbf{Step 1 (Expansion and cross-slot cancellation).}
Let $\mathcal C_k(M)$ place $M$ in slot $k$ and $T$ in all
other slots. For $k\ne\ell$, let
$\mathcal C_{k,\ell}(U,V)$ place $U$ in slot $k$, $V$ in
slot $\ell$, and $T$ elsewhere. All tensor factors retain
their original slot order. We first prove that every
$R\in\mathfrak W_N$ satisfies
\begin{align}
\Tr\bigl[R\,\mathcal C_{k,\ell}(D_i,D_j)\bigr]=0
\qquad(k\ne\ell).
\label{eq:indef-S6-cross}
\end{align}
In particular, $D_{i,N}=\sum_k\mathcal C_k(D_i)$ by
Eq.~\eqref{eq:indef-S6-TN-DN}. The support identities give
\begin{align}
T_N^+&=(T^+)^{\otimes N},&TT^+T&=T,\notag\\
D_iT^+T&=D_i,&TT^+D_j&=D_j.
\label{eq:indef-S6-slot-support-identities}
\end{align}
The product for each pair of slots is
\begin{align}
&\mathcal C_k(D_i)T_N^+\mathcal C_\ell(D_j)\notag\\
&=\begin{cases}
\mathcal C_k(D_iT^+D_j),&k=\ell,\\
\mathcal C_{k,\ell}(D_i,D_j),&k\ne\ell.
\end{cases}
\label{eq:indef-S6-slot-product-cases}
\end{align}
In both cases, the identities in
Eq.~\eqref{eq:indef-S6-slot-support-identities} simplify the
occupied slots, while $TT^+T=T$ restores $T$ in every
remaining slot. Hence
\begin{align}
&D_{i,N}T_N^+D_{j,N}\notag\\
&\stackrel{(a)}{=}\sum_{k,\ell=1}^N
\mathcal C_k(D_i)T_N^+\mathcal C_\ell(D_j)\notag\\
&\stackrel{(b)}{=}\sum_{k=1}^N
\mathcal C_k(D_iT^+D_j)
+\sum_{k\ne\ell}\mathcal C_{k,\ell}(D_i,D_j).
\label{eq:indef-S6-product-expansion}
\end{align}
Here (a) expands $D_{i,N}$ and $D_{j,N}$ using
Eq.~\eqref{eq:indef-S6-TN-DN}; (b) applies
Eq.~\eqref{eq:indef-S6-slot-product-cases} to each pair.

We next prove Eq.~\eqref{eq:indef-S6-cross}.
Because $\Tr_BD_i=0$ and $D_i=PD_iP$, the operators
$T+sD_i$ and $T+tD_j$ are CPTP for sufficiently small
real $s,t$. Indeed, the partial traces remain $I_A$,
and positivity is preserved on the support of $T$.
Fix $k\ne\ell$. Let $Z_k(s,t)=T+sD_i$,
$Z_\ell(s,t)=T+tD_j$, and $Z_m(s,t)=T$ otherwise.
The independent-slot normalization
Eq.~\eqref{eq:indef-all-CPTP} holds throughout a neighborhood
of $(s,t)=(0,0)$; hence
\begin{align}
0&\stackrel{(a)}{=}
\left.\frac{\partial^2}{\partial s\,\partial t}
\Tr\!\left[R\bigotimes_{m=1}^NZ_m(s,t)\right]
\right|_{s=t=0}\notag\\
&\stackrel{(b)}{=}
\Tr\bigl[R\,\mathcal C_{k,\ell}(D_i,D_j)\bigr].
\label{eq:indef-S6-cross-derivative}
\end{align}
Here (a) differentiates the constant normalization value
one; (b) selects the coefficient of $st$ from the two
independently perturbed slots. Thus
Eq.~\eqref{eq:indef-S6-cross} holds, and the second sum in
Eq.~\eqref{eq:indef-S6-product-expansion} has zero trace
against $R$.

\medskip
\noindent\textbf{Step 2 (One-slot normalization).}
To evaluate the surviving term from slot $k$ using a trace
on the one-slot Hilbert space, define
\begin{align}
W_k:=\Tr_{\mathrm{rest}(k)}\!\left[
(I_k\otimes T_{\mathrm{rest}(k)}^{1/2})R
(I_k\otimes T_{\mathrm{rest}(k)}^{1/2})\right].
\label{eq:indef-S6-reduced-W}
\end{align}
We prove that there is a density operator $\sigma_k$ with
\begin{align}
W_k=I_B\otimes\sigma_k,
\qquad \sigma_k\geq0,\quad\Tr\sigma_k=1.
\label{eq:indef-S6-sigma}
\end{align}
In this expression $I_k$ acts on slot $k$, while
$T_{\mathrm{rest}(k)}$ is the tensor product of the other
$N-1$ copies of $T$, with factors placed in their original
order. The sandwich construction shows $W_k\geq0$.
For every one-slot CPTP Choi operator $Z$,
\begin{align}
\Tr(W_kZ)
&\stackrel{(a)}{=}\Tr\bigl[R\,\mathcal C_k(Z)\bigr]
\stackrel{(b)}{=}1.
\label{eq:indef-S6-reduced-W-normalization}
\end{align}
Here (a) follows from the partial-trace definition of
$W_k$ and cyclicity of the trace; (b) uses
Eq.~\eqref{eq:indef-all-CPTP} with $Z$ in slot $k$ and
$T$ elsewhere. Write $d_B:=\dim\mathcal H_B$. The $N=1$
case of Theorem~\ref{thm:indef-N-normalization} gives
$\Tr W_k=d_B$ and $\Pi_1W_k=0$. Since $\Pi_1=Q_1$
by Eq.~\eqref{eq:indef-PiN}, we obtain
\begin{align}
0&\stackrel{(a)}{=}Q_1W_k
\stackrel{(b)}{=}W_k-\frac{I_B}{d_B}\otimes\Tr_B W_k.
\label{eq:indef-S6-one-slot-Q}
\end{align}
Here (a) uses the one-slot normalization condition, and
(b) uses the definition of $Q_1$ in
Eq.~\eqref{eq:indef-PQ}. Now define
$\sigma_k:=d_B^{-1}\Tr_B W_k$. Because $W_k\geq0$,
we have $\sigma_k\geq0$, and
\begin{align}
\Tr\sigma_k
&\stackrel{(a)}{=}\frac{\Tr W_k}{d_B}=1.
\label{eq:indef-S6-one-slot-sigma-trace}
\end{align}
Here (a) is the partial-trace identity, and the last
equality uses $\Tr W_k=d_B$. Finally,
Eq.~\eqref{eq:indef-S6-one-slot-Q} gives
$W_k=I_B\otimes\sigma_k$, completing
Eq.~\eqref{eq:indef-S6-sigma}.

\medskip
\noindent\textbf{Step 3 (Additivity).}
We prove Eq.~\eqref{eq:indef-S6-additivity} by evaluating
its $(i,j)$ entry. The calculation is
\begin{align}
F^{(N)}(R)_{ij}
&\stackrel{(a)}{=}\sum_{k=1}^N
\Re\Tr\bigl[R\,\mathcal C_k(D_iT^+D_j)\bigr]\notag\\
&\stackrel{(b)}{=}\sum_{k=1}^N
\Re\Tr(W_kD_iT^+D_j)\notag\\
&\stackrel{(c)}{=}\sum_{k=1}^N
\Re\Tr[(I_B\otimes\sigma_k)D_iT^+D_j]\notag\\
&\stackrel{(d)}{=}\sum_{k=1}^N F(\sigma_k)_{ij}.
\label{eq:indef-S6-additivity-calculation}
\end{align}
Here (a) uses the expansion
\eqref{eq:indef-S6-product-expansion} and cancellation
\eqref{eq:indef-S6-cross}; (b) follows from the definition
\eqref{eq:indef-S6-reduced-W}; (c) is
Eq.~\eqref{eq:indef-S6-sigma}; and (d) is the definition
\eqref{eq:indef-S6-information}, with the inverse replaced
by $T^+$ when necessary. This proves the matrix equality
\eqref{eq:indef-S6-additivity}, hence part (1).

\medskip
\noindent\textbf{Step 4 (Product realization).}
Fix a density operator $\sigma$ and choose
\begin{align}
R_\sigma^{(N)}:=(I_B\otimes\sigma)^{\otimes N}.
\label{eq:indef-S6-product}
\end{align}
To establish part (2), we prove that the operators $W_k$
from Eq.~\eqref{eq:indef-S6-reduced-W}, evaluated at
$R=R_\sigma^{(N)}$, satisfy
\begin{align}
W_k=I_B\otimes\sigma\qquad(1\leq k\leq N).
\label{eq:indef-S6-product-W-goal}
\end{align}
For independent CPTP Choi operators $Z_1,\ldots,Z_N$,
\begin{align}
&\Tr\!\left[R_\sigma^{(N)}\bigotimes_{k=1}^NZ_k\right]\notag\\
&\stackrel{(a)}{=}\prod_{k=1}^N\Tr[(I_B\otimes\sigma)Z_k]\notag\\
&\stackrel{(b)}{=}\prod_{k=1}^N\Tr[\sigma\Tr_BZ_k]
=1.
\label{eq:indef-S6-product-normalization}
\end{align}
Here (a) uses tensor-product factorization; (b) uses
$\Tr_BZ_k=I_A$ and $\Tr\sigma=1$. Hence
$R_\sigma^{(N)}\in\mathfrak W_N$ by
Theorem~\ref{thm:indef-N-normalization}.
For this product, Eq.~\eqref{eq:indef-S6-reduced-W} gives
\begin{align}
W_k
&\stackrel{(a)}{=}(I_B\otimes\sigma)
\prod_{\ell\ne k}\Tr[(I_B\otimes\sigma)T]\notag\\
&\stackrel{(b)}{=}I_B\otimes\sigma.
\label{eq:indef-S6-product-W-calculation}
\end{align}
Here (a) takes the partial trace over the other slots;
(b) uses $\Tr[(I_B\otimes\sigma)T]=1$ for CPTP $T$.
This proves Eq.~\eqref{eq:indef-S6-product-W-goal}.
Applying Eq.~\eqref{eq:indef-S6-additivity} with
$\sigma_k=\sigma$ for all $k$ then proves
Eq.~\eqref{eq:indef-S6-product-realization}.
\end{proof}

\subsection{Proofs of the scaling theorem and its consequences}

\begin{proof}[Proof of Theorem~\ref{thm:indef-S6-scaling}]
For $R\in\mathfrak W_N$, part (1) of
Lemma~\ref{lem:indef-S6-slot-additivity} gives
$\sigma_1,\ldots,\sigma_N$. Their average
$\bar\sigma=N^{-1}\sum_k\sigma_k$ is a density operator, and
linearity of $F$ gives
\begin{align}
F^{(N)}(R)&\stackrel{(a)}{=}\sum_{k=1}^NF(\sigma_k)
\stackrel{(b)}{=}NF(\bar\sigma).
\label{eq:indef-S6-theorem-average}
\end{align}
Here (a) is Eq.~\eqref{eq:indef-S6-additivity}, and (b)
uses linearity of $F$. Thus every matrix on the left of
Eq.~\eqref{eq:indef-S6-matrix-set} occurs on its right.
Conversely, Eq.~\eqref{eq:indef-S6-product-realization}
shows that every $NF(\sigma)$ occurs as
$F^{(N)}(R_\sigma^{(N)})$ with
$R_\sigma^{(N)}\in\mathfrak W_N$; this proves the
opposite inclusion.

Since $N>0$, Eq.~\eqref{eq:indef-S6-theorem-average}
also gives $F^{(N)}(R)>0$ exactly when
$F(\bar\sigma)>0$. The matrix-set equality therefore
identifies the admissible domains in
Lemma~\ref{lem:indef-S6-information-reduction}, and
\begin{align}
J_{6,\mathrm{gen}}^{(N)}(G)
&\stackrel{(a)}{=}
\inf_{\substack{\sigma\geq0,\ \Tr\sigma=1\\F(\sigma)>0}}
\Tr\!\left[G(NF(\sigma))^{-1}\right]\notag\\
&\stackrel{(b)}{=}\frac1N
\inf_{\substack{\sigma\geq0,\ \Tr\sigma=1\\F(\sigma)>0}}
\Tr[GF(\sigma)^{-1}]
\stackrel{(c)}{=}\frac{J_{6,\mathrm{gen}}^{(1)}(G)}{N}.
\label{eq:indef-S6-theorem-scaling-proof}
\end{align}
Here (a) combines Eq.~\eqref{eq:indef-S6-matrix-set}
with Lemma~\ref{lem:indef-S6-information-reduction};
(b) uses $(NF)^{-1}=N^{-1}F^{-1}$; and (c) applies the
same reduction with $N=1$. This proves both claims.
\end{proof}

\begin{proof}[Proof of Proposition~\ref{prop:indef-S6-singular-extension}]
The support condition gives
$D_{i,N}=P_ND_{i,N}P_N$ for $P_N=P^{\otimes N}$, and
$T_N^+=(T^+)^{\otimes N}$.  Thus
Lemma~\ref{lem:indef-S6-information-reduction} applies
without changing the optimization over normalization
operators $R\in\mathfrak W_N$.  It yields
\begin{align}
J_{6,\mathrm{gen}}^{(N)}(G)
=\inf_{\substack{R\in\mathfrak W_N\\F^{(N)}(R)>0}}
\Tr\!\left[G F^{(N)}(R)^{-1}\right].
\end{align}
The support identities also meet the assumptions of both
parts of Lemma~\ref{lem:indef-S6-slot-additivity}. Part (1)
gives $F^{(N)}(R)=NF(\bar\sigma)$ for the average of
its one-slot states, and part (2) realizes every $NF(\sigma)$
using $R_\sigma^{(N)}$. These give the two inclusions in
Eq.~\eqref{eq:indef-S6-matrix-set}, with pseudoinverses.
Substituting this set equality into the displayed infimum
and using its $N=1$ version gives
\begin{align}
J_{6,\mathrm{gen}}^{(N)}(G)
&=\inf_{\substack{\sigma\geq0,\ \Tr\sigma=1\\F(\sigma)>0}}
\Tr\!\left[G(NF(\sigma))^{-1}\right]
=\frac{J_{6,\mathrm{gen}}^{(1)}(G)}{N}.
\end{align}
This proves the proposition.
\end{proof}

\section{Proof of the adaptive comb reduction} \label{subsec:proof-adaptive-comb-reduction}
The proof follows the same two directions as the proof of
Theorem~1 of Ref.~\cite{Hayashi2024}.  The forward direction maps a
state-side conic variable for a fixed sensor comb to a process-side
variable.  The reverse direction requires three steps.  First, the
full-rank approximation reduces the process-side optimization to
positive-definite normalization blocks.  Second, canonical adaptive
factorization converts such a block into a sensor comb and an
invertible congruence.  Third, cone-preserving correspondence transfers
the process-side variable to the induced state-side program.  We begin
with the full-rank approximation.

\subsection{Full-rank approximation}

\begin{lemma}[Full-rank approximation]
\label{lem:adaptive-full-rank-approximation}
For $h=1,2,3,4,5$, the value of $J_{h,\mathrm{ad}}^{(N)}(G)$ is unchanged if the optimization
is restricted to operators satisfying the defining cone,
adaptive-normalization, and local-unbiasedness conditions and having a
positive-definite normalization block.  Although $J_{h,\mathrm{ad}}^{(N)}(G)$ is defined by a
minimum over the original domain, the restriction to positive-definite
normalization blocks is expressed as an infimum because the optimum may
lie on the singular boundary.  More precisely,
the relation
\begin{align}
J_{h,\mathrm{ad}}^{(N)}(G)
&=\inf\bigl\{\operatorname{Tr}[(G\otimes\bar T)X]:\notag\\
&\quad X\in {\cal S}_{BA}^h,\quad X_{0,0}>0,\notag\\
&\quad X\text{ satisfies }\eqref{eq:adaptive-X-normalization},\notag\\
&\qquad\qquad\qquad\quad \eqref{c2-G}\bigr\}.
\label{eq:full-rank-restriction}
\end{align}
holds for $h=1,\ldots,4$, and 
the same relation holds by replacing 
${\cal S}_{BA}^h$ by 
${\cal S}_{BA}^5(\bar T)$ for $h=5$.

\end{lemma}

\begin{proof}
The proof regularizes an arbitrary feasible variable while preserving its cone membership, adaptive normalization, and local unbiasedness, and then shows convergence of the objective value.  Let
$X
=
\sum_{r,s=0}^{d}
|r\rangle\langle s|\otimes X_{r,s}$
be arbitrary in the original optimization domain of $J_{h,\mathrm{ad}}^{(N)}(G)$, and set
$R_X:=X_{0,0}$.

We first choose a positive-definite normalization operator satisfying
the adaptive conditions.  Namely, define
$R^\circ
:=
\frac{I_{{\cal K}_B\otimes{\cal K}_A}}{\dim{\cal K}_A}$.
Then $R^\circ>0$ and
\begin{align}
\operatorname{Tr}R^\circ
=
\dim{\cal K}_B.
\label{eq:canonical-full-rank-trace}
\end{align}
Moreover, because every reduced operator of $R^\circ$ is proportional
to the identity, it satisfies
\begin{align}
(R^\circ)^{(k)}
=
\frac{I_{(B,k)}}{d_{B,k}}
\otimes
\operatorname{Tr}_{(B,k)}(R^\circ)^{(k)},
\qquad k=1,\ldots,N.
\label{eq:canonical-full-rank-adaptive}
\end{align}

Each $0<\varepsilon<1$ determines the diagonal operator on ${\cal R}$
\begin{align}
D_\varepsilon
:=
\sqrt{1-\varepsilon}\,|0\rangle\langle0|
+
\frac{1}{\sqrt{1-\varepsilon}}
\sum_{i=1}^{d}|i\rangle\langle i|.
\label{eq:D-epsilon}
\end{align}
We regularize $X$ by
\begin{align}
X_\varepsilon
:=
(D_\varepsilon\otimes I)X(D_\varepsilon\otimes I)
+
\varepsilon
|0\rangle\langle0|\otimes R^\circ.
\label{eq:adaptive-full-rank-regularization}
\end{align}
In block form,
\begin{align}
(X_\varepsilon)_{0,0}
&=
(1-\varepsilon)R_X
+
\varepsilon R^\circ,
\label{eq:regularized-00-block}
\\
(X_\varepsilon)_{0,i}
&=
X_{0,i},
\qquad i=1,\ldots,d,
\label{eq:regularized-0i-block}
\\
(X_\varepsilon)_{i,0}
&=
X_{i,0},
\qquad i=1,\ldots,d,
\label{eq:regularized-i0-block}
\\
(X_\varepsilon)_{i,j}
&=
\frac{1}{1-\varepsilon}X_{i,j},
\qquad i,j=1,\ldots,d.
\label{eq:regularized-ij-block}
\end{align}

We verify in turn the cone condition, adaptive normalization, positive
definiteness of the normalization block, and local unbiasedness.  The
The first term in
Eq.~\eqref{eq:adaptive-full-rank-regularization} is obtained from $X$
by a congruence acting only on ${\cal R}$, and the second term belongs
to the same cone.  For $h=3$, the two required positivity conditions
are preserved separately.  In fact, since $D_\varepsilon$ is real
diagonal in the basis defining partial transpose,
\begin{align}
X_\varepsilon^{\Gamma_{\cal R}}
&=(D_\varepsilon\otimes I)X^{\Gamma_{\cal R}}
(D_\varepsilon\otimes I)
+\varepsilon|0\rangle\langle0|\otimes R^\circ\geq0.
\label{eq:adaptive-full-rank-PPT}
\end{align}
Together with $X_\varepsilon\geq0$, this proves the PPT condition.
Thus $X_\varepsilon$ belongs to ${\cal S}_{BA}^h$ for
$h=1,2,3,4$.  For $h=5$, positivity is preserved, and the additional
antisymmetric constraints remain valid because all parameter-parameter
blocks are multiplied by the same factor $(1-\varepsilon)^{-1}$.

Equation~\eqref{eq:regularized-00-block} gives the normalization block
\begin{align}
R_\varepsilon
:=
(X_\varepsilon)_{0,0}
=
(1-\varepsilon)R_X
+
\varepsilon R^\circ.
\label{eq:regularized-normalization-block}
\end{align}
Since the adaptive normalization conditions and the total-trace
condition are affine, Eqs.~\eqref{eq:adaptive-causal-normalization},
\eqref{AD}, \eqref{eq:canonical-full-rank-trace}, and
\eqref{eq:canonical-full-rank-adaptive} imply
the condition \eqref{eq:adaptive-X-normalization}.
Furthermore,
we have
$R_\varepsilon
\geq
\varepsilon R^\circ
>0$.
Thus, the normalization block of $X_\varepsilon$ is positive definite.
Therefore, $X_\varepsilon$ belongs to the restricted optimization
domain on the right-hand side of
Eq.~\eqref{eq:full-rank-restriction}.

The local-unbiasedness constraints depend only on the $(0,i)$ and
$(i,0)$ blocks.  Equations~\eqref{eq:regularized-0i-block} and
\eqref{eq:regularized-i0-block} therefore give
$
\operatorname{Tr}[W_{i,j}X_\varepsilon]
=
\operatorname{Tr}[W_{i,j}X]
=
\delta_{i,j}$, which implies the condition \eqref{c2-G}.
Finally, the objective depends only on the parameter-parameter blocks.
Eq.~\eqref{eq:regularized-ij-block} implies
$\operatorname{Tr}[(G\otimes\bar T)X_\varepsilon]
=
\frac{1}{1-\varepsilon}
\operatorname{Tr}[(G\otimes\bar T)X]$, which yields
$\lim_{\varepsilon\downarrow0}
\operatorname{Tr}[(G\otimes\bar T)X_\varepsilon]
=
\operatorname{Tr}[(G\otimes\bar T)X]$.
Since $X$ was arbitrary in the original optimization domain of $J_{h,\mathrm{ad}}^{(N)}(G)$
and the objective values of $X_\varepsilon$ converge to that of $X$,
Eq.~\eqref{eq:full-rank-restriction} follows.
\end{proof}

We next convert a positive-definite process-side normalization block into a sensor comb. This construction is needed for the reverse direction of Theorem~\ref{thm:adaptive-comb-reduction}: starting from a process-side variable with normalization block $R$, we must identify an adaptive sensor comb $L_R$ whose induced state model can be related to the original process-side model. The following lemma constructs such a comb together with an invertible map $V_R$ that represents the link product as an ordinary congruence. The following correspondence lemma then uses this map to transfer the conic variable and all its constraints between the process and state sides. Constructing $L_R$ alone does not establish equality of the conic optima: one must also verify that the induced maps preserve the cone, local-unbiasedness equations, and objective values in both directions.

\subsection{Canonical adaptive factorization}

\begin{lemma}[Canonical adaptive factorization]
\label{lem:canonical-adaptive-factorization}
Let $R>0$ satisfy Eqs.~\eqref{eq:adaptive-causal-normalization}
and \eqref{AD}.  Set
$L_R':=d_{B,N}^{-1}\operatorname{Tr}_{B,N}R$.  Then
\begin{align}
R=I_{B,N}\otimes L_R'.
\label{eq:R-final-factorization}
\end{align}
Let
${\cal H}_{\rm pre}:=(\bigotimes_{k=1}^{N}{\cal H}_{A,k})
\otimes(\bigotimes_{k=1}^{N-1}{\cal H}_{B,k})$,
let ${\cal H}_{\rm aux}$ be a copy of ${\cal H}_{\rm pre}$, and let
$J:{\cal H}_{\rm aux}\to{\cal H}_{\rm pre}$ be the basis-identification
unitary.  Using the vectorization convention
$|A\rangle\rangle:=\sum_\mu A|e_\mu\rangle\otimes|\bar e_\mu\rangle$,
define
\begin{align}
B_R:=\sqrt{(L_R')^{T_{\rm pre}}},
\qquad
L_R:=|B_R\rangle\rangle\langle\langle B_R|.
\label{eq:BR-definition}
\end{align}
Then $L_R$ is a sensor comb satisfying Eqs.~\eqref{AD0T} and
\eqref{ADT}.  Moreover,
$V_R:=I_{B,N}\otimes\overline{B_R}J$ is an invertible map from
${\cal H}_{B,N}\otimes{\cal H}_{\rm aux}$ onto
${\cal K}_B\otimes{\cal K}_A$ and satisfies
\begin{align}
V_RV_R^\dagger=R,
\qquad
V_R^\dagger\Lambda V_R=\Lambda*L_R
\label{eq:VR-link-identity}
\end{align}
for every operator $\Lambda$ on ${\cal K}_B\otimes{\cal K}_A$.
\end{lemma}

\begin{proof}
The proof first verifies that $L_R$ satisfies the sensor-comb normalization and then establishes the two identities in Eq.~\eqref{eq:VR-link-identity}.  The factorization \eqref{eq:R-final-factorization} follows from the
$k=N$ instance of
Eq.~\eqref{eq:adaptive-causal-normalization}.  The remaining recursive
conditions imply that $L_R'$ satisfies the deterministic normalization
conditions for the first $N-1$ stages: the identities for
$k=1,\ldots,N-1$ descend directly after the final output factor has
been removed.  Since these conditions contain
only partial traces and identities in the fixed product basis, they
are invariant under $T_{\rm pre}$.  Hence $(L_R')^{T_{\rm pre}}$
satisfies the same deterministic normalization conditions.  Using
\begin{align}
\operatorname{Tr}_{\rm aux}
|B_R\rangle\rangle\langle\langle B_R|
=
B_RB_R^\dagger
=
(L_R')^{T_{\rm pre}},
\end{align}
we obtain Eqs.~\eqref{AD0T} and \eqref{ADT} for $L_R$.

Under the link-product convention
\begin{align}
\Lambda*L_R
:=
\operatorname{Tr}_{\rm pre}
\left[
(\Lambda^{T_{\rm pre}}\otimes I_{\rm aux})
(I_{B,N}\otimes L_R)
\right],
\label{eq:link-product-convention-factorization}
\end{align}
a direct matrix-element calculation gives
$\Lambda*L_R
=
V_R^\dagger\Lambda V_R$.
Furthermore, we have
\begin{align}
&V_RV_R^\dagger
=
I_{B,N}\otimes
\overline{B_R}B_R^T
=
I_{B,N}\otimes
\overline{B_RB_R^\dagger}
\notag\\
=&
I_{B,N}\otimes
\overline{(L_R')^{T_{\rm pre}}}
=
I_{B,N}\otimes L_R'
=R,
\end{align}
where the penultimate equality uses the Hermiticity of $L_R'$.
Together with $\Lambda*L_R=V_R^\dagger\Lambda V_R$, this proves Eq.~\eqref{eq:VR-link-identity} and completes the proof.
\end{proof}

The preceding lemma constructs, from a positive-definite adaptive normalization block $R$, a comb $L_R$ and an invertible map $V_R$ satisfying Eq.~\eqref{eq:VR-link-identity}. To use this construction in the reverse direction of Theorem~\ref{thm:adaptive-comb-reduction}, we must show that congruence by $V_R$ transfers the state-side conic program to the process-side conic program without changing its normalization, local-unbiasedness conditions, objective value, or cone membership. The following lemma establishes this correspondence.

\subsection{Cone-preserving correspondence}

\begin{lemma}[Cone-preserving correspondence]
\label{lem:adaptive-cone-correspondence}
Let $R>0$ satisfy the adaptive normalization conditions, and let
$L_R$ and $V_R$ be defined by
Lemma~\ref{lem:canonical-adaptive-factorization}.  For an operator
$Y
=
\sum_{r,s=0}^{d}
|r\rangle\langle s|\otimes Y_{r,s}$
on
${\cal R}_C\otimes{\cal H}_{B,N}\otimes{\cal H}_{\rm aux}$, define
\begin{align}
{\cal V}_R(Y)
:=
(I_{\cal R}\otimes V_R)
Y
(I_{\cal R}\otimes V_R^\dagger).
\label{eq:forward-congruence}
\end{align}
The strict positivity of $R$ makes $V_R$ invertible.  For an operator $X$ on
${\cal R}_C\otimes{\cal K}_B\otimes{\cal K}_A$, define
\begin{align}
{\cal W}_R(X)
:=
(I_{\cal R}\otimes V_R^{-1})
X
(I_{\cal R}\otimes V_R^{-\dagger}).
\label{eq:reverse-congruence}
\end{align}
Then ${\cal V}_R$ and ${\cal W}_R$ are mutually inverse.  They give a
bijection between state-side variables normalized by
$Y_{0,0}=I$ and process-side variables normalized by $X_{0,0}=R$.
They preserve the local-unbiasedness constraints, the objective value,
and each of the cones used for $h=1,2,3,4,5$.
\end{lemma}

\begin{proof}
~\par

\noindent{\bf Step 1: Preservation of the affine constraints and objective.}  
This step proves that the two congruence maps identify the normalization blocks, local-unbiasedness constraints, and objective values on the state and process sides.

The inverse relations follow immediately from the invertibility of
$V_R$.  
The normalization-block identity
$X_{0,0}=
V_RY_{0,0}V_R^\dagger$ gives
\begin{align}
Y_{0,0}=I
\quad\Longleftrightarrow\quad
X_{0,0}=V_RV_R^\dagger=R.
\label{eq:normalization-correspondence}
\end{align}

Trace cyclicity and
Eq.~\eqref{eq:VR-link-identity} give
\begin{align}
&\operatorname{Tr}
\left[
\frac{|0\rangle\langle i|+|i\rangle\langle0|}{2}
\otimes\bar F_j^{(N)}
\;{\cal V}_R(Y)
\right]
\notag\\
&=
\operatorname{Tr}
\left[
\frac{|0\rangle\langle i|+|i\rangle\langle0|}{2}
\otimes
(V_R^\dagger\bar F_j^{(N)}V_R)
\;Y
\right]
\notag\\
&=
\operatorname{Tr}
\left[
\frac{|0\rangle\langle i|+|i\rangle\langle0|}{2}
\otimes
(\bar F_j^{(N)}*L_R)
\;Y
\right].
\label{eq:LU-correspondence}
\end{align}
Thus, the state-side and process-side local-unbiasedness constraints
are equivalent.

Likewise,
\begin{align}
\operatorname{Tr}
\left[
(G\otimes\bar T){\cal V}_R(Y)
\right]
&=
\operatorname{Tr}
\left[
(G\otimes V_R^\dagger\bar T V_R)Y
\right]
\notag\\
&=
\operatorname{Tr}
\left[
(G\otimes(\bar T*L_R))Y
\right].
\label{eq:objective-correspondence}
\end{align}
Hence, the objective value is preserved in both directions.

It remains to verify the cone conditions.  The transpose
$T_{\rm pre}$ appearing in the link product has already been absorbed
into $B_R$ in Eq.~\eqref{eq:BR-definition}.  Therefore, neither
${\cal V}_R$ nor ${\cal W}_R$ applies a transpose to $X$ or $Y$.
Both maps are ordinary congruences acting only on the physical factor.

\noindent{\bf Step 2: Preservation of the five cone conditions.}  
This step proves, separately for $h=1,2,3,4,5$, that both congruence maps preserve the relevant cone and hence establish a bijection between the state-side and process-side feasible sets.  The case $h=1$ concerns the separable cone.  Suppose that
$Y=\sum_\alpha A_\alpha\otimes Q_\alpha$, where
$A_\alpha\geq0$ and $Q_\alpha\geq0$.  Congruence preserves
positive semidefiniteness, each
$V_RQ_\alpha V_R^\dagger$ is positive semidefinite, and hence
${\cal V}_R(Y)=\sum_\alpha A_\alpha\otimes
V_RQ_\alpha V_R^\dagger$ is separable.  Conversely, a separable
operator $X=\sum_\alpha A_\alpha\otimes P_\alpha$ satisfies
${\cal W}_R(X)=\sum_\alpha A_\alpha\otimes
V_R^{-1}P_\alpha V_R^{-\dagger}$.  Every
$V_R^{-1}P_\alpha V_R^{-\dagger}$ is positive semidefinite.  Therefore,
${\cal W}_R(X)$ is separable.  Thus, the separable cone is preserved
in both directions.

The case $h=2$ additionally requires the real block structure and
vector positivity.  The block identity
$({\cal V}_R(Y))_{r,s}=V_RY_{r,s}V_R^\dagger$ shows that
Hermiticity and the symmetry $Y_{r,s}=Y_{s,r}$ are preserved.  Since
$\langle u|{\cal V}_R(Y)|u\rangle
=\langle (I_{\cal R}\otimes V_R^\dagger)u|Y|
(I_{\cal R}\otimes V_R^\dagger)u\rangle$ for every vector $|u\rangle$
in the full tensor-product space, vector positivity is also preserved.
Conversely, the block identity
$({\cal W}_R(X))_{r,s}=V_R^{-1}X_{r,s}V_R^{-\dagger}$ preserves the
same block conditions, and the analogous quadratic-form identity with
$V_R^{-\dagger}$ preserves vector positivity.  Hence ${\cal V}_R$ and
${\cal W}_R$ preserve the cone for $h=2$.

The case $h=3$ requires both positivity of the operator itself
and positivity after partial transpose on the parameter factor
${\cal R}_C$.  Both congruence maps preserve the first condition.
The map $V_R$ acts only on the physical
factor, so partial transpose commutes with the congruence and gives
$({\cal V}_R(Y))^{\Gamma_{\cal R}}
=(I_{\cal R}\otimes V_R)Y^{\Gamma_{\cal R}}
(I_{\cal R}\otimes V_R^\dagger)$.  This congruence preserves
positivity.  The inverse identity with $V_R^{-1}$ then shows that positivity of
$({\cal V}_R(Y))^{\Gamma_{\cal R}}$ also implies positivity of
$Y^{\Gamma_{\cal R}}$.  Thus, the PPT condition is preserved in both
directions.  The transpose $T_{\rm pre}$ used in the link product does
not enter this argument.

The case $h=4$ requires positive semidefiniteness and the block
conditions defining ${\cal B}''$.  Congruence by $V_R$ preserves
positive semidefiniteness and these block conditions.  Since the
inverse map is the congruence by $V_R^{-1}$, it preserves the same
conditions in the reverse direction.

The case $h=5$ has the same positivity and block conditions as
$h=4$, together with the antisymmetric trace constraints.  These
constraints satisfy
\begin{align}
&\operatorname{Tr}\left[(X_{i,j}-X_{j,i})\bar T\right]
\notag\\
&=\operatorname{Tr}\left[(Y_{i,j}-Y_{j,i})
V_R^\dagger\bar T V_R\right]
=\operatorname{Tr}\left[(Y_{i,j}-Y_{j,i})(\bar T*L_R)\right]
\label{eq:HN-constraint-correspondence}
\end{align}
whenever $X={\cal V}_R(Y)$.  Hence a state-side zero constraint implies
the corresponding process-side zero constraint.  Every
process-side operator satisfies $X={\cal V}_R({\cal W}_R(X))$.  The same
identity applied to $Y={\cal W}_R(X)$ therefore proves the reverse implication.
Therefore, ${\cal S}^5(\bar T*L_R)$ and
${\cal S}_{BA}^5(\bar T)$ correspond bijectively.  This completes the
proof for all five cones.
\end{proof}

The preceding lemmas provide the two constructions needed for Eq.~\eqref{thm-rhs}.  The forward direction uses the adjoint link map to send a state-side feasible point for a fixed sensor comb to the process side.  The reverse direction uses full-rank approximation, canonical factorization, and inverse congruence to send a process-side feasible point to the state side.  The proof below carries out these constructions and then combines the resulting inequalities.

\subsection{Proof of the reduction theorem}

\begin{proof}[Proof of Theorem~\ref{thm:adaptive-comb-reduction}]
~\par

\noindent{\bf Step 1: Map each fixed-comb state-side feasible point to a process-side feasible point.}

To establish the first inequality, a state-side feasible point for a fixed comb must yield a process-side feasible point with the same objective. Fix $L\in Comb$, and let
$Y
=
\sum_{r,s=0}^{d}
|r\rangle\langle s|\otimes Y_{r,s}
$ satisfy the cone, normalization, and local-unbiasedness conditions
used in the definition of
$S_h
\left[
\bar T*L,
\left(
\bar F_j^{(N)}*L
\right)_j
\right]$.

This fixed sensor comb defines the linear map $\Gamma_L$ through
\begin{align}
\operatorname{Tr}
\left[
\Lambda\Gamma_L(Z)
\right]
=
\operatorname{Tr}
\left[
(\Lambda*L)Z
\right]
\label{eq:forward-comb-duality}
\end{align}
for every process operator $\Lambda$ and every operator $Z$ on the
final output system.  Thus, $\Gamma_L$ is the adjoint, with respect to
the trace pairing, of linking a process operator with the fixed comb
$L$.  Apply this map blockwise and define
\begin{align}
X:=
({\rm id}_{\cal R}\otimes\Gamma_L)(Y)
=
\sum_{r,s=0}^{d}
|r\rangle\langle s|
\otimes
\Gamma_L(Y_{r,s}).
\label{eq:forward-proof-X}
\end{align}

The identity $Y_{0,0}=I$ gives the normalization block of $X$ as
$X_{0,0}
=
\Gamma_L(I)$.
For every normalized process Choi operator $\Lambda$,
Eq.~\eqref{eq:forward-comb-duality} gives
$\operatorname{Tr}
\left[
\Lambda X_{0,0}
\right]
=
\operatorname{Tr}
\left[
\Lambda*L
\right]
=
1$.
The normalization identities of the sensor comb therefore imply that
$X$ satisfies
\eqref{eq:adaptive-X-normalization}.

For every $i,j=1,\ldots,d$,
Eq.~\eqref{eq:forward-comb-duality} gives
\begin{align*}
\operatorname{Tr}
\left[
W_{i,j}X
\right]
=
\operatorname{Tr}
\left[
\frac{
|0\rangle\langle i|
+
|i\rangle\langle0|
}{2}
\otimes
\left(
\bar F_j^{(N)}*L
\right)
Y
\right]
=
\delta_{i,j}.
\end{align*}
Thus, $X$ satisfies the process-side local-unbiasedness conditions \eqref{c2-G}.

To verify complete positivity explicitly, let ${\cal H}_{\rm c}$
be the tensor product of the connected slot spaces, namely all input
spaces and all output spaces except ${\cal H}_{B,N}$.  The positive
comb operator $L$ acts on ${\cal H}_{\rm c}\otimes{\cal H}_{\rm aux}$;
write $L=\sum_\alpha|\ell_\alpha\rangle\langle\ell_\alpha|$.
In the bases fixed for the link product, reshape each vector into
$V_\alpha:{\cal H}_{\rm c}\to{\cal H}_{\rm aux}$ by
$\langle\mu|V_\alpha|a\rangle
:=\langle a,\mu|\ell_\alpha\rangle$, and set
$K_\alpha=I_{B,N}\otimes V_\alpha$, with the slot factors put in
their prescribed order.  Contracting the connected indices in
Eq.~\eqref{eq:adaptive-output-state}, followed by the trace duality
\eqref{eq:forward-comb-duality}, gives
\begin{align}
\Lambda*L&=\sum_\alpha K_\alpha\Lambda K_\alpha^\dagger,
\notag\\
\Gamma_L(Z)&=\sum_\alpha K_\alpha^\dagger ZK_\alpha.
\label{eq:forward-comb-Kraus}
\end{align}
Thus $\Gamma_L$ is completely positive.  Its blockwise extension
preserves the separable cone for $h=1$, the real block structure and
positive semidefiniteness for $h=2$, both the original and
partial-transpose positivity conditions for $h=3$, and
positive semidefiniteness for $h=4$.
For $h=5$, Eq.~\eqref{eq:forward-comb-duality} also preserves the
additional antisymmetric trace constraints.  Hence, $X$ belongs to
the process-side cone used in the definition of $J_{h,\mathrm{ad}}^{(N)}(G)$.

Finally, write $G=\sum_{r,s=1}^{d}g_{r,s}|r\rangle\langle s|$. Using the blockwise definition of $X$ in Eq.~\eqref{eq:forward-proof-X}, followed by the defining duality relation \eqref{eq:forward-comb-duality} with $\Lambda=\bar T$ and $Z=Y_{r,s}$, we obtain 
\begin{align} 
&\operatorname{Tr}[(G\otimes\bar T)X] = \sum_{r,s=1}^{d} g_{s,r} \operatorname{Tr} \left[ \bar T\Gamma_L(Y_{r,s}) \right] 
\notag\\
&= \sum_{r,s=1}^{d} g_{s,r} \operatorname{Tr} \left[ (\bar T*L)Y_{r,s} \right] 
= \operatorname{Tr} \left[ \left( G\otimes(\bar T*L) \right)Y \right]. \label{eq:forward-proof-objective} 
\end{align}
It follows that
\begin{align}
J_{h,\mathrm{ad}}^{(N)}(G)
\leq
S_h
\left[
\bar T*L,
\left(
\bar F_j^{(N)}*L
\right)_j
\right]
\end{align}
for every $L\in Comb$.  Taking the infimum over $L$ gives
\begin{align}
J_{h,\mathrm{ad}}^{(N)}(G)
\leq
\inf_{L\in Comb}
S_h
\left[
\bar T*L,
\left(
\bar F_j^{(N)}*L
\right)_j
\right].
\label{eq:adaptive-upper-direction}
\end{align}
This completes the forward direction.

\noindent{\bf Step 2: Map each full-rank process-side feasible point to an induced state-side feasible point.}

This step maps every full-rank process-side feasible point $X$ to a sensor comb $L_{R_X}$ and a feasible point $Y$ of the induced state-side program with the same objective value.  Comparing the state-side infimum with the objective of $X$ and then optimizing over $X$ will give the required lower bound on $J_{h,\mathrm{ad}}^{(N)}(G)$.  By
Lemma~\ref{lem:adaptive-full-rank-approximation}, we may restrict the
process-side optimization to variables with positive-definite
normalization blocks.  Let
\begin{align}
X
=
\sum_{r,s=0}^{d}
|r\rangle\langle s|\otimes X_{r,s}
\label{eq:reverse-proof-X}
\end{align}
satisfy the process-side cone, adaptive-normalization, and
local-unbiasedness conditions, and set
$R_X
:=
X_{0,0}
>
0$.
Lemma~\ref{lem:canonical-adaptive-factorization} associates with
$R_X$ a sensor comb $L_{R_X}$ and an invertible operator $V_{R_X}$
such that
\begin{align}
V_{R_X}V_{R_X}^{\dagger}
&=
R_X,
\label{eq:reverse-proof-normalization}
\\
V_{R_X}^{\dagger}\Lambda V_{R_X}
&=
\Lambda*L_{R_X}
\label{eq:reverse-proof-link}
\end{align}
for every process operator $\Lambda$.

Define the state-side operator by the inverse congruence
\begin{align}
Y
:=
{\cal W}_{R_X}(X)
:=
\left(
I_{\cal R}\otimes V_{R_X}^{-1}
\right)
X
\left(
I_{\cal R}\otimes V_{R_X}^{-\dagger}
\right).
\label{eq:reverse-proof-Y}
\end{align}
Writing
$Y
=
\sum_{r,s=0}^{d}
|r\rangle\langle s|\otimes Y_{r,s}$,
its blocks are explicitly
\begin{align}
Y_{r,s}
=
V_{R_X}^{-1}
X_{r,s}
V_{R_X}^{-\dagger},
\qquad
r,s=0,\ldots,d.
\label{eq:reverse-proof-blocks}
\end{align}
In particular, Eq.~\eqref{eq:reverse-proof-normalization} yields
\begin{align}
Y_{0,0}
=
V_{R_X}^{-1}
R_X
V_{R_X}^{-\dagger}
=
V_{R_X}^{-1}
V_{R_X}V_{R_X}^{\dagger}
V_{R_X}^{-\dagger}
=
I.
\label{eq:reverse-proof-state-normalization}
\end{align}
Hence, $Y$ satisfies the state-side normalization constraint.

For every $i,j=1,\ldots,d$, cyclicity of the trace and
Eq.~\eqref{eq:reverse-proof-link} give
\begin{align}
&\operatorname{Tr}
\left[
\frac{
|0\rangle\langle i|
+
|i\rangle\langle0|
}{2}
\otimes
\left(
\bar F_j^{(N)}*L_{R_X}
\right)
Y
\right]
\notag\\
&=
\operatorname{Tr}
\left[
\frac{
|0\rangle\langle i|
+
|i\rangle\langle0|
}{2}
\otimes
\bar F_j^{(N)}
X
\right]
=
\delta_{i,j}.
\label{eq:reverse-proof-local-unbiasedness}
\end{align}
Thus, $Y$ satisfies the state-side local-unbiasedness constraints
\eqref{c2-G}.

Lemma~\ref{lem:adaptive-cone-correspondence} shows that the inverse
congruence in Eq.~\eqref{eq:reverse-proof-Y} preserves the relevant
cone condition.  More precisely,
\begin{align}
Y
&\in
{\cal S}^{h},
\qquad
h=1,2,3,4,
\label{eq:reverse-proof-cones-1234}
\end{align}
while, for $h=5$,
\begin{align}
Y
\in
{\cal S}^{5}
\left(
\bar T*L_{R_X}
\right).
\label{eq:reverse-proof-cone-5}
\end{align}
Thus, $Y$ belongs to the cone used in the definition of the
state-side conic program.

Equations~\eqref{eq:reverse-proof-state-normalization},
\eqref{eq:reverse-proof-local-unbiasedness},
\eqref{eq:reverse-proof-cones-1234}, and
\eqref{eq:reverse-proof-cone-5} show that $Y$ satisfies all the
constraints in the state-side conic program, which implies
\begin{align}
&S_h
\left[
\bar T*L_{R_X},
\left(
\bar F_j^{(N)}*L_{R_X}
\right)_j
\right]
\stackrel{(a)}{\leq}
\operatorname{Tr}
\left[
\left(
G\otimes
\left(
\bar T*L_{R_X}
\right)
\right)
Y
\right] \label{BNE}.
\end{align}
This particular choice of $L_{R_X}$ satisfies 
\begin{align}
&\inf_{L\in Comb}
S_h
\left[
\bar T*L,
\left(
\bar F_j^{(N)}*L
\right)_j
\right]
\notag \\
\le &S_h
\left[
\bar T*L_{R_X},
\left(
\bar F_j^{(N)}*L_{R_X}
\right)_j
\right]
\stackrel{(a)}{\leq}
\operatorname{Tr}
\left[
\left(
G\otimes
\left(
\bar T*L_{R_X}
\right)
\right)
Y
\right] \notag \\
\stackrel{(b)}{=}&
\operatorname{Tr}
\left[
\left(
G\otimes
V_{R_X}^{\dagger}\bar T V_{R_X}
\right)
Y
\right]
\notag\\
=&
\operatorname{Tr}
\left[
(G\otimes\bar T)
\left(
I_{\cal R}\otimes V_{R_X}
\right)
Y
\left(
I_{\cal R}\otimes V_{R_X}^{\dagger}
\right)
\right]
\stackrel{(c)}{=}
\operatorname{Tr}
\left[
(G\otimes\bar T)X
\right],
\label{eq:reverse-proof-pointwise-bound}
\end{align}
where 
$(a)$, $(b)$, and $(c)$ follow from 
Eq.~\eqref{BNE},
Eq.~\eqref{eq:reverse-proof-link}, 
and Eq.~\eqref{eq:reverse-proof-Y}.

Taking the infimum of the right-hand side over all process-side
operators $X$ with positive-definite normalization blocks, and then
using Lemma~\ref{lem:adaptive-full-rank-approximation}, yields
\begin{align}
J_{h,\mathrm{ad}}^{(N)}(G) \ge
\inf_{L\in Comb}
S_h
\left[
\bar T*L,
\left(
\bar F_j^{(N)}*L
\right)_j
\right].
\label{eq:adaptive-lower-direction}
\end{align}
This completes the reverse direction.

\noindent{\bf Step 3: Combine the two inequalities.}

Combining Eqs.~\eqref{eq:adaptive-upper-direction} and
\eqref{eq:adaptive-lower-direction}, we conclude that
\begin{align}
J_{h,\mathrm{ad}}^{(N)}(G)
=
\inf_{L\in Comb}
S_h
\left[
\bar T*L,
\left(
\bar F_j^{(N)}*L
\right)_j
\right]
\end{align}
for $h=1,\ldots,5$, which proves Eq.~\eqref{thm-rhs}.
\end{proof}

\bibliography{ref3}

\end{document}